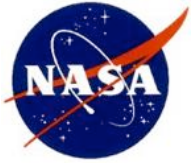

# Projected Near-Earth Object Discovery Performance of the Large Synoptic Survey Telescope

*Center for Near-Earth Object Studies*

*Steven R. Chesley and Peter Vereš*

**April 2017**

# Contents

## EXECUTIVE SUMMARY

This report describes the methodology and results of an assessment study of the performance of the Large Synoptic Survey Telescope (LSST) in its planned efforts to detect and catalog near-Earth objects (NEOs).

LSST is a major, joint effort of the US National Science Foundation and the Department of Energy, with significant support from private donors. The project has a number of key science goals, and among them is the objective of cataloging the solar system, including NEOs. LSST is designed for rapid, wide-field, faint surveying of the night sky, and thus has an 8.4 m primary mirror, with 3.2 Gigapixels covering a $9.5\,\mathrm{deg}^2$ field of view. The system is projected to reach a faint limit of $V \simeq 25$ in a 30-second exposure visit to a given field and perform nearly 2.5 million visits in its 10-year survey.

The baseline LSST survey approach is designed to make two visits to a given field in a given night, leading to two possible NEO detections per night. These nightly pairs must be linked across nights to derive orbits of moving objects. However, the presence of false detections in the data stream leads to the possibility of high rates of false tracklets, and the ensuing risk that the resulting orbit catalog may be contaminated by false orbits. NEO surveys to date have successfully eliminated this risk by making 3–5 visits per night to obtain confirming detections so that the single-night string of detections has a high reliability. The traditional approach is robust, at the expense of reduced sky coverage and a diminished discovery rate. The baseline LSST approach, in contrast, is potentially fragile to large numbers of false detections, but maximizes the survey performance.

One of our key objectives was to investigate this fragility by conducting high-fidelity linkage tests on a full-density simulated LSST detection stream. We also sought to quantify the overall performance of LSST as an NEO discovery system, under the hypothesis that the NEO detections arising from the baseline LSST survey observing cadence can be successfully linked.

We used the latest instantiation of the LSST baseline survey and the most current NEO population model to derive the fraction of NEOs detected and cataloged by LSST from among the source population. As a part of this we developed a high-fidelity detection model that accurately represented the LSST focal plane and implemented a smooth degradation in detection efficiency near the limiting magnitude, rather than the usual step function. The study carefully modeled losses from trailed detections associated with fast moving objects, and we investigated other minor effects, such as telescope vignetting, and asteroid colors and light curves.

For the linking tests, we included all major sources of detections for a single selected observing cycle (full moon to full moon), leading to 66 million detections, of which 77% were false detections, 23% were main-belt asteroids, and only 0.14% were NEOs. Using only a single 8-core workstation we were able to successfully link 94% of potential NEO discoveries in the detection stream. We are confident that with appropriately-sized computation resources, some algorithmic improvements and careful tuning of the linking algorithms the linking efficiency can be significantly improved. In this simulation, 96% of objects in the NEO catalog were correctly linked. Of the 4% that involved erroneous linkages, almost all comprised detections of two distinct main-belt asteroids. This situation, known as "main-belt confusion" is fundamental to the NEO search problem and is readily resolved as the main-belt asteroid catalog becomes filled in over time. Despite the 77% rate of false detections, less than 0.1% of derived NEOs in this simulation included false detections. From these results, within the study hypotheses, we conclude that the two visits-per-night observation cadence can be successful in cataloging NEOs. This conclusion does assume a certain rate of false positives, but is unlikely to be sensitive to increases by factors of a few in the false detection rate, given the significant computational resources allocated by LSST for the problem.

Our simulations revealed that in 10 years LSST would catalog $\sim 60\%$ of NEOs with absolute magnitude $H < 22$, which is a proxy for 140 m and larger objects. This results neglects linking losses and the contribution of any other NEO surveys. Including our worst-case linking efficiency we reach a overall performance assessment of 55% completeness of NEOs with $H < 22$. We estimate that survey mis-modeling could account for systematic errors of up to 5%. We find that restricting the evaluation metric to so-called Potentially Hazardous Asteroids (PHAs) increases the completeness by 3–4%, and that including the benefits of past and expected future NEO survey activity increases completeness at the end of the baseline LSST survey by 15-20%. Assembling these results leads to a projection that by the end of the baseline LSST survey the NEO catalog will be $80 \pm 5\%$ complete for PHAs with $H < 22$.

As described in detail in the report, these results are largely consistent with other results obtained independently; indeed the small 1–2% variation among independent estimates of NEO completeness is remarkable and reassuring. The results above require pairs of observations in three distinct nights over no more than 12 days. A maximum linking interval of 20 days, for which high linking efficiency has not been demonstrated, leads to a 2–3% improvement in completeness. We also tested a special-purpose LSST cadence designed to enhance the NEO discovery rate, but our results show little improvement over the baseline for a ten-year survey. Surveying longer does provide an increase in completeness, by roughly 2% per year for a lone LSST and 1% per year when including contributions from other surveys. Thus, in our judgement, the $H < 22$ PHA catalog can be expected to approach $\sim 85\%$ completeness, but not 90%, after 12 years of LSST operation.

## 1. INTRODUCTION

The Large Synoptic Survey Telescope (LSST, Figure 1) is a next generation all-sky survey telescope currently being constructed atop Cerro Pachón in Chile. LSST is funded by the National Science Foundation, the department of Energy and private contributors. After expected first light in 2020 and two years of commissioning, it will start its mission of systematic surveying of the entire accessible sky.

The main features of the telescope are an 8.4m primary mirror, wide-field optimized optical pathway, 3.2 Gigapixel camera, alt-azimuthal mount, dome, and data storage, distribution and processing facilities. LSST will provide 10 million alerts and 15 Terabytes of data per night, almost in real time. A single 30-second exposure with a 9.5 square degree field of view will achieve 24.5 magnitude depth in r-band (Table 1). In contrast to comparably large telescopes that use multiple instruments and focus on a single target for hours, LSST will scan ∼ 6,000 square degrees per night.

The main science goal of LSST is the study and understanding of Dark Matter and Dark Energy achieved through mapping and measuring of weak gravitational lensing. LSST will also detect an unprecedented number of transient optical events such as novae and supernovae, will map the stars of the Milky Way and inventory the Solar System—from nearby asteroids to distant comets and trans-neptunian objects (TNOs). LSST has been projected to discover more than 5 million main-belt asteroids (MBAs), 100,000 near-Earth objects (NEOs) and 40,000 TNOs and increase the number of known Solar System objects by an order of magnitude (Jones et al. 2016). In total, LSST will produce 10 million time-domain events per night and will transmit the events within 60 seconds of observation. Its final catalog will contain 37 billion objects, 7 trillion observations and 30 trillion measurements and will require 100 teraflops computing power requirement and 15 petabytes of storage. Because of the immensity of the data set, LSST must work in an automated regime for identification, linking and orbit determination of Solar System objects.

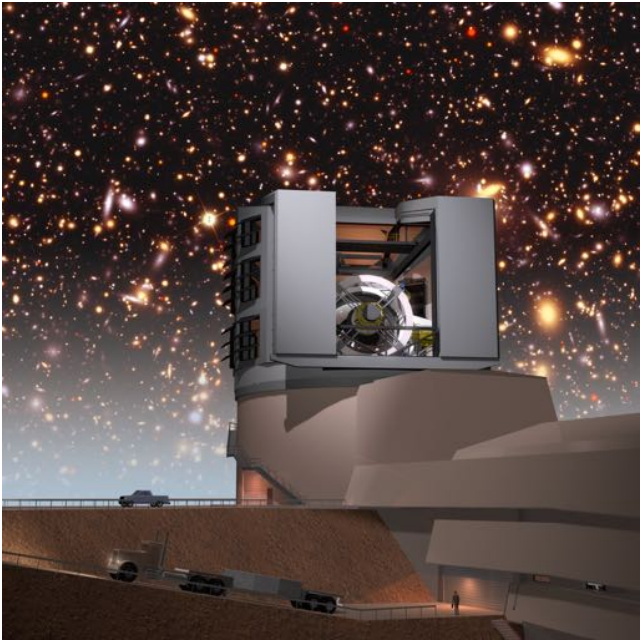

**Figure 1.** 3D artistic rendering of the LSST telescope with deep sky background. (Image credit: LSST)

**Table 1.** LSST by the numbers

| Location | Cerro Pachón, Chile |
|---|---|
| Primary mirror | 8.4 m |
| Effective Aperture | 6.67 m |
| Focal Length | 10.3 m |
| Wavelength | 320–1060 nm |
| Filters | u, g, r, i, y, z |
| Mount | Alt-Az |
| Field diameter | 3.5 deg |
| Camera | 3.2 Gigapixel |
| Pixel size | 0.2 arcsec |
| Exposure time | 30 s |
| Start year | 2022 |

Cataloging the NEO population has been a NASA objective for about two decades. In 1998 the Spaceguard goal was set to find 90% of NEOs larger than 1 km within 10 years. After development of multiple dedicated survey telescopes and almost two decades of search, 15,000 NEOs have been found, with about 900 larger than 1 km, and the Spaceguard goal has been met. In light of new technologies and challenges related to mitigation of smaller hazardous asteroids, potential asteroid mining, asteroid sample return, asteroid redirection and crewed missions to NEOs, a Congressional mandate was set for NASA to discover 90% of NEOs larger than 140 m. The current dedicated surveys, though productive, will not approach that goal for many decades. It has been suggested that LSST, as the largest survey telescope in the world, may potentially reach this target; however, a comprehensive and detailed study has been lacking.

The primary objective of this study was to provide an independent, high-fidelity estimate of the performance of LSST for discovery of NEOs. The LSST project has made some estimates of its NEO search performances (Ivezić et al. 2007; Jones et al. 2007, 2015); however, these studies were met with skepticism from some in the NEO community. This is in part because LSST will not be an NEO-optimized survey and also the automated NEO processing pipeline has not been tested in such a large scale environment. In addition to the usual survey modeling approach, where only the NEO detections are investigated for visibility in the field, we conducted a detailed simulation of the linking process—from creating detections in the fields, applying all potential observational and detection constraints, injecting false detections and MBAs, building tracklets, tracks and linking the tracks into derived orbits. A central element of the study was to use realistic numbers of real and false detections, managing false tracklets and linkages and analyzing the performance and accuracy of the pipeline under the full computational load. For the simulations, we employed and modified the Pan-STARRS Moving Object Processing System (MOPS, Denneau et al. 2013), which was originally developed for an LSST-class survey. Even though it has been used by several existing telescopes with slight modification, the pipeline has never been run in all stages with real data, from detections to automated orbit derivation. An LSST version of MOPS is under development.

An important result of the study was the NEO completeness, i.e., the fraction of the NEO population cataloged by LSST. Another key objective was to test whether automated linking and orbit determination would succeed in a realistic scenario. In addition to efficiency, we studied the accuracy of survey products, e.g., false linkages, and the computational cost. Other elements included analysis of uncertainties and detailed analysis of detection models to understand sensitivity in overall detection efficiency.

## 1.1. *Study plan and process overview*

The question of cataloging is tied to a definition of a discovery. Traditionally, an NEO is *discovered* when it is observed on several nights and its orbit can be derived. Then, a provisional designation is announced by the Minor Planet Center for the individual object. This commonly requires coordinated effort and follow-up from multiple observing sites: the discovery site submits a few positions of a NEO candidate and others recover it within a few days. However, the number of LSST NEO candidates announced daily will be dramatically larger than the current load and far beyond the follow-up capabilities of available telescopes. Nevertheless, the LSST cadence will be driven by multiple science objectives and will not be optimized toward NEO discovery. Moreover, one-night LSST NEO candidates will be faint and out of reach of most other telescopes and will often be spurious due to false detections or mis-linked objects. Instead of the traditional reporting of single-night NEO candidates, LSST will rely on its own linking and will report well-determined orbits with detections on multiple nights. Therefore, this study assessed the entire process: from detecting a given NEO on a first night to identifying and connecting multiple detections of the same object in a single night into a tracklet and finally linking multiple tracklets in three distinct nights into an orbit.

A key metric is the NEO catalog completeness obtained by LSST at the end of its operation. Completeness can be derived as the differential completeness at a given size or as integral completeness of all objects larger than a given size. For this study, the primary metric for LSST NEO performance is $C_{H<22}$, the integral NEO discovery completeness for absolute magnitude $H < 22$, which corresponds to approximately 140 m in diameter for a spherical shape and geometric albedo of 0.14 (mean NEO albedo, Stuart & Binzel (2004)). The completeness will be assessed in all stages of the discovery process—for individual detections, single night tracklets and multi-night tracks.

The question of completeness in an idealized case can by answered in a simplified, *low-density* simulation, where only a small fraction of the NEO population is used. There is no background noise or other asteroids that could cause linking confusion. This approach allows a quick assessment of a 10-year survey's completeness with dramatically reduced computation time, but it does not stress the survey system's ability to process potentially large numbers of detections, tracklets, tracks and orbits in the presence of noise and confusion. The low-density simulations thus assume that the linking will be successful. Such low-density simulations are convenient for testing individual observational models and multiple LSST cadences. In addition to studying LSST's NEO completeness when operating alone, the low-density simulations were used to study the NEO discovery rate when existing surveys and LSST are working together, and what fraction of the NEO catalog should already be discovered when LSST starts to operate.

However, our low-density simulations do not address a major challenge of LSST's NEO survey, which is the false detection rate and other types of transient sources in the field that will affect the survey's efficiency and accuracy. The number of simulated detections entering the pipeline must match the expected data load of the real LSST and the pipeline must be able to process detections, create tracklets and tracks and derive orbits. Thus our *high-density* simulations were designed with a realistic density of false detections and MBAs, in addi-tion to NEOs. High-density simulations are computationally expensive, and so we simulated only one month of LSST fields, enough to demonstrate that all stages of the pipeline work operate successfully and to determine the linking efficiency and losses in the process.

This study relied on inputs from LSST, such as the planned observing schedule, detailed properties of the focal plane, limiting magnitudes of fields, magnitude losses due to the fast motion of NEOs, and false detection rates. We also implemented an LSST-specific calculation of signal-to-noise, astrometric and photometric uncertainties and color indices for converting filter magnitudes to a common pass-band. The effects of light curve variation and varying spectral types were taken into account as well. Variations of inputs had to be taken into account as a source of uncertainties. The resulting completeness is presented as a function of $H$, epoch and applied parameters.

## 1.2. *Terminology and definitions*

**Detection:** - An individual detection in a single field, with derived on-sky position (RA, DEC), magnitude, signal-to-noise ratio and location and brightness uncertainties. A synthetic detection is generated from an asteroid ephemeris computed from a provided orbit for a given epoch and field. There are also two types of false detections generated in a field — random noise and image difference artifacts.

**Tracklet:** A set of two or more detections on the same night, aligned along a line or great circle. For our study, the time separation between the first and last detection of a tracklet was allowed to range from 5 minutes to 2 hours. We also required the rate of motion to be in the range (0.05–2.0°/day).

**Track:** A set of three or more tracklets observed on at least three distinct nights. Tracks can be made of any kinds of tracklets. By this definition, the shortest track is 2 days long. The upper limit of track length for this study was 12 or 20 days in low-density simulations and 12 days in high-density simulations.

**Efficiency (Completeness):** $C_{H<H_0}$, the fraction of objects in the source population with $H < H_0$ that are found (in detections, tracklets, tracks or orbits) by the survey. May also be a diameter limited completeness $C_{D>D_0}$.

**Accuracy:** The purity or reliability of the derived catalog (e.g., in detections, tracklets, tracks or orbits).

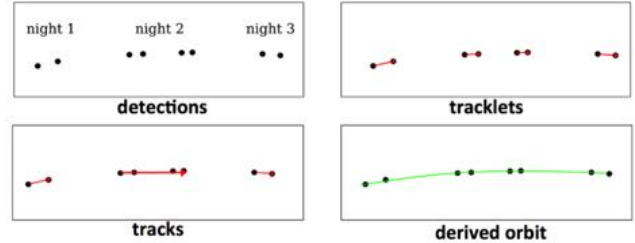

**Figure 2.** Schematic diagram of the detections, tracklets, tracks and derived orbits for a single object observed on 3 nights.

**NEO - Near-Earth Object:** An asteroid or a comet on an orbit with perihelion distance $q < 1.3$ au.

**PHA - Potentially Hazardous Asteroid:** An NEO with MOID $< 0.05$ au and absolute magnitude diameter $H < 22$. For 14% albedo, this is approximately equivalent to $D > 140$ m.

**MOID - Minimum Orbit Intersection Distance:** Minimum separation between two Keplerian orbital ellipses. In the present context, the MOID represents the closest distance that an asteroid can possibly pass to the Earth, irrespective of the timing constraints.

**MOPS - Moving Object Processing System:** A large software package for simulating and processing synthetic and real asteroid data. It works in three main stages: 1) generating synthetic detections for orbits in survey fields (SYNTH stage), 2) generating tracklets (TRACKLET stage) from the detections, and 3) generating tracks and orbits from the tracklets (LINKOD stage).

**MBA - Main Belt Asteroid:** Asteroids with orbits between Mars and Jupiter.

**Low-density simulation:** A computationally less expensive simulation run through only the SYNTH and TRACKLET stages of MOPS, with a small number of orbits (3000) all having $H = 0$. Tracks and size dependent outputs are made in post-processing. Observational constraints can be added in post-processing.

**High-density simulation:** A computationally intensive simulation run through the MOPS SYNTH, TRACKLET and LINKOD stages, with a realistic full-density of synthetic detections, false detections and $H$ distributions, providing detections, tracklets, tracks and derived orbits.

**CLEAN:** A tracklet, track or orbit consisting of detections from a single synthetic object.

**BAD:** A tracklet, track or orbit consisting of at least one false detections.

**MIXED:** A tracklet, track or orbit consisting of detections from two or more synthetic objects.

**NONSYNTH:** A tracklet, track or orbit consisting of entirely of false detections.

**Not CLEAN:** A erroneously linked tracklet, track or orbit that is either BAD, MIXED or NONSYNTH.

### 1.3. *External Collaboration*

This study represents a substantially independent effort by the authors and was conducted at JPL. There was, however, substantial and extensive interaction with LSST project members at the University of Washington (UW) in order to ensure that the JPL study inputs and assumptions conformed to project requirements and plans. This information sharing included validation of selected components of the simulation software.

Separately, the original study plan called for independent processing of JPL-produced data products, detection lists in particular, through collaborations with UW and also with Caltech's IPAC. JPL did deliver detection lists to both external teams, and engaged in some subsequent coordination to ensure that the data were in acceptable form. While few results from this data exchange have been shared to date, we understand that the data have been used nonetheless, and we anticipate sharing of additional data sets in the future. Work is ongoing at IPAC and results will be reported separately.

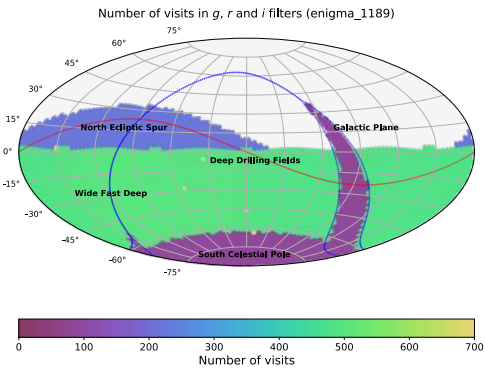

**Figure 3.** Sky map of number of visits with g, r, or i filters for the ten-year `enigma_1189` OpSim survey. The various proposals are evident by the differing number of visits. Note that the Deep Drilling proposal covers only a few fields, each with thousands of visits, although the color scale cuts off at 700 visits. Most of the time is spent on the Wide-Fast-Deep proposal. (Image courtesy of L. Jones, Univ. Washington.)

## 2. MODELING LSST: KEY INPUTS AND ASSUMPTIONS

This section describes the details of several key elements of our LSST simulations. The following subsections describe the field-by-field information assumed for the various LSST surveys, the asteroid population models employed, and the models for the LSST focal plane and asteroid detections. The section closes with a description of our approach for injecting false detections into the LSST data stream.

### 2.1. Survey Fields

We used the outputs of the Operations Simulator (OpSim, Delgado et al. 2014), the LSST-developed application that plans and schedules the field selection over the 10-years of the planned survey. The fields were selected in an optimized process following the science objectives, telescope design, and observational conditions, such as airmass, cloud coverage and sky background variation due to conditions measured at the LSST site. The OpSim output is a list of individual fields with the mid-exposure time, right ascension and declination of the field center, orientation of the field with respect to North, limiting magnitude, filter, airmass, seeing, exposure time, etc. The scheduling is driven by five different "proposals" that drive the survey (Table 2, Figure 3). Most of the time is spent on the so-called Wide-Fast-Deep proposal that covers 18,000 square degrees in the Southern hemisphere and uses approximately 85% of the survey time in a two-visit-per-night cadence (Figure 3). The rest is divided between the South Celestial Pole, the Galactic Plane and the North Ecliptic Spur, where Solar System science is prioritized. A few individual fields are imaged with an extreme visit pattern, the so-called

Deep Drilling Cosmology proposal. Because the areal coverage of the Deep Drilling proposal is negligible with respect to other proposals and its nightly revisit rate is extreme, which could create problems for MOPS, we deleted visits under this proposal from our simulations.

The OpSim observing conditions are derived from historical records from Cerro Tololo and Cerro Pachon. Both seeing and cloud cover are randomly sampled from a distribution that is consistent with the site and season. More details are reported by Delgado et al. (2014).

This study revolved around two distinct 10-year baseline surveys—the earlier `enigma_1189` and the later `minion_1016`. Both surveys have similar field selection, revisit rates (Figure 4), spatial coverage (Figure 5), and limiting magnitude (Figure 6). The `enigma_1189` survey begins on January 1, 1994, and the `minion_1016` survey on January 1, 2022. There are also two variations of the baseline `enigma_1189` survey with a different revisit rate—`enigma_1271` targeted three visits per night and `enigma_1266` sought 4 visits/night. The `minion_1016` survey was improved with a new sky brightness model that led to ∼ 0.25 mag reduction in sensitivity.

Another OpSim survey used in this study was the 15-year `astro_1016` survey (abbreviated from the full simulation designation `astro_lsst_01_1016`), which mostly follows the 10-year `minion_1016` survey, but adds five extra years and prioritizes observations of the Northern Ecliptic Spur. Thus `astro_1016` is enhanced for NEO surveying. Histograms of solar elongations and ecliptic latitude of the field centers show that the opposition area coverage is significantly increased (Figure 5). The `astro_1016` survey also begins on January 1, 2022.

The key feature of the LSST baseline survey is that most of the fields are visited twice per night, with a small fraction observed three or more times per night and a small fraction observed only once per night (singleton fields). Singleton fields are not suitable for NEO search and were neglected in the simulations. Approximately 17% of visits are singletons, mostly observed in u or y-bands. The total number of fields and other basic properties of the individual surveys are listed in Table 3.

For the baseline observing cadence, the final visit of a field on a single night is within 2 hours of the first in 94% of cases, and the median transient time interval (TTI) between consecutive visits is 21 minutes (Figure 4). Each field is observed in a given LSST filter (u,g,r,i,z,y). Table 4 lists the limiting magnitudes for the various filters for the key surveys considered in this study. In the MOPS simulation, magnitudes are computed in V-band and transformed to the appropriate filter for the given field according to color transformations described in a following subsection.

**Table 2.** The distribution of survey time among the science objectives in two 10-year surveys (`enigma_1189`, `minion_1016`) and the NEO-enhanced 15-year survey (`astro_1016`). The `astro_1016` survey allocated more time to the North Ecliptic Spur for the last 5 years.

| Proposal | enigma_1189 | minion_1016 | astro_1016 |
|---|---|---|---|
| Wide-Fast-Deep | 85.4% | 85.1% | 68.8% |
| Northern Ecliptic Spur | 6.4% | 6.5% | 24.4% |
| Deep Drilling | 4.3% | 4.5% | 4.3% |
| South Celestial Pole | 2.1% | 2.2% | 1.4% |
| Galactic Plane | 1.7% | 1.7% | 1.1% |

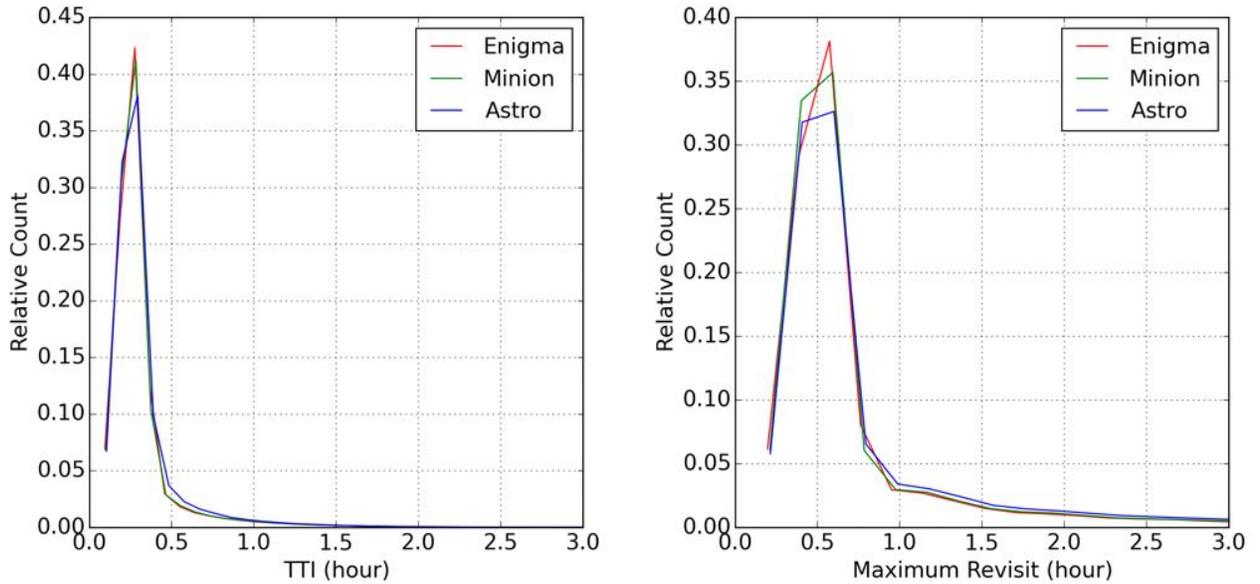

**Figure 4.** Mean time between individual visits of the same field per night (left - transient time interval, TTI) and time span of the revisit between first and last visit of the field in a night (right). Does not include singletons and Deep Drilling.

**Table 3.** Comparison of simulated LSST surveys. Used fields neglect singletons and Deep Drilling.

| Survey name | Visits | Duration Years | Fields total | Used fields | Mean limiting mag V | Effective sky coverage Number of footprints |
|---|---|---|---|---|---|---|
| enigma_1189 | 2 | 10 | 2,469,307 | 1,943,901 | 23.63 | 805,128 |
| minion_1016 | 2 | 10 | 2,447,931 | 1,914,170 | 23.40 | 789,564 |
| astro_1016 | 2 | 15 | 3,725,498 | 2,995,827 | 23.37 | 1,130,394 |
| enigma_1271 | 3 | 10 | 2,438,990 | 1,932,391 | 23.65 | 463,636 |
| enigma_1266 | 4 | 10 | 2,417,999 | 1,928,921 | 23.65 | 309,154 |

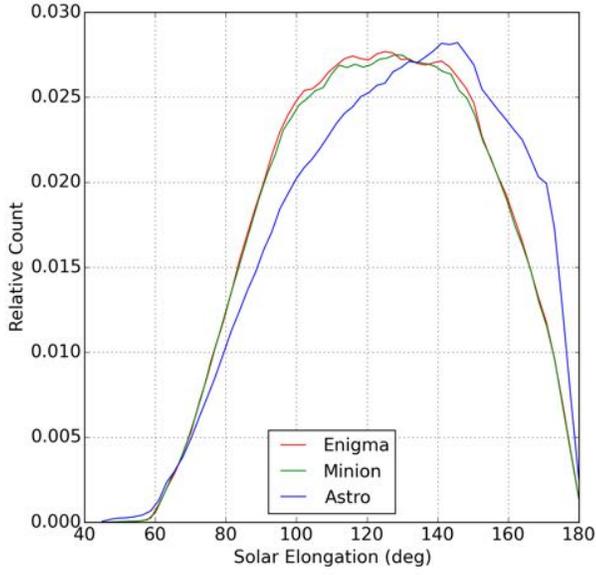

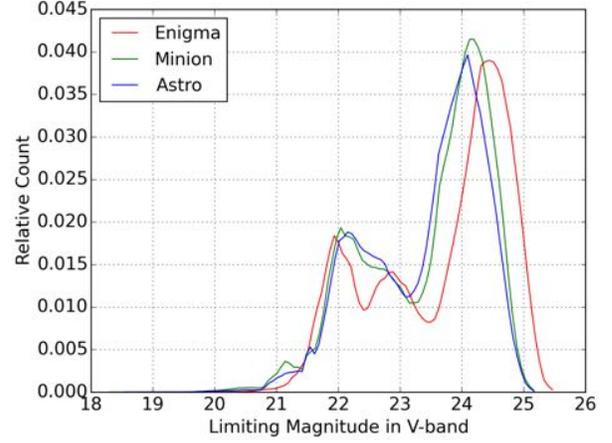

**Figure 6.** Limiting magnitude in V-band in fields used in simulated LSST surveys. Does not include singletons and Deep Drilling.

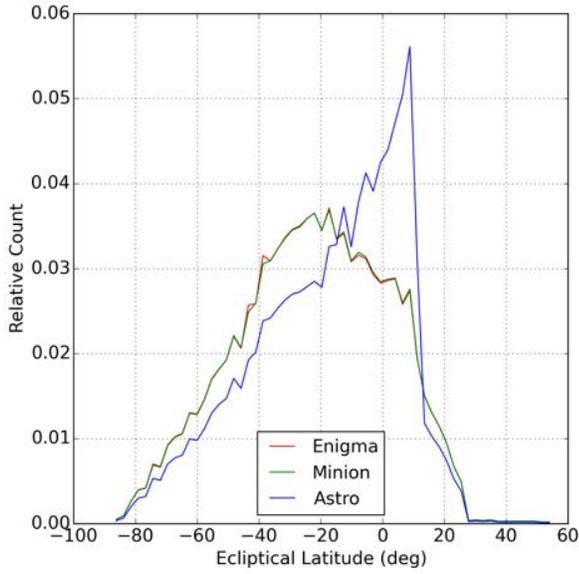

**Figure 5.** Distribution of solar elongation (top) and ecliptic latitude (bottom) of fields used in simulated LSST surveys. Does not include singletons and Deep Drilling. The NEO-enhanced `astro_1016` survey shows increased coverage near opposition and near the ecliptic plane.

**Table 4.** SNR=5 limiting magnitudes ($m_5$) of three OpSim Surveys and time spent in individual filters.

| Filter | Average $m_5$ | Max $m_5$ | Time spent (%) |
|---|---|---|---|
| | `enigma_1189` | | |
| r | 24.38±0.36 | 25.21 | 22 |
| i | 23.66±0.38 | 24.57 | 22 |
| z | 22.44±0.42 | 23.89 | 20 |
| y | 21.49±0.25 | 22.11 | 18 |
| g | 24.67±0.37 | 25.53 | 10 |
| u | 23.75±0.37 | 24.66 | 8 |
| | `minion_1016` | | |
| r | 24.09±0.33 | 24.91 | 22 |
| i | 23.34±0.37 | 24.33 | 22 |
| z | 22.28±0.41 | 23.71 | 20 |
| y | 21.56±0.23 | 22.21 | 18 |
| g | 24.38±0.35 | 25.24 | 10 |
| u | 23.09±0.33 | 23.99 | 7 |
| | `astro_1016` | | |
| r | 24.00±0.37 | 24.91 | 23 |
| i | 23.25±0.37 | 24.28 | 23 |
| z | 22.19±0.41 | 23.71 | 22 |
| y | 21.56±0.24 | 22.21 | 15 |
| g | 24.30±0.36 | 25.24 | 10 |
| u | 23.10±0.33 | 23.99 | 6 |

## 2.2. Asteroid Population Models

### 2.2.1. Near-Earth Objects

This study tested two NEO population models. Each model is represented by a set of Keplerian orbits with absolute magnitudes $H$ defined in Johnson's V-band following the models' respective size-frequency distribution.

The first NEO model that we used is from an earlier work (Bottke et al. 2002; Grav et al. 2011). "Bottke's" model consists of 268,896 orbits with $H < 25$. After our study was already underway, a newer model was published by Granvik et al. (2016), which we refer to as "Granvik's" model. It has 801,959 orbits, again down to $H < 25$. Even though Granvik's model has three times as many orbits, the number of $H < 23$ orbits is about the same as in the Bottke NEO model (Figure 7). On the other hand, Granvik's model lacks objects with $H < 17$. For our simulations, this feature can be omitted because these large NEO are few and mostly already discovered. The main difference is in the slope of the population above $H > 23$, where Granvik's population describes the observational data better than Bottke's early estimate. Histograms of the orbital elements (Figure 8) show the similarity between Bottke's and Granvik's NEO populations. Bottke's orbital element distribution does not depend on $H$, while Granvik's population does. The slight differences between the distributions of orbital elements are evident in Figure 9. Compared to Granvik, Bottke predicted more objects at low perihelion distances and larger eccentricities, while Granvik, on the other hand, shows an excess at larger perihelion distance and inclinations.

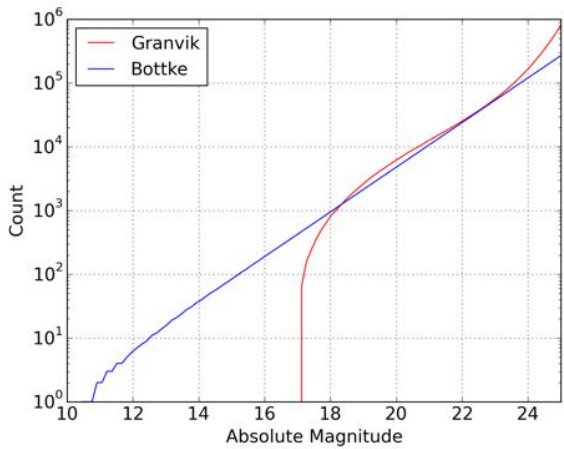

**Figure 7.** Cumulative distribution of Bottke's and Granvik's NEO populations.

### 2.2.2. Main Belt Asteroids

Although Main Belt asteroids (MBAs) are not the target of this study, they are the most numerous population of Solar System objects that will be seen by LSST, and they represent a source of background noise and possible confusion for NEO identification. In our LSST simulations, we used the Grav et al. (2011) model of the main-belt population (Figure 10). This population contains 13,883,361 orbits and is the most robust population model available to date.

In the Grav MBA model, the cumulative distribution slope is equal to $\alpha = 0.28 \pm 0.01$ for H between 16 and 20. However, the population was created for a PanSTARRS4-like survey that had a limiting magnitude of $m_V = 24.5$, and so the population is truncated to remove MBAs that are fainter than $m_V = 24.5$ when at perihelion and at opposition. This truncation results in an artificial break, seen in Figure 10, in the Grav population size-frequency distribution at $H \simeq 21$.

To investigate how this break affects the areal density of MBAs in the LSST survey simulation, we compared the simulated MBA density in LSST fields to the predicted number density by Gladman et al. (2009) who had observed MBAs with the 3.8-meter Mayall telescope at Kitt Peak National Observatory in 2001 within the so-called SKADS survey. SKADS detected asteroids in a fixed 8.4 deg² patch of sky around the opposition point in the Johnson-Cousins R-band down to limiting magnitude of 23.0–23.5 on six nights spanning an 11-night baseline. Based on Gladman et al. (2009), the debiased cumulative number of MBAs follows the equation

$$N(> H) = C * 10^{\alpha H} \qquad (1)$$

where $\alpha = 0.30 \pm 0.02$. This slope $\alpha$ was derived for $15 < H < 18$, with an assumed validity to at least $H = 20$. Gladman et al. (2009) derived the areal density of MBAs as

$$N(< m_R) = 210 * 10^{0.27 * (m_R - 23)} \qquad (2)$$

where $N(< m_R)$ is the cumulative number of asteroids per square degree brighter than $m_R$. The derived detection efficiency was 98% at $m_R = 17$.

To compare with LSST's expected number density of MBAs, we selected LSST fields with solar elongation greater than 178° and within 1° from the ecliptic from a subset of three months (observing cycles 28-30) with a generated full-density MBA population, yielding 27 fields. This simulation was run with fill factor of 90.88%, fading and color transformation assuming all asteroids are of a spectroscopic S type. There was a slight difference in the definition of detection efficiency. LSST

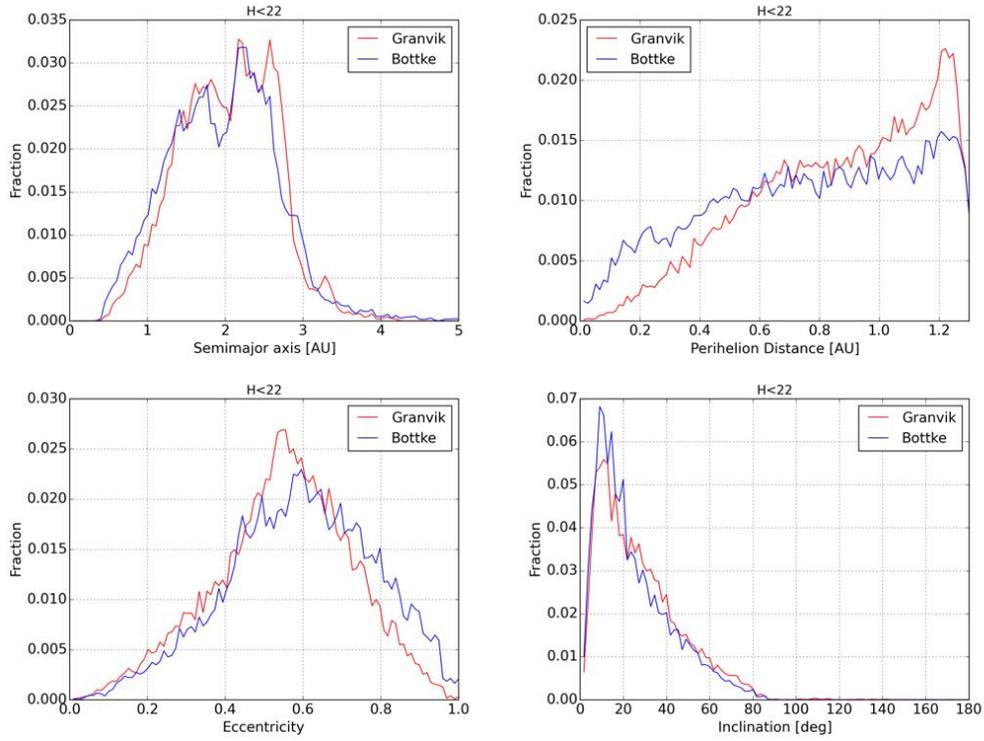

**Figure 8.** Histograms of orbital elements of Bottke's and Granvik's NEO populations.

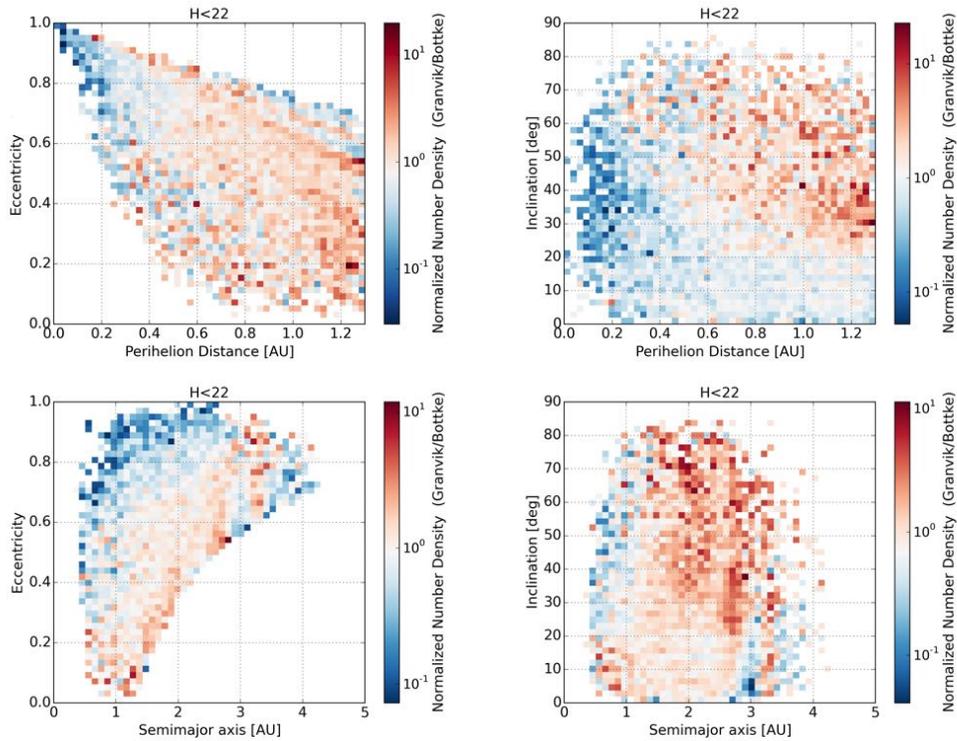

**Figure 9.** Comparison of orbital element distribution between Bottke's and Granvik's NEO populations. These density plots reflect the ratio of the normalized density between the respective models.

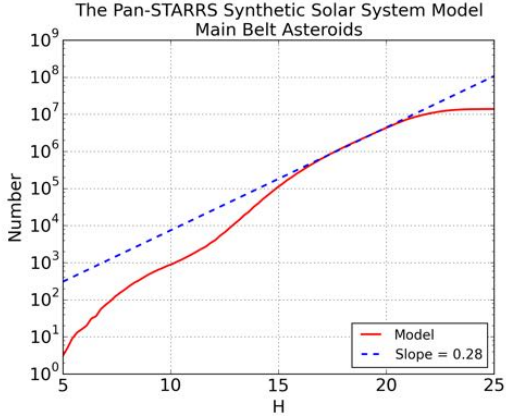

**Figure 10.** Cumulative distribution of MBAs as a function of absolute magnitude by Grav et al. (2011). The model slope change at $H \sim 21$ is an artifact of designing a population of Pan-STARRS4 accessible MBAs.

detections are subject to our so-called fading function (Section 2.4.3) that reduces detection efficiency as

$$\epsilon(m) = \frac{\epsilon_0}{1 + e^{\frac{m-m_5}{w}}} \qquad (3)$$

where $\epsilon_0$ is the detection efficiency, $m$ magnitude, $m_5$ the limiting magnitude defined for $SNR = 5$ and $w = 0.1$ the width of the fading function. SKADS defined its detection efficiency by

$$\epsilon(m) = \frac{\eta_0 - c(m-17)^2}{1 + e^{\frac{m-m_5}{w}}} \qquad (4)$$

where, based on observations, $\eta \approx 0.98$ and $c \approx 0.005$. Here $c$ measures the strength of the quadratic drop and the remaining parameters are the same as in the previous equation.

Additionally, there are several sources of uncertainties that needed to be included into the estimate of the MBA density:

a) A different slope of the population. $\alpha = 0.28$ and 0.30 for Grav and Gladman, respectively.

b) The transformation from the LSST bands and the SKADS R-band to V-band. The term $V - R$ in SKADS was $0.37 \pm 0.15$ mag, leading to relative uncertainty of about 9% in areal density when transforming to V-band.

c) The scaling of the detection efficiency. This study used a different model than SKADS for fading.

Figure 11 shows the number density of MBAs near opposition as a function of limiting magnitude of the field in V-band based on the SKADS survey (red) and the simulated LSST survey with the synthetic Grav MBA population (blue). Note that at $m_5 > 24.5$ the simulated MBA density drops because of the artificially truncated Grav's population. For the fraction of fields with a limiting magnitude fainter than 24.5 in V-band (Table 5), the MBA density was underestimated by up to 12% in the worst case. The density of MBAs decreases significantly as a function of ecliptic latitude (Figure 12).

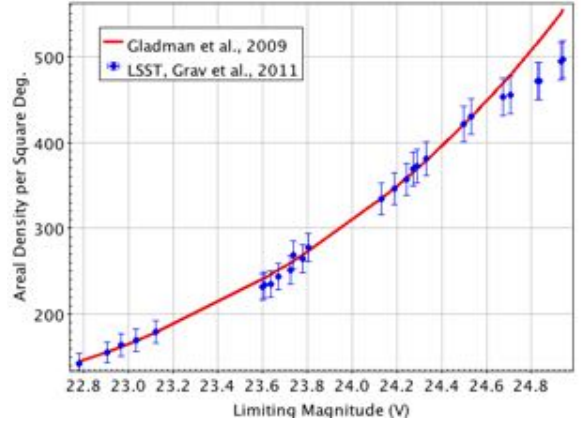

**Figure 11.** Number of MBAs per square degree based on Gladman et al. (2009) and this work (based on Grav et al. (2011) population).

**Table 5.** Fraction of fields having limiting magnitude (V) > 24.5 in simulated surveys.

| Survey name | Percentage |
|---|---|
| enigma_1189 | 19% |
| minion_1016 | 9% |
| astro_1016 | 7% |
| enigma_1271 | 27% |
| enigma_1266 | 27% |

### 2.3. Focal plane model

The LSST camera consists of 21 platforms called rafts, each consisting of a $3 \times 3$ array of 9 CCD chips, yielding a total of 189 CCDs. Each chip comprises 4096x4096 10-micron pixels, and so the total number of active pixels is 3,170,893,824. The mosaic layout of the camera is depicted in Figure 13. Because there are gaps between chips within the $3 \times 3$ rafts and also between the rafts, some fraction of the focal plane is not useable. In this work we use LSST specifications of the chips and gaps (Table 6). Including the chip and raft gaps, the active fraction of the focal plane, called fill factor $F$, can be derived as $(189 \times A_{\mathrm{CCD}})/(21 \times A_{\mathrm{RAFT}})$ where $A_{\mathrm{CCD}} = X_{\mathrm{CCD}} * Y_{\mathrm{CCD}}$ and $A_{\mathrm{RAFT}} = X_{\mathrm{RAFT}} \times Y_{\mathrm{RAFT}}$. Here, $X_{\mathrm{RAFT}}$ and $Y_{\mathrm{RAFT}}$ denote lengths in the horizontal and vertical direction of a single raft, including CCD extents ($X_{\mathrm{CCD}}$, $Y_{\mathrm{CCD}}$), intra-CCD gaps ($G_{\mathrm{CCD}}$) and half of a raft gap ($G_{\mathrm{RAFT}}$) all of the way around the edge of the raft. Thus

$$X_{\mathrm{RAFT}} = 3 * X_{\mathrm{CCD}} + 2 * G_{\mathrm{CCD}_X} + G_{\mathrm{RAFT}_X}$$
$$Y_{\mathrm{RAFT}} = 3 * Y_{\mathrm{CCD}} + 2 * G_{\mathrm{CCD}_Y} + G_{\mathrm{RAFT}_Y}. \qquad (5)$$

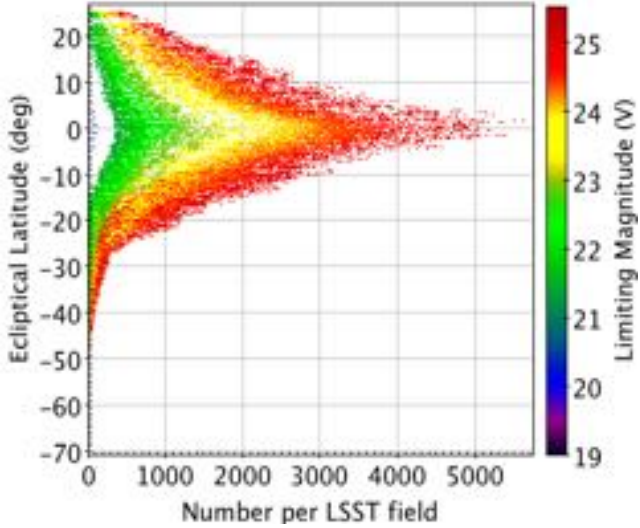

**Figure 12.** Number of detected MBAs per LSST field as a function of limiting magnitude (V) and ecliptic latitude.

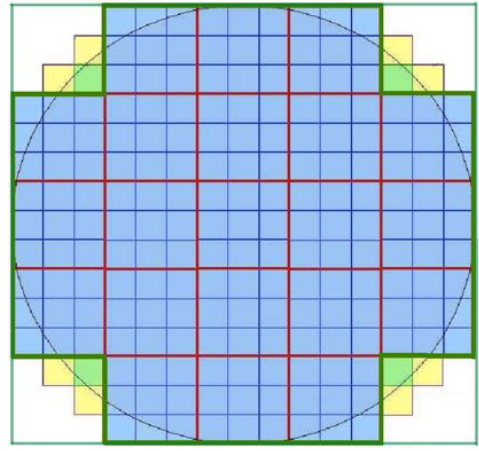

**Figure 13.** Model of the LSST focal plane. Red lines represent raft gaps, blue lines chip gaps. Active chips are depicted by light blue color.

**Table 6.** Dimension of LSST chips and gaps

| Type | Length [arcsec] |
|---|---|
| $X_{\mathrm{CCD}}$ | 800.0 |
| $Y_{\mathrm{CCD}}$ | 814.4 |
| $G_{\mathrm{CCD}_X}$ | 45.0 |
| $G_{\mathrm{CCD}_Y}$ | 30.6 |
| $G_{\mathrm{RAFT}_X}$ | 50.0 |
| $G_{\mathrm{RAFT}_Y}$ | 35.6 |

The total active area with the provided values is equal to $9.50\,\mathrm{deg}^2$, whereas the total raft area yields $10.45\,\mathrm{deg}^2$, resulting in a fill factor $F = 0.9089$. Gaps can be simulated by the exact pixel mask or by a statistical treatment of detections. The pixel mask approach is computationally more expensive, because it requires building and matching up the fields with the 3.2 billion of pixels. This work used the probabilistic approach, where the fill factor represents the probability of a detection to be detected in a single frame. To simulate the field, we employed a square layout equal to 25 rafts with the area of 12.445 square degrees and then applied a mask for the 4 corner rafts to obtain the above mentioned $10.45\,\mathrm{deg}^2$. Finally, 90.89% of detections were randomly selected to form the detection list.

LSST utilizes an altitude-azimuthal mount and the camera is able to rotate, and thus the fields are not generally aligned with the local RA-DEC frame. In fact, due to desired dithering, each exposure is observed in a randomized field orientation (e.g., Figure 14). Figure 15 shows the distribution of position angles within the `enigma_1189` survey in 10 years and the variation $\Delta_{\mathrm{rot}}$ of the position angle within multiple visits of the same field in a night. The average $\Delta_{\mathrm{rot}}$ is 5°, with median 2.4°.

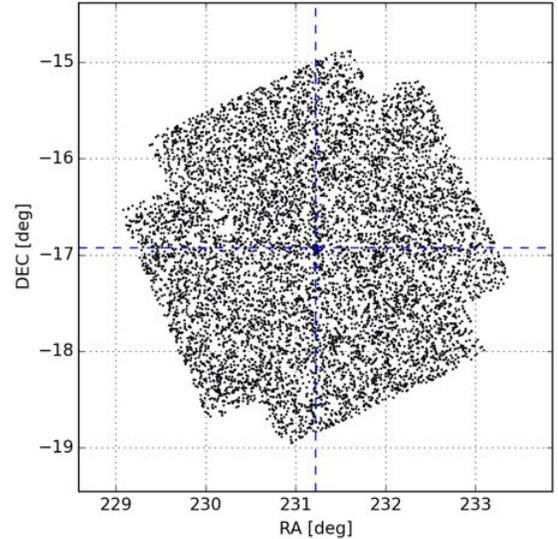

**Figure 14.** Detections modeled in an example LSST field rotated by a position angle of 66°.

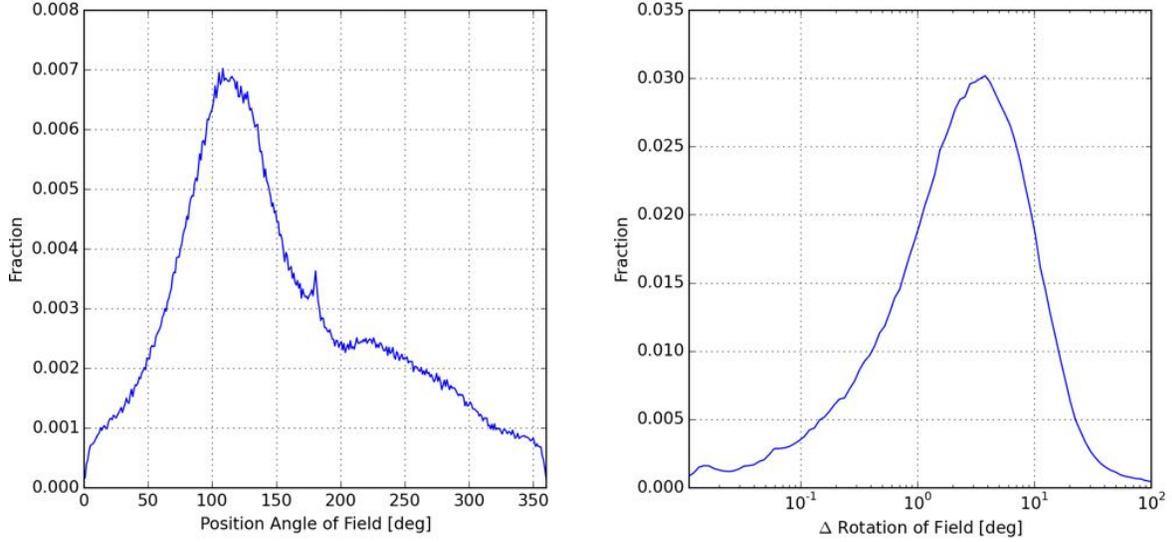

**Figure 15.** Histograms of position angle of the fields (left) and the change in rotation $\Delta_{\rm rot}$ of the same field in a single night (right).

### 2.4. Asteroid Detection Model

#### 2.4.1. Detection threshold

We generated synthetic detections from the NEO and MBA population models (see Section 2.2) by propagation of the orbits to the epochs of the OpSim fields. The propagation used JPL's small body codes with the DE405 planetary ephemerides, which includes all eight planets, plus Pluto and the Moon, as perturbing bodies. We did not use any asteroids as perturbers.

Each ephemeris-based position in the field was subsequently altered by adding realistic astrometric and photometric errors based on the computed signal-to-noise ratio (SNR). The limiting magnitude of the field $m_5$ is defined for SNR=5 and is presumed constant across the field. (While sensitivity in a given image may be reduced in areas with higher background noise, the affected region is small, especially after bright source masking as described in Sec. 2.5.2, and the performance impact is presumed negligible.) The SNR is computed[1] from the difference between the computed magnitude $m$ and $m_5$ as

$$SNR = \frac{1}{\sqrt{(0.04 - \gamma).\chi + \gamma\chi^2}} \tag{6}$$

where $\gamma = 0.038$ and $\chi = 10^{0.5(m-m_5)}$. Then, photometric uncertainty is derived as

$$\sigma_m = 2.5 \log_{10}\left(1 + \frac{1}{SNR}\right) \tag{7}$$

and the computed $m$ is combined with an error drawn from a normal distribution with a mean of zero and variance $\sigma_m^2$.

We have assumed that LSST astrometry is measured relative to a post-Gaia star catalog and so absolute systematic errors are negligible, while relative errors have a floor at 10 mas. The astrometric error $\sigma_{\rm astr}$ for any detection is therefore computed as quadrature combination of 10 mas and the ratio of the seeing $\Theta$ and SNR

$$\sigma_{\rm astr}^2 = (10\,{\rm mas})^2 + \left(\frac{\Theta}{SNR}\right)^2. \tag{8}$$

However, asteroids are moving targets and—depending on the rate of motion—their shape deviates from the stellar PSF, and is in fact a convolution of the motion and the PSF. The faster the object moves, the larger the astrometric error. Therefore, if the trail length $L > \Theta$, the seeing term of $\Theta$ is replaced by a geometric mean of seeing and the trail length as

$$\Theta' = \sqrt{\Theta L}. \tag{9}$$

To obtain realistic astrometry, we combine the original astrometry with an astrometric error term drawn from a normal distribution with a zero mean and variance of $\sigma_{\rm astr}^2$. Figure 16 shows histograms of astrometric uncertainties, in both linear and log-scale. The latter shows that there are two populations of NEA detections, those with high SNR and therefore low uncertainty, around 10 mas, and another centered around 100 mas from low SNR detections, which presumably also includes most of the objects with relatively fast rates of motion. The median astrometric error obtained for NEOs is 47 mas.

---

[1] See https://github.com/lsst/mops_daymops/blob/master/python/add_astrometric_noise.py.

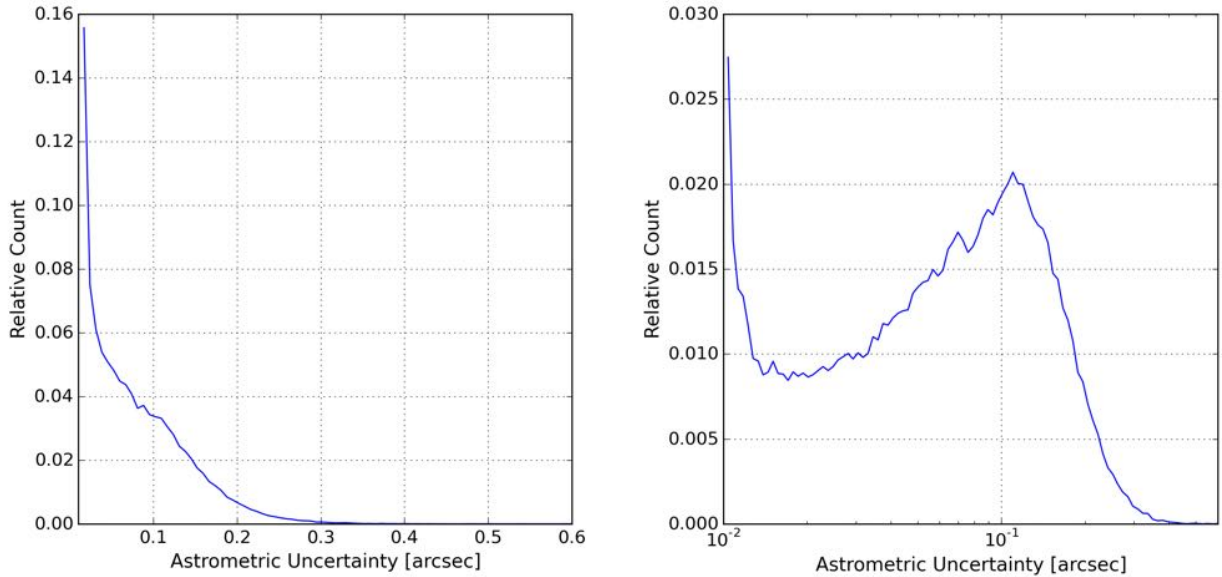

**Figure 16.** Distribution of astrometric uncertainties of NEOs - normal scale (left) and logarithmic scale (right).

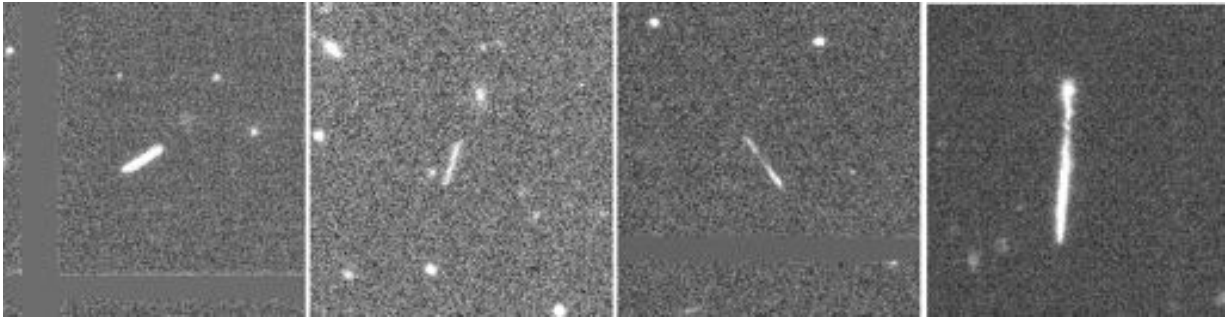

**Figure 17.** Collection of trailed Near-Earth asteroids observed by the Pan-STARRS1 survey. Left to right: 2016 PP27, 2016 CJ31, 2015 BE551, 2016 SA2. The last two trails exhibit brightness variation due to a rapid rotation that is seen within the short exposure time of 40 seconds.

### 2.4.2. *Trailing and detection losses*

LSST will observe the sky with a sidereal tracking rate, and so the static sources will be detected as point-spread-functions (PSFs); however, asteroids will move in the images within the exposure time (Figure 17). For instance, in a 30-second single exposure a typical NEA moving at $0.64°$/day will move by 0.8 arcsec (Figure 18). The expected mean seeing at the LSST site is 0.85 arcsec (Figure 19), therefore, instead of a PSF, most NEAs will appear trailed (Figure 17), with the fastest objects being the smallest and closest to the Earth. The detected trail may be described as a convolution of a PSF and a line (Vereš et al. 2012). The flux of the moving source is spread along multiple pixels and the per-pixel signal decreases as a function of its apparent velocity. The longer the trail, the fainter is the peak signal. Also, the signal-to-noise ratio (SNR) decreases due to trailing as the trailed detection contains the sum of noise from a greater number of pixels.

We consider two types of magnitude losses. The first one is the the loss that happens when a Gaussian or a PSF-like filter is used to identify the sources in the image. The source function is traditionally modeled according to the static and well defined sources like unsaturated stars[2], therefore, if the model finds a trail, it only captures a fraction of the flux in it (Figure 20). Also, it tends to derive multiple sources for a trail that follow the direction of the motion. We call this magnitude loss the *detection loss* and it is described by the function

$$\Delta m_{\text{detect}} = 1.25 \log_{10}\left(1 + cx^2\right) \qquad (10)$$

---

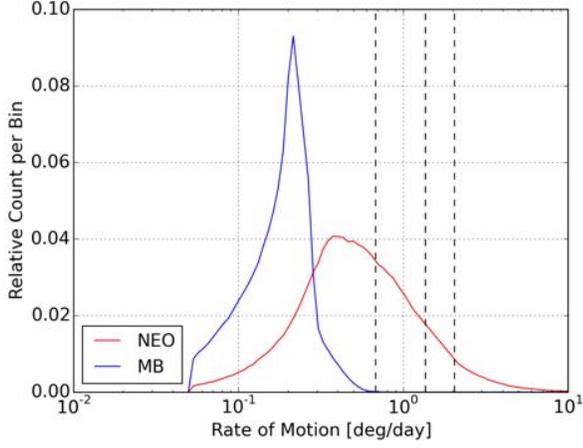

**Figure 18.** Normalized histogram of rate of motion of main belt and Near-Earth asteroids detected by LSST in OC 28 of `enigma_1189`. Dashed lines represent 1, 2 and 3-PSF motion in a 30-second exposure and at an average seeing of 0.86 arcsec.

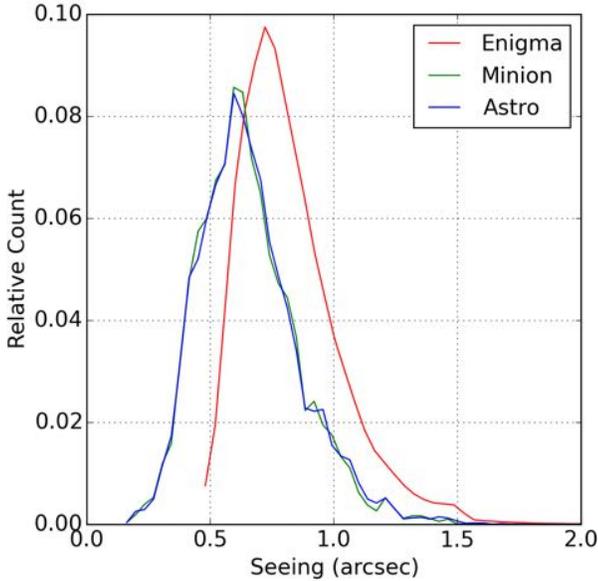

**Figure 19.** Normalized histogram of seeing in three simulated surveys, `enigma_1189`, `minion_1016`, and `astro_1016`.

where $c = 0.42$ and $x = (vt_{exp})/\Theta$ is the trail length in units of seeing $\Theta$ disks. Here $v$ is the rate of motion, $t_{exp}$.

After LSST detects a source, it calculates the SNR with a number of algorithms, including a trail fitting kernel, which leads to *SNR loss*, our second type of trailing magnitude loss. Here the SNR of the trail is calculated from the source flux and the noise of the background from the entire trail. The greater the area of the source, the total background noise, which decreases the overall SNR of the detection. The magnitude loss due to the SNR trailing penalty can be described by the function

$$\Delta m_{SNR} = 1.25 \log_{10} \left( \frac{1 + ax^2}{1 + bx} \right) \quad (11)$$

where $a = 0.76$, $b = 1.16$ and $x$ is again the normalized trailing factor. As shown in Figure 21, the detection losses are a factor of two worse than the SNR losses, which immediately implies that significantly trailed detections found by LSST will have SNR $\gg 5$.

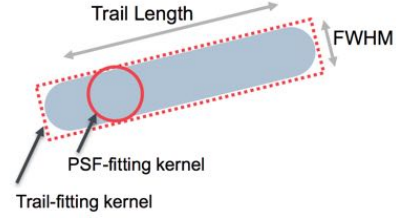

**Figure 20.** The profile of the trail and the PSF kernel.

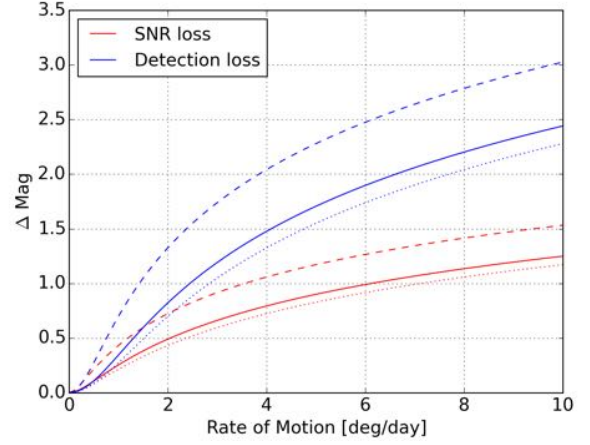

**Figure 21.** SNR and detection loss as a function of rate of motion for a 30-second exposure and average seeing of 0.86 arcsec. The magnitude loss strongly depends on seeing. Dotted line represent a substandard seeing (1.0 arcsec), dashed line a superb seeing (0.6 arcsec).

### 2.4.3. *Detection Fading*

In survey simulations the limiting magnitude is often used as a step function cutoff that determines strictly which detections are visible in the simulated field. However, in real surveys, the detection limit is actually better represented by a function that gradually fades near the limiting magnitude to capture the probability of a given detection. The function can be described as

$$\epsilon(m) = \frac{F}{1 + e^{\frac{m - m_5}{w}}} \quad (12)$$

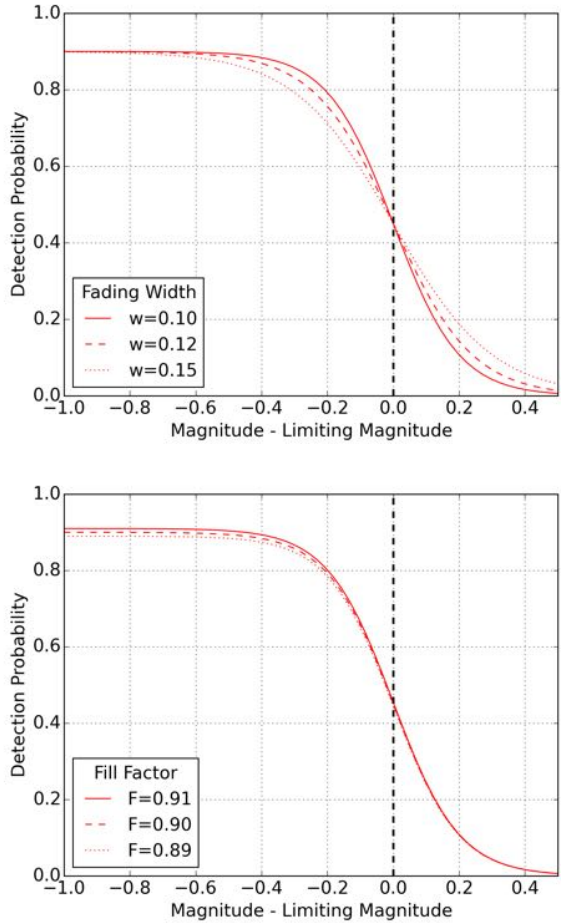

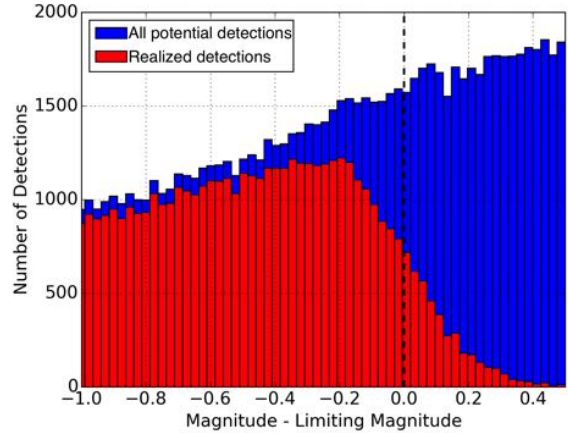

**Figure 23.** Histogram of relative brightness for all NEOs in the field (blue) in one night of the LSST survey relative to limiting magnitude $m_5$ (dashed line) and NEOs actually detected after fading and fill factor are applied (red). The cumulative number of NEOs detected with fading is slightly greater than with the $m_5$ step function scenario.

**Figure 22.** Probability of a detection to be detected, with a fixed fill factor of 90% and variable width (top) and fixed width $w = 0.1$ and variable fill factor (bottom). The $m_5$ limiting magnitude is equal to a 50% probability detection, neglecting fill factor.

where $\epsilon(m)$ is the probability of the detection, $F$ the fill factor, $m$ the magnitude of the detection, $m_5$ the $m_5$ limiting magnitude of the field and $w = 0.1$ the width of the fading function. The fading function is depicted in Figure 22. Because the limiting magnitude is unique per field, the fading function applies on a per-field basis. The importance of using the fading function lies in the fact that it actually allows detection of asteroids fainter than the $m_5$ limit. Given the size-frequency distribution of the source population, this tends to slightly increase the number of detections in the field with respect to a scenario where only the $m_5$ step function was used to determine detection.

### 2.4.4. Asteroid Colors

The absolute magnitudes of model asteroids are provided in the Johnson V-band and are used to calculate the apparent magnitudes. However, LSST will observe in 5 distinct filters in visible and near infrared (u,g,r,i,z,y) and the transformation to V-band depends on the spectral characteristics of the asteroid. Even though distributions of asteroid colors in the main belt are relatively well understood to $H < 18$ (DeMeo & Carry 2014), sampling of the NEA population is rather sparse. For instance, the SDSS database (Carvano et al. 2010) contains only 174 NEA and the Asteroids Lightcurve Database (Warner et al. 2009) currently contains 1,115 NEA with identified spectral type. That is only a small fraction of the NEAs currently known (15,000 as of November 2016). Bottke's and Granvik's NEA populations, which we used in this work do not include albedo, diameter, or spectral type. We generated the distribution of spectral classes of NEA by the debiased distribution derived by Stuart & Binzel (2004) shown in Table 7. For simplification and magnitude transformation from LSST filters to Johnson V-band, we divide these classes into two groups: S types (Q, S and high-albedo X types) and C types (C, D and low-albedo X types). With this scheme, the numbers of C and S type asteroids are similar. Also, the generated colors do not depend on $H$ or orbital elements. The magnitude transformation model used in the study is presented in Table 8. The tabulated color indices are based on the LSST filter bandpasses, with mean reflectance spectra from DeMeo et al. (2009) and the Chance & Kurucz (2010) model for the solar spectrum (Lynne Jones, private communication).

**Table 7.** NEA Spectral Classes by Stuart & Binzel (2004). Note, that X-class consist of two distinct subpopulations based on albedo.

| Class | Fraction of Population |
|-------|------------------------|
| X     | 0.17, 0.17             |
| S     | 0.23                   |
| D     | 0.18                   |
| Q     | 0.14                   |
| C     | 0.10                   |

**Table 8.** Magnitude transformation from Johnson's V-band to LSST filter system.

| Class | V-u    | V-g    | V-r   | V-i   | V-z   | V-y   |
|-------|--------|--------|-------|-------|-------|-------|
| C     | -1.614 | -0.302 | 0.172 | 0.291 | 0.298 | 0.303 |
| S     | -1.927 | -0.395 | 0.255 | 0.455 | 0.401 | 0.406 |

#### 2.4.5. *Light Curve Variation*

The apparent magnitudes of asteroids derived from the MOPS ephemerides do not reflect the amplitude variation due to the asteroid shapes, rotation and spin axis orientation. This effect could be significant when the sky is observed in a sparse temporal resolution, e.g., a few times per night. If the asteroid has an apparent brightness close to the limiting magnitude, a change in the brightness due to its rotation in the time interval between two LSST exposures can cause the asteroid to be visible in only one of the pair. Therefore, the tracklet will not be created and the asteroid would be lost on that particular night. On the other hand, some asteroids can be brighter than the ephemeris magnitude just because they were observed near the maximum of the light curve, leading to the possibility of finding objects nominally below the detection limit.

In this work we generated amplitudes and periods for the model populations based on the debiased model by Masiero et al. (2009). We extended this model so that it depends on the absolute magnitude $H$ (Figure 24) by the method described by Vereš et al. (2015). The simplified light-curve corrections are represented by a sine wave defined by the generated amplitudes and periods at the epochs of the time of observation. This approach does not reflect the real shape of asteroids nor an amplitude that depends on the phase angle or the spin axis orientation. Despite the simplification, the generated magnitudes should reflect the reality because they follow the debiased not the observed distribution. The main drawback lies in the fact that Masiero et al. (2009) observed

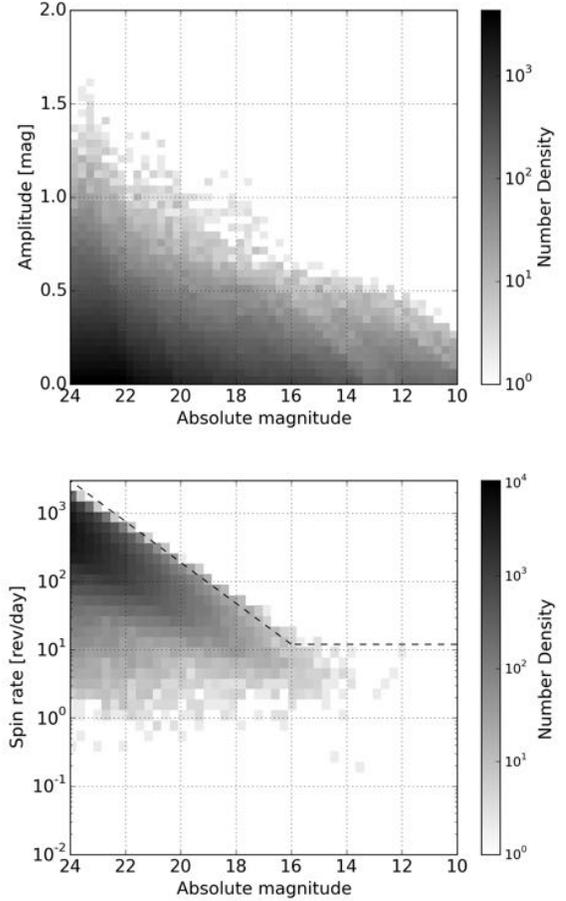

**Figure 24.** Density plots of light curve amplitudes (top) and rotation periods (bottom) versus absolute magnitude for generated NEA populations. The dashed curve in the bottom panel indicates the strength boundary proposed by Masiero et al. (2009).

MBAs near opposition at very low phase angles, while NEA phase angles vary widely. At large phase angles and differing geometry, the resulting light curve is altered and the resulting amplitudes may be smaller.

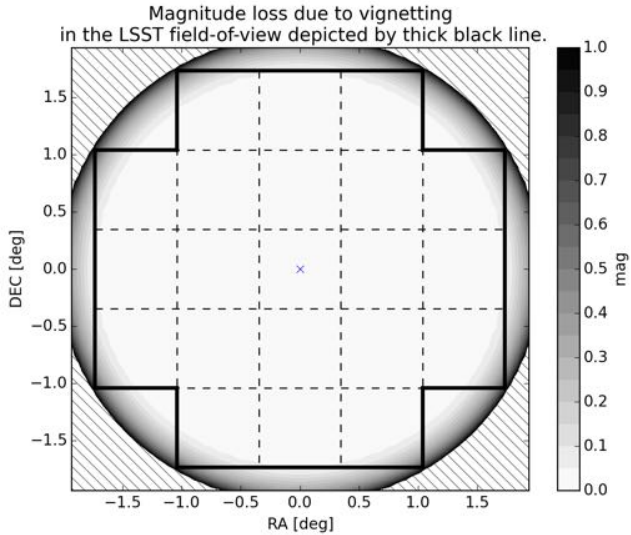

**Figure 25.** Contour plot showing magnitude loss due to vignetting in the LSST focal plane depicted by thick black line. Dashed lines represent individual rafts.

### 2.4.6. *Vignetting*

Optical and mechanical pathways with lenses and mirrors cause vignetting. Vignetting decreases brightness, especially at great distance from the optical axis of the system. Large detectors and wide fields are prone to vignetting even though specialized optical elements and optimized wide-field systems like LSST reduce the effect significantly. The LSST vignetting model depends only on the distance from the center of the field. Figure 25 shows the effect of vignetting causing magnitude loss in the LSST focal plane. Figure 26 show that 93% of the collecting area has 0.1 magnitude or less penalty due to vignetting. The same figure shows that the magnitude loss is significant only far from the field center and will only affect the most distant corners of the detector. Therefore, vignetting should not have a significant effect on survey performance.

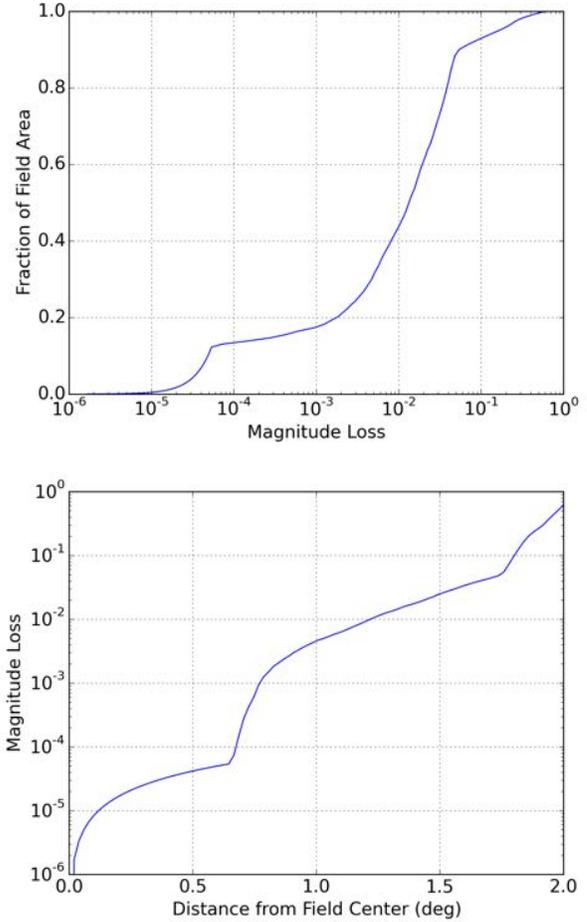

**Figure 26.** Fraction of affected focal plane area as a function of vignetting magnitude loss (top). Magnitude loss due to vignetting as a function of distance from the field center (bottom).

### 2.5. False-positive models

The LSST transient detection data stream will include many detections that are not associated with solar system objects, and the objective of linking only real LSST detections of moving objects to form tracks and orbits represents a significant challenge. There are three broad categories of non-solar system detections that are expected from LSST. The first two are spurious detections arising from random noise and image differencing artifacts, both of which will enter the MOPS input stream and which we discuss below. The third category of LSST transient detections arise from real astrophysical phenomena (e.g., variable stars, supernovae, etc.) that appear in the same location in multiple instances. Such astrophysical transients will be filtered out of the MOPS input stream by virtue of their stationary appearance and thus will not affect the asteroid linking problem.

#### 2.5.1. Random Noise

In addition to real astrophysical sources in the static and difference images, LSST's CCD images will produce two types of false detections. The first source comes from random fluctuations in the sky background and from detector noise, which are both driven by gaussian statistics at the individual pixel level. The number $N_{>\eta}$ of these *random* sources above a given signal-to-noise threshold $\eta$ in the CCD image where Gaussian noise is convolved with a Gaussian PSF follows the formula by Kaiser (2004)

$$N_{>\eta} = \frac{S}{2^{5/2}\pi^{3/2}\sigma_g^2} \eta e^{-\eta^2/2}, \tag{13}$$

where $S$ is the total number of pixels in the focal plane array, $\sigma_g \simeq \Theta/2.35$, and $\Theta$ is the FWHM seeing measured in pixels (Figure 27).

Obviously, the number of random false positives strongly depends on the seeing (Figure 27), and the better the seeing, the larger the number of random false positives. The average `enigma_1189` seeing of 0.80 arcsecond leads to 650 random false positives with $SNR > 5$ in one LSST image.

We generated random false positives in random x-y positions in the field. The number of random false positives for a given field was selected from a normal distribution with a mean and variance of $N_{>5}$ from Equation 13. Magnitudes were assigned to the generated random noise as follows: We generated a random number $p$ from a uniform distribution [0,1]. This number corresponds to the normalized cumulative distribution $N(>\eta)/N_{TOTAL}$. Then $\eta = \sqrt{\eta_0^2 - 2\log(1-p)}$ which could be directly transformed to a magnitude as $V = V_{LIM} - 2.5\log(\eta/\eta_0)$ where $V_{LIM}$ is the $m_5$ lim-iting magnitude at $\eta_0 = 5$. In our case, the threshold of the system is equal to $\eta_0 = 5$. The number density of random false positives has a strong dependence on $\eta$; therefore, most of the random noise sources will be near the the limiting magnitude (Figure 30).

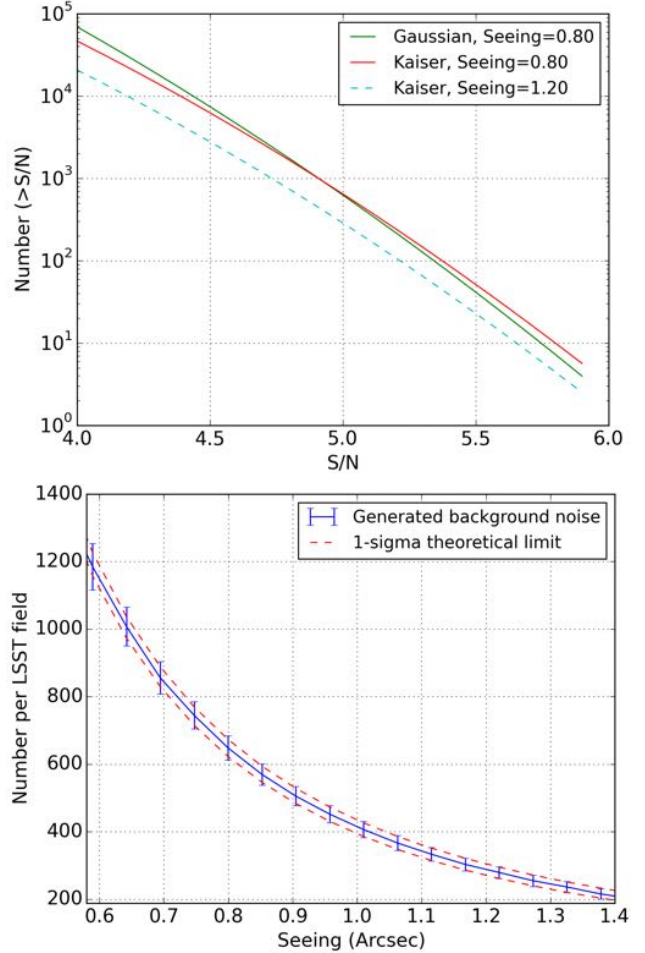

**Figure 27.** The counts of random noise (top) in LSST field as a function of signal-to-noise ratio. Similarity to normal distribution is demonstrated by the green line. The theoretical and generated numbers of random noise (bottom) in LSST fields.

#### 2.5.2. Difference Image Artifacts

The second source of false detections are called *artifacts*, which arise from differencing a field image with a fiducial image of the static sky that has been derived from a stack of several (or a great many) images of the same field over some time period. This differencing technique removes stationary objects so that only transient sky phenomena, including moving objects, appear as detections in the difference image. However, registration errors across the field can leave dipole-shaped artifacts in the difference image at the location of a static source. Artifacts may also originate from a poor convolution kernel, variable seeing across the field, stray light in the optical system or reflections in the lenses. Artifacts are often concentrated around bright sources due to saturation or diffraction spikes, and masking around these sources can be an efficient means of substantially reducing the rate of artifacts. Although an improved optical configuration and machine learning can remove many of these false artifacts, some number will always remain in the data stream.

For this study we assumed the estimated density of differencing artifacts derived by Slater et al. (2016), who used actual imagery obtained by the Dark Energy Camera (DECAM) on Cerro Tololo (Flaugher et al. 2015) and processed them with a nascent version of the LSST image processing pipeline (Figure 28). Slater et al. report that the primary result of their study is that "the LSST pipeline is capable of producing a clean sample of difference image detections, at roughly the 200-400 per square degree level." (See also a summary presentation by LSST at http://ls.st/azy.) This is their final result, but our study used a preliminary estimate as the point of departure for our linking simulations. This earlier estimate allowed for roughly 90-380 artifacts per square degree, and we took the geometric mean of this range as the starting point, which leads to 185/deg$^2$ or 1777 artifacts per LSST field. Slater et al. (2016) did find far higher concentrations of artifacts near bright stationary sources, which they eliminated by masking the area around them, thus allowing the reported low artifact density. Following their result, we modeled bright source masking by reducing the effective fill factor by 1%.

To seed the detection list with artifacts, we selected the number of artifacts in each field according to a gaussian distribution with mean and variance 1777 and distributed them randomly across the field. Thus our artifact rate was roughly 3× the rate from random noise in typical seeing (Figure 29), and about half of the upper bound derived by Slater et al. (2016) from processing actual DECam data.

Our model for difference artifacts is independent of observing conditions such as seeing and field density. However, we note that the most dense regions of the galactic plane are relegated to the Galactic Plane proposal, which happens to be mostly covered by a single-visit-per-night cadence, and is anyway only a few percent of observing time. See Fig. 3 and Table 2. If we remove all Galactic Plane proposal fields from enigma_1189 there is a negligible effect (0.2%) on NEO completeness $C_{H<22}$. Thus our linking and completeness results do not require or assume operation in star fields with extreme density.

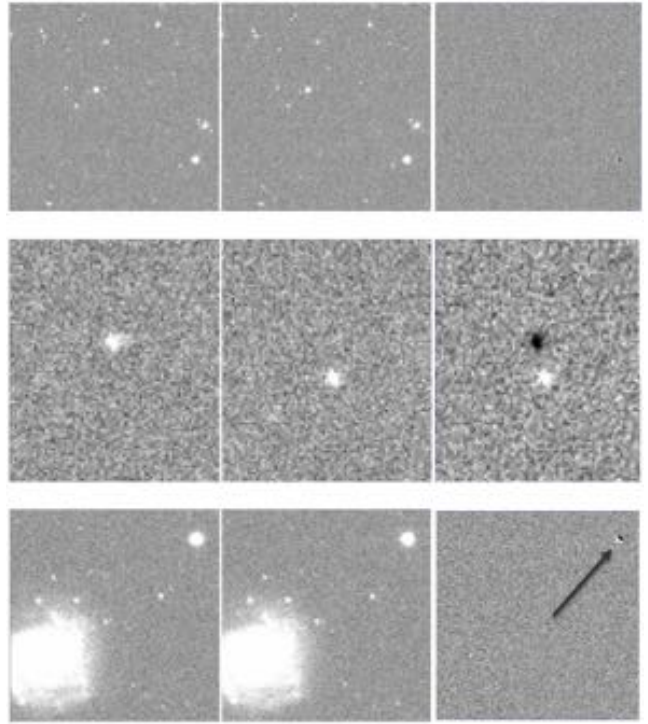

**Figure 28.** Example of LSST image processing pipeline applied to DECAM images (Slater et al. 2016). The first two columns represent images taken at two different times and the third column is the difference image. The first row does not show any moving object and the resulting difference image does not contain any sources. The second row contains a moving object; the difference image shows it as a positive and a negative source. The third row shows an example artifact of a badly subtracted static source (a "dipole") in the upper right corner.

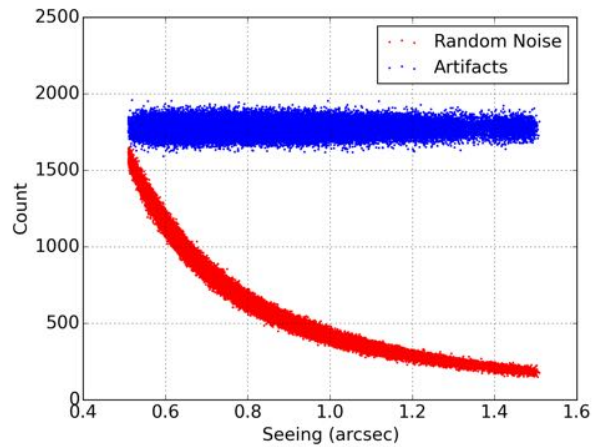

**Figure 29.** Random noise and artifact counts per individual field as a function of seeing during one month of the LSST survey.

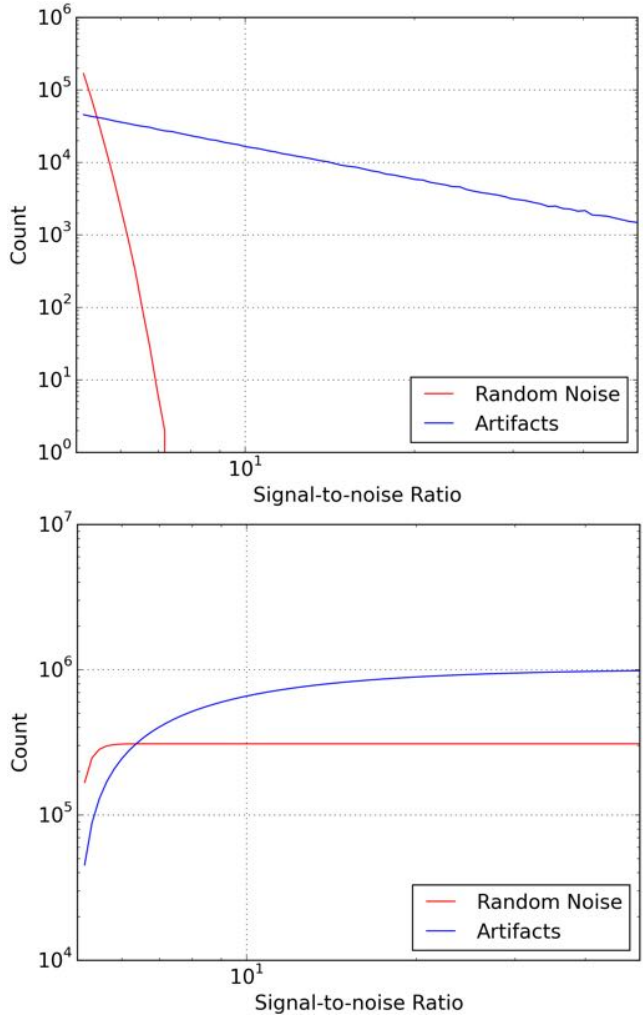

**Figure 30.** Histogram (top) and cumulative distribution (bottom) of random noise and artifacts on one night of the LSST survey.

Following the Slater et al. (2016) report we model the SNR distribution of differencing artifacts as $\propto \mathrm{SNR}^{-2.5}$ distribution of the static sources: $\eta = \eta_0 (1/(1-p)^2)^{1/3}$ where $p$ is a randomly generated number from a uniform distribution [0,1]. (See Figure 30.) The magnitude of a simulated artifact is then derived according to $V = V_{LIM} - 2.5 \log(\eta/\eta_0)$ where $V_{LIM}$ is the $m_5$ limiting magnitude at $\eta_0 = 5$. Artifacts have much shallower dependence on $\eta$, and therefore tend to be far brighter than random noise sources. Roughly half of modeled artifacts have SNR $> 10$, while virtually none of the random false detections had SNR $> 7$.

The brightness distribution of artifacts suggests that at least some potential false tracklets that include artifacts can be immediately eliminated by enforcing consistency in the photometry. However, according to Figure 30, about 90% of artifacts have SNR $< 20$, and if a bright artifact with SNR $= 20$ is paired with a faint asteroid detection having SNR $= 5$ the magnitude difference will be $\Delta m = 2.5 \log_{10} \frac{20}{5} \simeq 1.5$ mag. As it happens, MOPS limits the photometric variation among tracklet components to $\Delta m < 1.5$ mag by default, which suggests that few false tracklets in our simulation have been eliminated in this way. This criteria could be made more strict, which would reduce the false tracklet rate at the risk of removing real objects that are actually more interesting by virtue of a large light-curve amplitude. Note that Figure 24 indicates some asteroids can have amplitudes larger than 1.5 mag, and in rare cases amplitudes higher than 2 mag have been reported. Thus, as a rule, the photometric consistency requirement should be as relaxed as much as feasible in order to avoid eliminating real tracklets. We suspect that this requirement can be dropped altogether without significantly impacting linking performance.

We note that our study neglects the possibility that artifacts are spatially correlated in RA-DEC, which could introduce difficulties in the linking process whereby artifacts could reappear near the same RA-DEC location and mimic the motion of asteroids. RA-DEC correlation among artifacts could possibly arise from two causes, either camera defects or stationary sources. For LSST, the rotational dithering of the camera serves to break the correlations that might arise from defects in the instrument, most of which would already be masked in processing. Masking of the area around bright stationary sources serves to eliminate them as a source of artifacts. Jones et al. (2017) report that the rate of correlated detections in the DECam data stream was low enough to be negligible for our purposes, only $\sim 2/\mathrm{deg}^2$. This no-correlation assumption is at variance with the Pan-STARRS1 experience, but appears to be well justified for LSST.

## 3. RESULTS

### 3.1. *Detection efficiency*

We first investigate LSST completeness results for the ten-year survey, neglecting linking losses. This subsection discusses the tracklet generation approach and explores the effects upon the completeness metric of several detection modeling details.

#### 3.1.1. *Methodology*

The list of detections for a given night is submitted to the `makeTracklets` component of MOPS. A tracklet is created for a detection in the first image if there is a second detection in its vicinity in the second image. The radius of the search circle is defined by the lower and upper velocity thresholds of `makeTracklets`, which were set to 0.05 and 2.0°/day in this study. If there are more possible connections in the circle, in addition to the CLEAN tracklet, consisting of the same object, a MIXED tracklet consisting of two objects or a BAD tracklet that includes a false detection, is created as well. Increasing the upper velocity limit increases the number of false tracklets. In some simulations, for velocities of 1.2–2.0°/day, we used the velocity information from the trail to limit the search area for companion detections. At 1.2°/day, a detection will have a non-PSF shape and its length will be 1.8 times the PSF for the average 0.86 arcsec seeing, and so its length and orientation can be determined. Thus, instead of a large circular search area, smaller regions consistent with the anticipated velocity and direction of the trails are searched. Any matching detections must have a mutually compatible trail length and direction. See Figure 31 for a graphical depiction.

Neglecting tracklet linking inefficiencies, having only detection and tracklet lists is enough to predict tracks. Thus much can be learned from simulations that are not burdened by MBA and false detections. Moreover, to perform quick studies of the 10-year survey detection and tracklet efficiency, we decreased the number of NEO orbits to 3000 and set $H = 0$ for all of them, to ensure that they would be included in the detection list if they were in the field. The differential detection efficiency is then derived from the list of simulated detections or tracklets by adding a bin-by-bin H correction, which can then be accumulated according to the anticipated size-frequency distribution to derive the integral completeness. This approach is > 100× faster than a full-density NEO-only simulation. An additional advantage of this approach is that allows us to adjust individual parameters in post-processing and simulate their effects on overall performance. The process of the low-density scheme is depicted in Figure 32. Ephemerides of the 3000 representative NEO orbits were computed for all OpSim fields, providing a detection list for tracklets. False detections were not included. Post-processing readily creates tracks because all detections and tracklets are identified with the object by MOPS. We studied two scenarios for potential tracks: 12-day and 20-day tracks. The output of this quick simulation is an integral completeness that agrees well with the full-density NEO-only simulation (Figure 33).

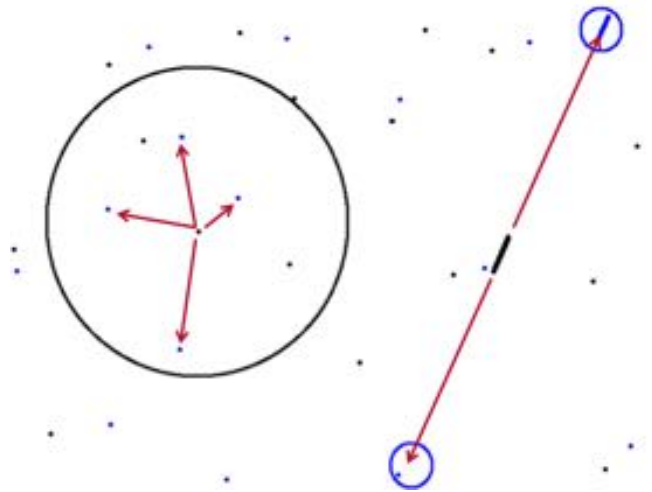

**Figure 31.** Schematic diagram for tracklet generation. Black dots represent detections from the first image, blue dots from the second one. The large black circle represents the upper velocity limit for creating tracklets without rate information (up to 1.2°/day). Arrows in that circle are all possible tracklets, connecting the first detections with all detections from the second image in the reach. Every detection in the image has such a circle and corresponding set of tracklets. If the detection is faster than 1.2°/day it will be trailed (on the right), and information on the trail length and orientation can be used to search a smaller area for its counterpart in the second image (in two separate regions because the direction of motion is unknown). The matching detection must also be a trail with similar length and orientation.

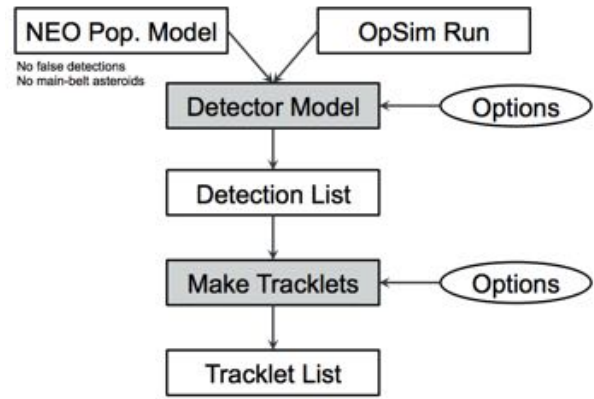

**Figure 32.** Schematic flow of the low-density simulation, from creation of the detection list to generating tracklets in a field. Tracks are created in postprocessing.

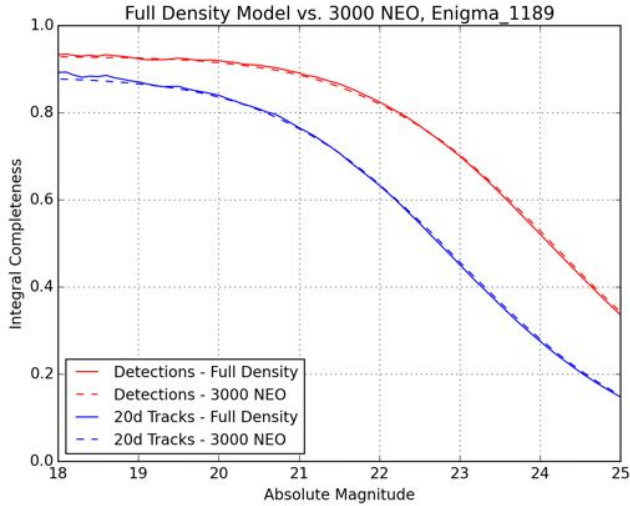

**Figure 33.** Comparison of completeness for full-density NEO model (solid) and a 3000 object random sample (dashed). The red curves represent objects detected in at least one detection in 10 years and the blue curves require a track of 3 tracklets within 20 days. Assumed 92% fill factor, with fading and trailing losses.

### 3.1.2. *Main Result and Model Dependencies*

The key output of the low-density simulation is the 10-year integral completeness for detections, tracklets and tracks as a function of absolute magnitude or time. In Sec. 3.7 we consider the contributions of other NEO search efforts in concert with LSST, but we emphasize that all results prior to that point in this report assume that LSST is operating alone and starting with an empty NEO catalog. Figure 34 reveals that over 80% of NEOs with $H < 22$ are detected at least once in the ten-year `enigma_1189` survey. That number drops to 76% for at least one tracklet in ten years, and 67% for three tracklets in ten years, irrespective of their timing. To actually consider an object discovered and cataloged in this study, we require three tracklets on distinct nights over no more than 20 days or 12 days, which leads to 61% or 58% complete, respectively, in Figure 34. The time history of completeness reveals that after ten years the rate of cataloging is still increasing at about 2% per year in $C_{H<22}$.

We tested the importance to $C_{H<22}$ of a number of our modeling assumptions as discussed in the following list. The relevance of each item for $C_{H<22}$ is tabulated in Table 9 and depicted in Figures 35 and 36.

- To compare the population models, we performed two low-density simulations with 3000 NEOs on

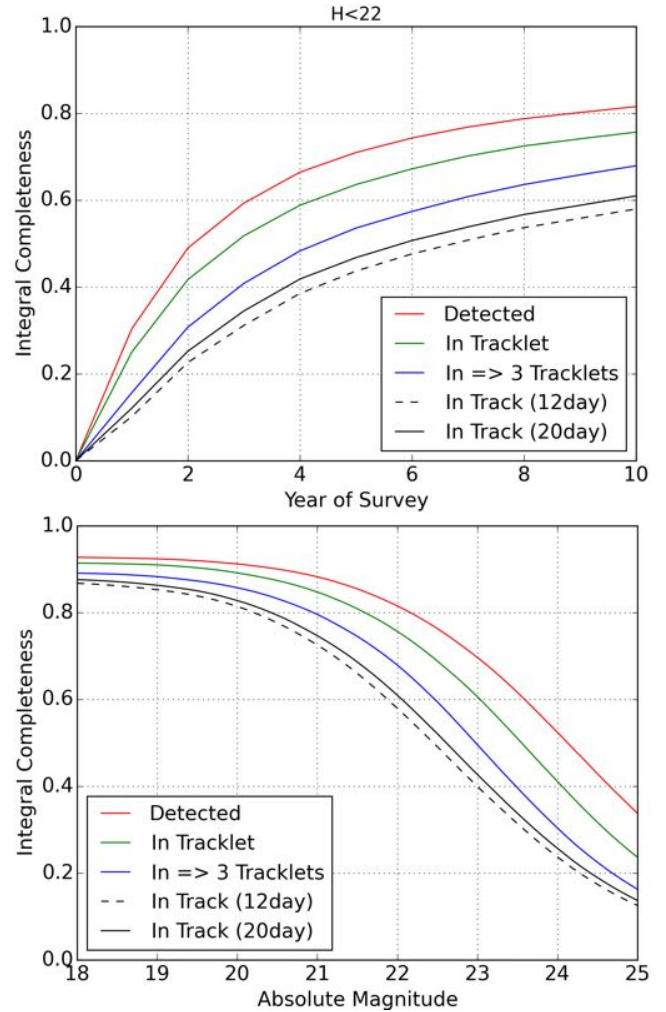

**Figure 34.** Detection, tracklet and track completeness in a low-density simulation as a function of time for $H < 22$ (top) and absolute magnitude for a ten-year survey (bottom) in the `enigma_1189` baseline survey.

`enigma_1189` fields for Bottke's and Granvik's NEO models. Granvik's population led to a slightly greater completeness $C_{H<22}$, but the efficiency is significantly greater for Bottke's population at $H < 25$ (Figure 35), primarily because most of Granvik's NEO are small and therefore much harder to be detected. Table 9 shows the percentage difference between the two models for detections, tracklets and tracks.

- Fill factor is one of the key effects that drives detection efficiency down. The low-density simulation focused on altering statistical fill factor by one percentage point. Dropping fill factor by one and two percentage points from 0.90 led to almost no loss for detections and small losses (0.5-0.9%) for 12- and 20-day tracks. (Figure 35, Table 9)

**Table 9.** $C_{H<22}$ completeness sensitivity due to individual parameters.

| Parameter | Penalty | Detection | Tracklet | 3 Tracklets | 12-day track | 20-day track |
|---|---|---|---|---|---|---|
| | | % | % | % | % | % |
| Population penalty | Bottke 's NEOs | 0 | 0 | 0 | 0 | 0 |
| | Granvik's NEOs | +0.7 | +1.0 | +1.4 | +1.7 | +1.6 |
| Fill Factor | 0.90 | 0 | 0 | 0 | 0 | 0 |
| | 0.89 | 0.0 | −0.1 | −0.2 | −0.5 | −0.4 |
| | 0.88 | 0.0 | −0.1 | −0.3 | −0.9 | −0.7 |
| Trailing losses | Off | 0 | 0 | 0 | 0 | 0 |
| | SNR loss | −1.0 | −1.3 | −1.6 | −1.8 | −1.8 |
| | Detection loss | −1.3 | −1.6 | −1.8 | −2.1 | −2.0 |
| Fading | Off | 0 | 0 | 0 | 0 | 0 |
| | On | +0.9 | −0.2 | −0.4 | −0.8 | −0.6 |
| Colors | S class | 0 | 0 | 0 | 0 | 0 |
| | C+S class | −0.3 | −0.3 | −0.4 | −0.5 | −0.5 |
| | C class | −0.6 | −0.7 | −0.9 | −1.2 | −1.2 |
| Light curve | Off | 0 | 0 | 0 | 0 | 0 |
| | On | +0.3 | +0.1 | −0.1 | −0.3 | −0.2 |
| Vignetting | Off | 0 | 0 | 0 | 0 | 0 |
| | On | −0.2 | −0.2 | −0.3 | −0.3 | −0.3 |
| Limiting magnitude | $m_5$ | 0 | 0 | 0 | 0 | 0 |
| | $m_5 - 0.2\,\mathrm{mag}$ | −1.7 | −2.1 | −2.7 | −3.2 | −3.1 |
| | $m_5 - 0.5\,\mathrm{mag}$ | −4.5 | −5.7 | −7.3 | −8.2 | −7.9 |
| Rate cut-off | $0.5°/\mathrm{day}$ | −9.5 | −9.6 | −12.3 | −13.6 | −13.3 |
| | $1.0°/\mathrm{day}$ | −2.0 | −1.1 | −1.6 | −1.9 | −1.7 |
| | $2.0°/\mathrm{day}$ | 0 | 0 | 0 | 0 | 0 |
| | $5.0°/\mathrm{day}$ | +0.5 | +1.1 | +1.6 | +0.9 | +1.1 |

- Trailing losses represent a major effect for NEOs; however, are negligible for distant asteroids like MBAs. NEO completeness is reduced by 1 percentage point for single detections and up to 2 percentage points for 12- and 20-day tracks. Even though detection losses cause much larger magnitude loss per a single detection, in a 10-year low-density simulation, detection losses are similar to SNR losses in completeness. (Figure 35, Table 9)

- Because of Malmquist bias, the fading model leads to more detections than a hard step function faint limit. However, for tracks fading actually decreased the completeness. This nonintuitive effect arises because at the faint limit the fading model behaves just as a very low (50%) fill factor. So, even though fading provided more detections, these originated from the faint end and below $m_5$ threshold, where the probability of a single detection is below 0.5. In case of a tracklet, this is less than $0.5^2$ and in case of tracks, that require 3 tracklets and minimum of 6 detections, the probability is only $0.5^6$ of geting cataloged. (Figure 35, Table 9)

- We tested three scenarios with NEOs being only S type, C type or a 50-50 mixture of S and C types, independent of orbital elements or $H$. The default class of NEO used in this work is S types. Switching to C types, led to a net decrease of detection efficiency. Nevertheless, the loss was rather small. Switching from S to mixture of C and S types led to a relative loss of 0.5% at $H < 22$ for 2-day tracks. If all asteroids were C types, then the com-

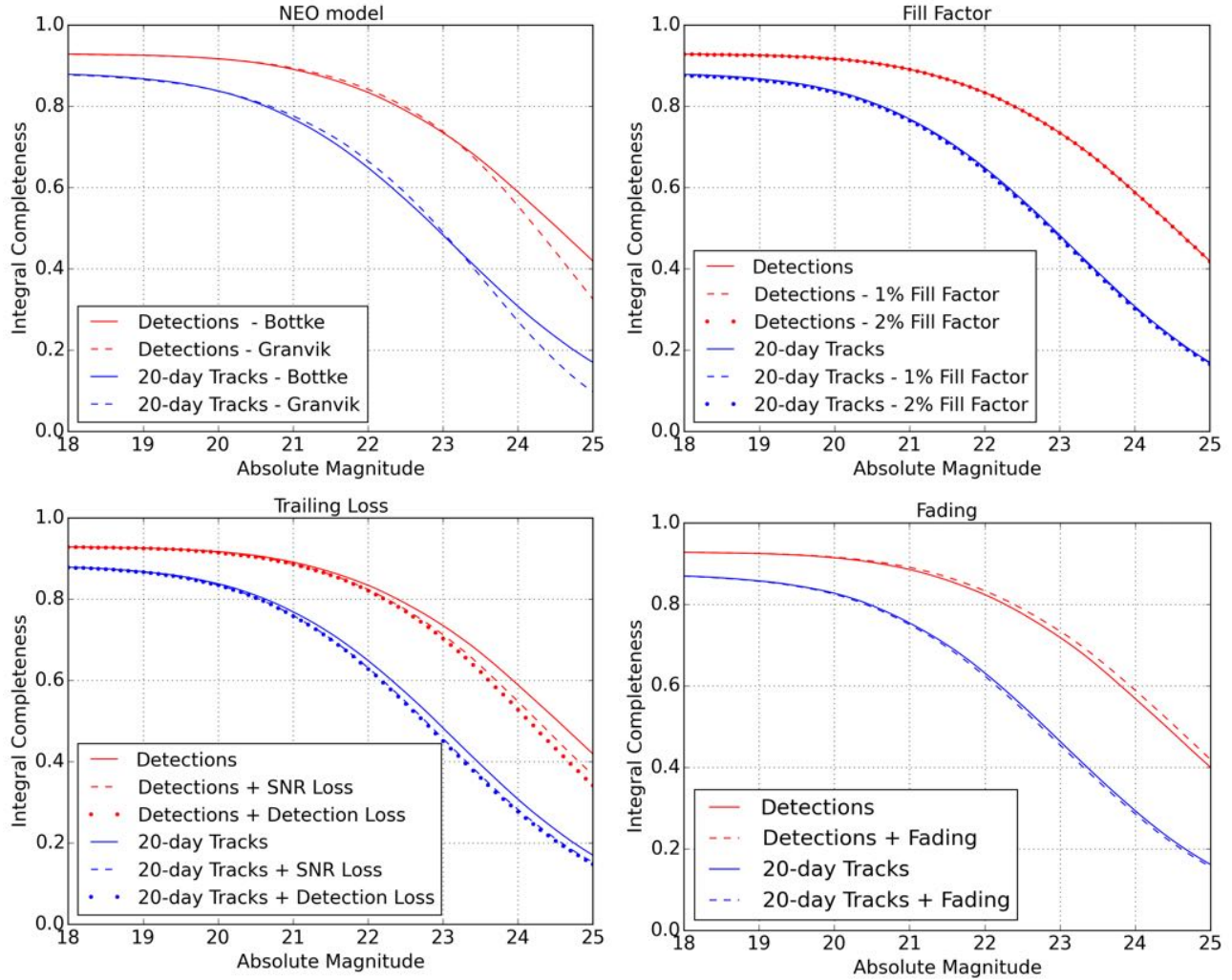

**Figure 35.** Effect of using two different NEO populations, effects of fill factor, trailing losses and fading on detections and tracks in a low-density simulation. Fill factor 0.90, `enigma_1189` with Bottke's NEOs.

pleteness loss will be slightly larger, at 1.2%.(Figure 36, Table 9)

- Like fading, the light curve variation provided somewhat more detections and yet fewer linkable tracks. For detections and tracklets, the completeness increased by 0.3% and 0.1% at $H < 22$. However, 12- and 20-day tracks showed that the light curve variability caused a decrease in completeness by 0.3% and 0.2%. This effect is similar to the detection gain due to fading and related combinatorics. The results showed that light curve variation has a negligible effect in completeness. (Figure 36, Table 9)

- As expected, vignetting plays only a minor role in the completeness of the survey. The NEO completeness penalty at $H < 22$ is only about 0.3% for the tracks. (Figure 36, Table 9)

- So far we have discussed several nominal modeling aspects of the baseline LSST survey. But we wish to also consider off-nominal performance of the LSST system. For instance, the limiting magnitude and seeing for fields is theoretical, even though it is based on long term observations from the summit of Cerro Pachón. What happens if LSST's limiting magnitude is overestimated? Systematic offsets in limiting magnitude could cause a drastic drop in completeness, possibly more than all previously mentioned parameters. For instance, if the limiting magnitude is only 0.2 mag shallower, the NEO ($H < 22$) completeness penalty is 3.1-3.2% for 12 and 20-day tracks. A 0.5 magnitude loss would lead to 7.9-8.2% completeness penalty for tracks. (Figure 36, Table 9)

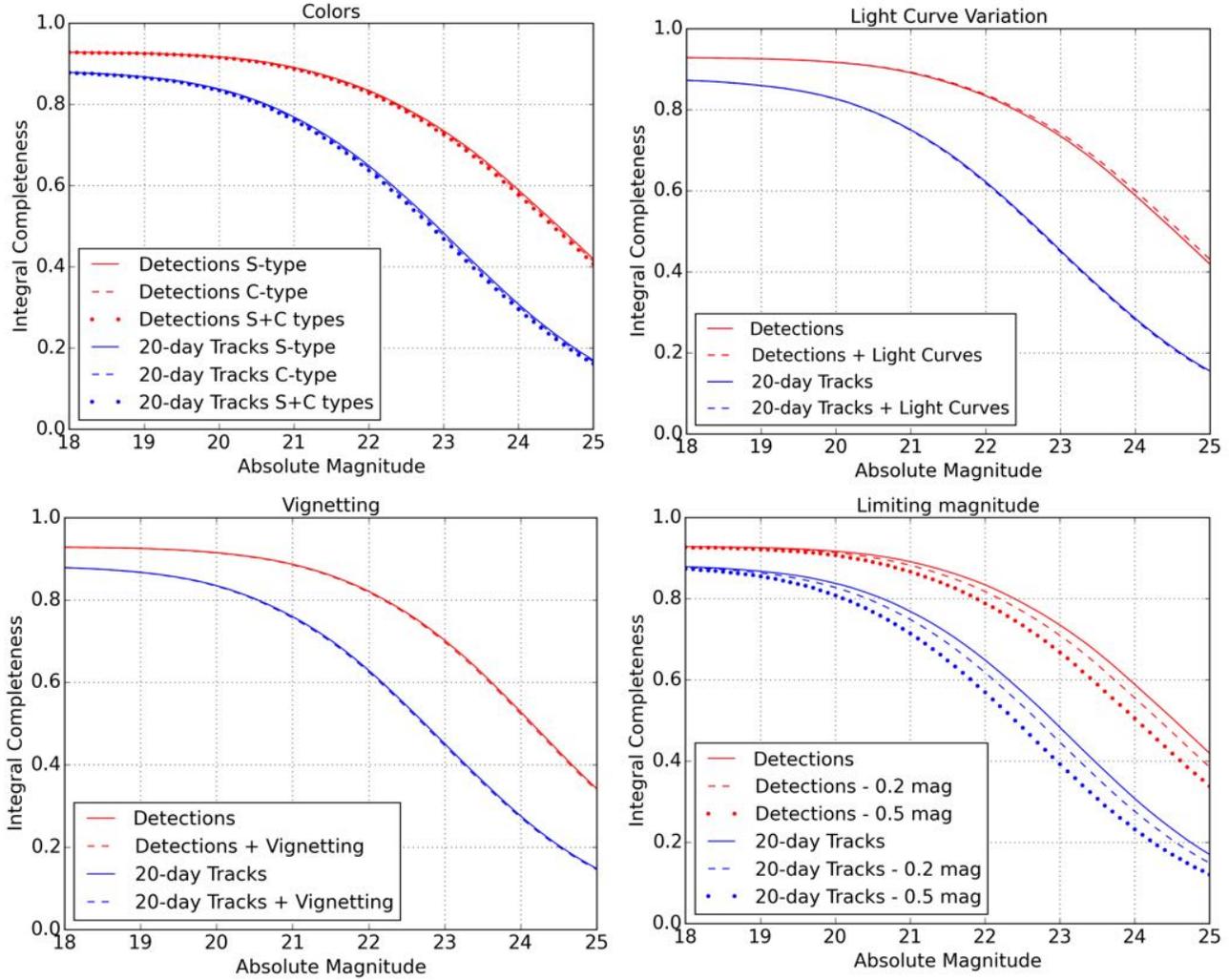

**Figure 36.** Effect of asteroid colors, light curve variation, vignetting and systematic penalty in limiting magnitude in a low-density simulation. Fill factor 0.90, `enigma_1189` with Bottke's NEOs.

Our low-density simulations provided a robust tool to test the effect of individual parameters and constraints on NEO completeness. Table 10 shows NEO completeness as individual parameters were added in the DETECTION and TRACKLET stages of MOPS. Low-density simulations do not use MOPS for creating tracks and linking orbits, but rather tracks are assembled in post-processing from list of tracklets and represent all potential tracks. Because this method derived all potential tracks, it can be used in comparison with the high-density simulations and running MOPS in LINKING stage to study the linking efficiency. For 12-day tracks $C_{H<22} = 65.9\%$ if no losses are applied. The completeness drops by 7.6% when fill factor, fading, trailing losses, vignetting, NEO colors and light curves are applied. Thus, in the 10 years of the `enigma_1189` survey, LSST will find 58.3% of NEOs with $H < 22$, assuming the linking works at a 100% efficiency.

**Table 10.** Effects of individual parameters on overall completeness $C_{H<22}$ for 12 and 20-day tracks in `enigma_1189`. All values $C_{H<22}$ and differences $\Delta$ are in percentage points. Assumed fill factor is 89%. Parameters are adding up in the table.

| Parameter | 12-day tracks | | 20-day tracks | |
|---|---|---|---|---|
| | $C_{H<22}$ | $\Delta$ | $C_{H<22}$ | $\Delta$ |
| None | 65.9 | | 68.1 | |
| + Fill Factor | 63.2 | -2.7 | 65.9 | -2.2 |
| + Fading | 62.4 | -0.8 | 65.3 | -0.6 |
| + Trailing Loss | 60.3 | -2.1 | 63.2 | -2.1 |
| + Vignetting | 59.9 | -0.4 | 62.9 | -0.4 |
| + Colors | 58.7 | -1.3 | 61.6 | -1.3 |
| + Light Curve | 58.3 | -0.3 | 61.2 | -0.3 |

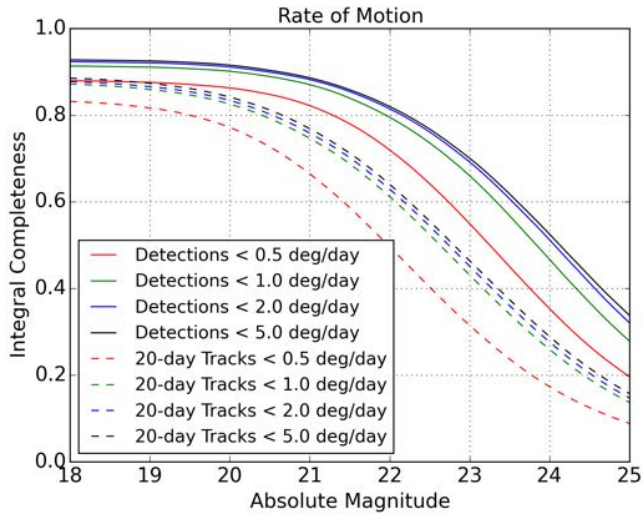

**Figure 37.** Effect of velocity thresholds on detections and tracks in a low-density simulation. Fill factor=0.90, for `enigma_1189`, with vignetting, light curves, C+S colors, detection losses and Bottke's NEO.

### 3.1.3. *Rate of Motion Cutoff*

This study enforced a tracklet velocity limit of 2.0°/day and a subset of that velocity range (1.2-2.0°/day) used a special treatment when creating a tracklet, using the length and orientation of the elongated detections. Figure 18 depicts the rates of motion of NEOs and MBAs. Linking complexity would be reduced for lower limiting velocities, but completeness would fall. Is increasing the velocity threshold worthwhile only to capture the few fast moving NEO that might be small but very close to the Earth? Figure 37 and Table 9 show how different velocity cutoffs for tracklets affect the NEO completeness. Decreasing the upper velocity limit to 0.5°/day would dramatically decrease $C_{H<22}$, by more than 13% for tracks. On the other hand, increasing the upper bound from 2.0 to 5.0°/day only had a slight benefit of about 1% in completeness for tracks. If making use of the trail velocity information for fast detections, increasing the upper velocity threshold is not likely to significantly increase the false tracklet rate, and it comes at a modest benefit.

### 3.1.4. *Arc Length of Discovered Objects*

Figures 38 and 39 document the arc length of NEOs with at least one 12-day track, assuming that other apparitions of the same object are successfully linked. While the mean arc length is tens of days for even the smallest discoveries (Figures 38, top panel), only 14% of the brighter ($H < 22$) NEO discoveries have less than a ten-day observed arc, and the median for such cases is about 3 years. The smaller objects, with $H > 22$, have a median observed arc around two weeks.

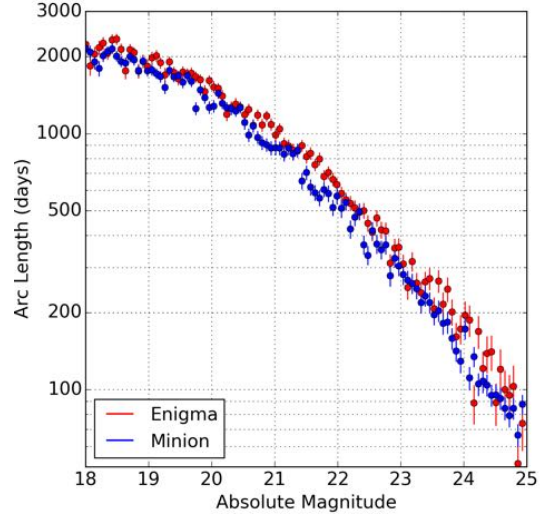

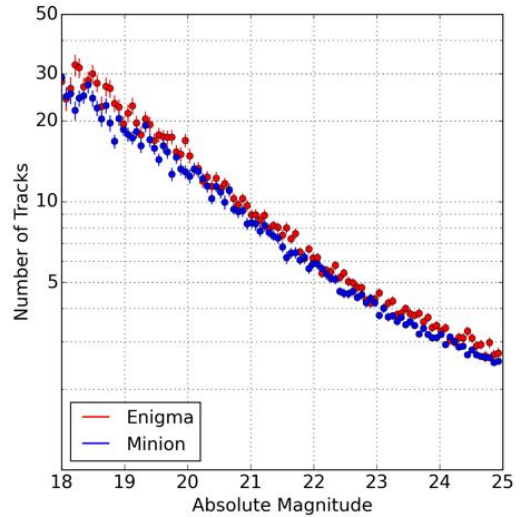

**Figure 38.** Mean arc length (top) and mean number of valid 12-day tracks (bottom) as a function of absolute magnitude for objects having at least one 12-day track over the ten-year `enigma_1189` and `minion_1016` surveys.

### 3.2. *Linking Efficiency*

A central question for this study is whether the linking of tracklets into tracks and orbits will prove successful with real LSST data. LSST MOPS will receive full-density lists of detections of moving and transient targets, including NEOs, MBAs and false detections. From these inputs, MOPS must create tracklets, tracks and orbits, despite the fact that the data stream is contaminated by potentially large numbers of false detections, which lead to high rates of false tracklets. To model the linking process with a realistic density, we studied three consecutive observing cycles from the 10-

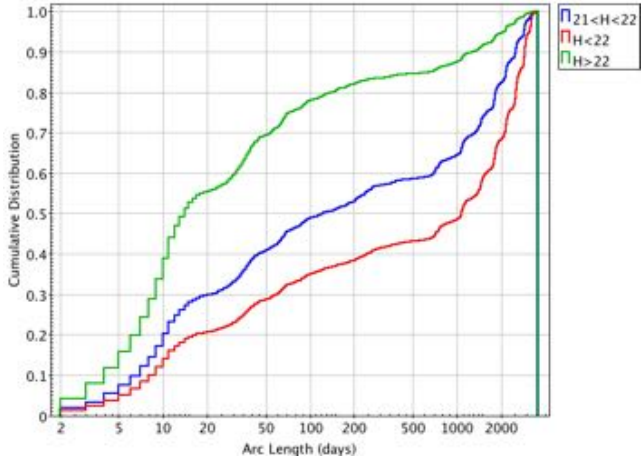

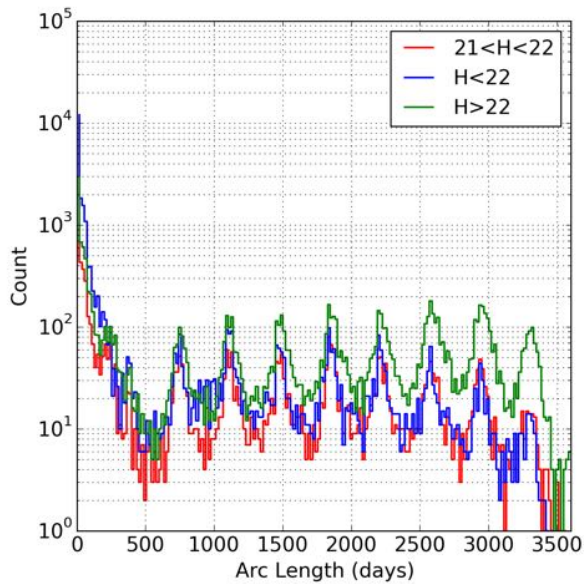

**Figure 39.** Cumulative distributions (top) and histograms (bottom) of arc length for objects having at least one 12-day track over the ten-year `enigma_1189` survey.

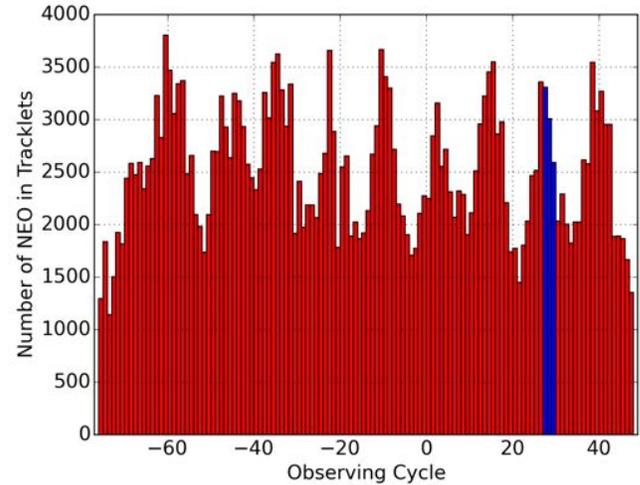

**Figure 40.** Number of NEO orbits from Bottke's population in observing cycles of the `enigma_1189` simulation. OC 28-30, which were used for full-density simulations, are highlighted in blue.

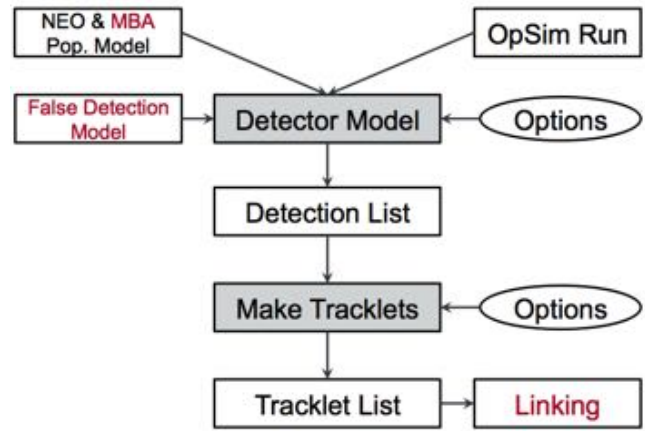

**Figure 41.** Schematic flow of a high-density simulation, from generation of a full-density detection list, through tracklet and track generation, to orbital determination. Elements in red are additions relative to the low-density approach of Figure 32. The Linking stage is depicted below in Figure 47.

year `enigma_1189` OpSim survey. A MOPS observing cycle (OC) has the length of a synodic month, from full moon to full moon. Our selected OCs were from a time of the year when more NEOs were observed (Figure 40). The seasonal variability of NEO detections is a consequence of the survey pattern and the locations of the ecliptic, the opposition point, and the Northern Ecliptic Spur in the survey. For instance the selected observing cycles (OC 28, 29 and 30) were for the months of May, June and July when the ecliptic has the largest altitude above the horizon around midnight and also the nights are the longest in the winter at the LSST site.

The high-density simulation (Figure 41) synthesized detections in the LSST fields from a full-density NEO model ($\sim 800,000$ synthetic objects from Granvik's model), an MBA model ($\sim 11$ million synthetic objects) and false detections (both random noise and differencing artifacts). The detection model assumed the specified filters and used the baseline fill factor with fading, and also included losses due to trailing and vignetting. Astrometric and photometric errors were included. The final detection lists were submitted to the `makeTracklets` routine, and tracklets were created. Finally, tracklets were submitted to the linking stage, the most challenging step.

### 3.2.1. Detection and tracklet volumes

The number of tracklets depends on the density of detections, which can be large (Figure 42). To understand the feasibility of the high-density simulation we gradually increased the number of detections in a single observing cycle (OC 28). The following steps are also summarized as Cases A-E in Table 11.

1. Initially, we only used NEO orbits from Bottke's model (Case A, Table 11). Switching to Granvik's NEO model increased the number of detection by 35% and tracklets by 55% (Case B). Because Granvik's NEO model is more current and has many more objects we used that population in the high-density simulations. At this stage, with only NEO orbits, nearly all tracklets were CLEAN, with only 4 MIXED tracklets (99.97% tracklet purity).

2. Adding the MBA population to Granvik's NEOs (Case C) increased the number of detections in one month to 15 million and the number of tracklets to 6 million. Most of the tracklets were for MBAs; however, about 17% of tracklets were MIXED, i.e., derived from different objects. The large number of MIXED tracklets was substantially reduced by taking advantage of trail-related velocity information when in the velocity range 1.2–2.0°/day (Case D). This increased the number of NEO tracklets by 20% and decreased the number of MIXED tracklets by factor of 5.

3. The next step added false detections from random noise to the full-density NEO and MBA detection list (Case E). This doubled the number of detections to 30 million, and so the synthetic to false detection ratio was about 1:1. However, the number of tracklets only increased from 6 million to 7.5 million. In this scenario tracklets were created up to the 2°/day limit without the use of velocity information. In addition to 1 million MIXED tracklets, the simulation generated about 700,000 BAD tracklets (i.e., those with both synthetic and false detections) and 600,000 NONSYNTH tracklets consisting solely of false detections.

4. The final, full-density simulation was achieved by also injecting differencing artifacts, which more than doubled again the total number of detections, to 66 million (Case F). Now, over 77% of detections were false, and so the ratio between synthetic and false detections was about 1:3.5. NEOs represent only 0.07% of the detection list. The full-density simulation was challenging for the tracklet stage, therefore we used trail-derived velocity information for tracklets created in the velocity range of 1.2–2.0°/day. Still, the total number of tracklets was very large, ∼ 11.9 million. Out of this sample, about 57% of tracklets were somehow erroneous, either including at least one false detection or detections of different objects. This simulation revealed that artifacts related to false positives create the majority of the linking challenge. Though we did not directly test it, the use of trail-related velocity information presumably leads to a dramatic reduction in the false tracklet rate for the full-density simulation.

**Table 11.** Number of detections and tracklets for OC 28 in various simulations. MIXED tracklets include detections from at least two distinct objects, BAD tracklets include detections from both false detections and moving objects, and NONSYNTH tracklets consist entirely of false detections.

| Case | NEO | MBA | False Det | Detections | | | | Tracklets | | | | | |
| | | | | Total | %NEO | %MBA | %False | Total | %NEO | %MBA | %MIXED | %BAD | %NONSYNTH |
|---|---|---|---|---|---|---|---|---|---|---|---|---|---|
| A | Bottke | No | None | 36k | 100 | 0 | 0 | 11k | 100 | 0 | 0 | 0 | 0 |
| B | Granvik | No | None | 49k | 100 | 0 | 0 | 17k | 100 | 0 | 0 | 0 | 0 |
| C | Granvik | Yes | None | 15M | 0.3 | 99.7 | 0 | 6.2M | 0.23 | 82.8 | 16.9 | 0 | 0 |
| D [a] | Granvik | Yes | None | 15M | 0.3 | 99.7 | 0 | 5.4M | 0.31 | 94.8 | 4.9 | 0 | 0 |
| E | Granvik | Yes | Random only | 30M | 0.2 | 50.6 | 49.2 | 7.5M | 0.19 | 68.2 | 14 | 9.7 | 7.9 |
| F [a] | Granvik | Yes | Random + artifacts | 66M | 0.1 | 22.6 | 77.3 | 12M | 0.14 | 42.7 | 2.2 | 6.1 | 48.8 |

[a] Tracklet generation used rate information from 1.2–2.0°/day. Otherwise rate information was ignored over entire range 0.05–2.0°/day.

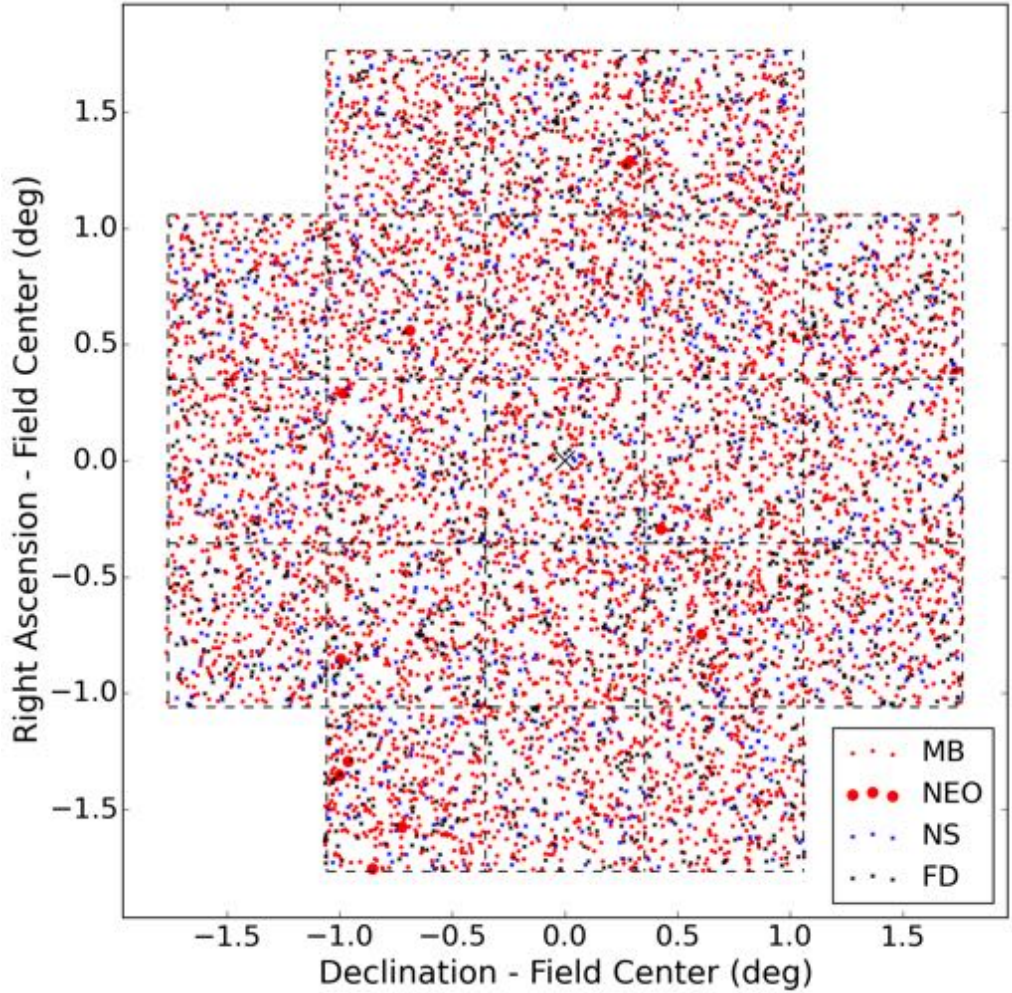

**Figure 42.** An example of a high-density LSST field from the `enigma_1189` survey, night 3052, field number 1891, r filter, $m_5 =$ 24.79, seeing 0.63 arcsec, airmass 1.04, field center at opposition-centered ecliptic coordinates (Lat., Long.) = $(2.91°, 1.26°)$. Thus the field is near opposition in excellent conditions. The various types of detections referenced in the legend are "MB"—main-belt asteroids, "NEO"—near-Earth objects, "NS"—false detections from random noise, and "FD"—false detections arising from image differencing artifacts.

### 3.2.2. *The Linking Process*

Automated linking of tracklets is a crucial element of LSST's NEO discovery pipeline. Without an automated linking stage, the NEO discovery rate would suffer and would rely heavily on follow-up observers, which will be impractical given the faint limit of most of the LSST detections. The MOPS linking algorithm connects tracklets from three distinct nights into candidate tracks that are subsequently tested through orbit determination. The process consists of the following four distinct steps:

1. *Assemble tracklet list.* The first step collects, for a given field, all of the tracklets from the last $N$ nights for which the earlier position and velocity project into the destination field. The forward mapping of tracklets is based on linear motion, and acceleration that leads to nonlinear motion is not accounted for. Thus some NEO tracklets may be neglected, especially those very near the Earth with a rapid change in geometry and observable acceleration.

The combinatoric challenges of linking strongly favor small $N$, but the objective of NEO completeness favors large $N$, which allows more missed detection opportunities. For LSST, $N$ usually ranges from 12-20, though 30 has been contemplated as a stretch goal. This study used $N = 12$ days for linking tests, consistent with our objective of understanding whether linkage could be at all successful in the presence of large numbers of false detections. NEO linkage of nearby objects is not likely to succeed for large $N$ unless MOPS is extended so that some plane-of-sky acceleration is allowed when assembling the field-by-field tracklet lists. This would likely lead to a modest increase in the NEO discovery rate at the expense of many more false tracklets and increased linking overhead.

2. *Assemble candidate track list.* The second step in linkage generates a list of candidate tracks based on the input tracklets. Generally, there are hundreds of available fields per night, each being processed in parallel. The `linkTracklets` algorithm is based on a kd-tree search (Kubica et al. 2007) that reduces the number of potential tracks to be tested from $n^2$ to $n \log n$, where $n$ is the number of tracklets available for linking on the given field. This saves a significant amount of computational resources, but the problem remains challenging.

`linkTracklets` has multiple tunable parameters, such as the minimum number of nights, the minimum number of detections, the minimum and maximum velocities, and some inscrutable kd-tree linking parameters (`vtree_thresh`, `pred_thresh`, `plate_width`). Different parameter values led to vastly different CPU and memory requirements, and markedly different numbers of candidate tracks. However, optimization of this stage is complex. The ideal parameter settings depended on the number of detections and varied from field to field. For instance, experiments with only synthetic NEO orbits led to 99% linking efficiency. Adding noise and MBAs and running tests for selected target fields and tracks and varying `linkTracklets` parameters led to inconclusive results because the correct parameters seemed to depend on the field, and optimizing on a full lunation was infeasible. We explored the optimization of the kd-tree parameters on a single, dense field in the middle of OC28. The total number of candidate tracks increased as a function of `vtree_thresh` and `pred_thresh`, and there was only a weak dependence on `plate_width`, at least for `plate_width` < 0.01 (Figure 55 in Appendix B). However, the most NEO tracks were derived for `plate_width` = 0.003 and `vtree_thresh` ≃ 0.003 (Figure 56 in Appendix B). Pushing the kd-parameters to obtain as many as possible NEO led to an extreme increase in the number of false candidate tracks (Figures 57–60 in Appendix B). Also, the memory and CPU load increased dramatically (Figures 61–62 in Appendix B).

This study was conducted with a single 8-core Linux workstation with 96 GB of memory (upgraded from 32 GB during the course of the study), and a crucial part of the challenge of linking was avoiding out-of-memory crashes. The final values utilized for the main linking simulation in this study were therefore a combination of feasibility and available computational resources (`vtree_thresh`, `pred_thresh`, `plate_width`)= (0.001, 0.001, 0.003). This corresponds to the lower left corner of the upper right plot in Figures 55–62. Better performance could have been obtained for, say, (`vtree_thresh`, `pred_thresh`, `plate_width`)= (0.003, 0.003, 0.003), but this would require use of a large cluster with more memory per core, something that will be readily available to LSST.

3. *Derive preliminary orbit.* The third step took the candidate tracks derived by `linkTracklets` and submitted them for Initial Orbit Determination (IOD). MOPS uses Gauss' method to generate potential initial orbits from the astrometry, and for each track the best fitting IOD is selected. Most false tracks were eliminated at this stage with no valid IOD.

4. *Perform differential corrections.* The fourth stage was Orbit Determination (OD), which used JPL OD routines to obtain converged orbits. This includes sophisticated fall-back logic to try to obtain 4- or 5-parameter fits if the 6-parameter orbit fit diverged. MOPS filtered out some false tracks at this stage based on rudimentary screening on post-fit residual statistics. As discussed below, MOPS's built-in orbit quality filtering is not strict and is agnostic regarding the expected errors in the as-trometry, and thus very few false orbits were rejected at this stage. All orbits that passed the MOPS default quality screening were added to the MOPS derived object table, which was the basis for understanding the overall linking performance.

### 3.2.3. Linking Performance

Linking tests were conducted on a single observing cycle (OC 28) of the `enigma_1189` baseline survey, with Granvik's NEO model, MBAs and the full false detection lists (Case F, Table 11). The NEO linking efficiency is defined as the number of unique NEOs present in the post-linking, derived-object catalog divided by the number of unique NEOs with possible 12-day tracks. The linking efficiency was 93.6% for $H < 22$ NEOs and 84.0% for all NEOs (i.e., $H < 25$). The lower efficiency for all NEOs arises from the fact that the vast majority of NEOs were of the smallest diameters, e.g., $23 < H < 25$. Also, smaller objects tend to have faster rates and greater acceleration because they are seen at closer geocentric distances, and they tend to shorter observability windows (Figures 38, 39). Note that the derived linking efficiency was for a single set of selected kd-tree parameters with a single 8-core workstation. With more powerful computational facilities and a more optimized (possibly on a per-field basis) kd-tree search, there is excellent reason to believe that the linking efficiency can be significantly improved.

Many derived NEO orbits stemmed from objects in the MBA input catalog. Table 12 shows the makeup of the 5348 NEO orbits (defined by $q < 1.3$ au) derived from OC 28 alone. Among these 5348 NEO orbits, 2222 originated from CLEAN linkages of actual NEOs, 1896 were CLEAN orbits associated with MBAs and 1230 were erroneous ("Not CLEAN") linkages. Nearly all of the erroneous linkages combined detections of different MBAs to form an NEO orbit; few were contaminated by false detections. At first blush this implies a purity of 77.0% in the NEO catalog, but we describe below why this apparently low accuracy is mostly a manifestation of an ineffective orbit quality screening applied by MOPS. Correct interpretation of the orbits and improved screening increases the accuracy to 96%. In contrast to the NEO orbits, Table 12 reveals that the MBA catalog has 99.8% purity already at this stage, without more refined filtering on orbit quality. Only 6 NEOs appear in the non-NEO orbit catalog, and most of these are borderline cases where $q \simeq 1.3$ au.

### 3.2.4. Orbit Quality Filtering

The large number of erroneous linkages that appear in the NEO orbit catalog stems from a weak orbit quality filter implemented by MOPS, which requires the post-fit

**Table 12.** Accuracy of derived orbits from OC 28. The derived classification assigns objects with $q \leq 1.3$ au as NEOs. The "Incorrect Class." column indicates the number of objects for which the source object and the derived object had a different classification based on perihelion distance $q$. "Not CLEAN" indicates erroneous linkage of observations from either false detections or multiple objects.

| Derived Classification | All | Incorrect Class. | Not CLEAN | Accuracy |
|---|---|---|---|---|
| NEO | 5348 | 1896 | 1230 | 77.0% |
| Non-NEO | 765,833 | 6 | 1635 | 99.8% |

RMS of astrometric residuals to be less than 0.4 arcsec, a criterion that is too readily met for astrometry having a median error less than 0.05 arcsec. Moreover, because the RMS is not normalized by the reported astrometric uncertainty, it fails to take into account the varying quality of astrometry within and between tracklets in a candidate track. The upshot of this approach is that most such erroneous linkages show residuals clearly inconsistent with the astrometric uncertainty, and yet they pass the MOPS quality control test. Rather than modifying MOPS and re-running the simulation, we post-processed the post-fit astrometric residuals, with their associated uncertainties, to derive the sum of squares of the normalized residuals for each orbit in the NEO catalog. This provided the so-called $\chi^2$ of residuals, from which it is straightforward from classical statistics to calculate the probability $p_{val}$ that the fit is valid, which is to say, the likelihood of getting a higher value of $\chi^2$ by chance. A higher post-fit $\chi^2$ naturally leads to a lower $p_{val}$ because the increased residuals reflect a poorer fit that has a lower probability.

Figure 43 depicts the distribution of $p_{val}$ among the 5348 cataloged NEO orbits. The histogram reveals that few erroneous linkages appear for $p_{val} > 0.25$ and that few NEOs appear for $p_{val} < 0.25$, thus we selected 25% as the $p_{val}$ cutoff for acceptable orbits. This criterion led to rejection of 7% of clean and 87% of not clean orbits. Most of the clean orbits that were filtered out were MBAs mis-classified as NEOs, 14% of which were filtered out. Only 2% of clean NEO orbits were removed by this filter. As tabulated in Table 13, more aggressive $p_{val}$ filtering—at the 50% or 90% level—is less effective at removing erroneous linkages, even as the loss of clean NEOs becomes unacceptable. Thus a modest modification of MOPS is necessary to allow a more statistically rigorous orbit quality filtering, but the rudimentary approach described here leads to a 96% purity (3816/3979, see Table 13) in the NEO catalog. In the context of ac-curacy, the clean MBAs that appear in the NEO orbit catalog are accounted as correctly linked, which is, in fact, the case.

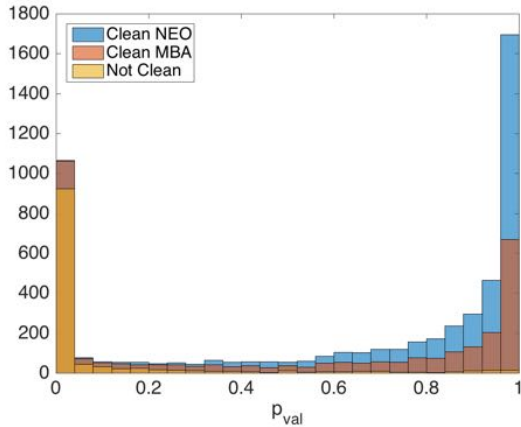

**Figure 43.** Histogram of postfit residual statistics of derived NEO orbits. In most cases, Not CLEAN NEO candidates can be easily distinguished.

**Table 13.** The number of cataloged NEO orbits of various classifications for varying values of the $p_{val}$ orbit quality filter. Here "Non-NEO" refers to MBAs that appear in the derived NEO catalog with $q < 1.3$ au.

| | $p_{val}$ cutoff | | | |
|---|---|---|---|---|
| Classification | 0% | 25% | 50% | 90% |
| All | 5348 | 3979 | 3636 | 2314 |
| CLEAN | 4118 | 3816 | 3532 | 2279 |
| Not CLEAN | 1230 | 163 | 104 | 35 |
| w/False Detection | 35 | 3 | 1 | 1 |
| CLEAN NEO | 2222 | 2180 | 2062 | 1375 |
| CLEAN MBA | 1896 | 1636 | 1470 | 904 |
| Not CLEAN NEO | 2 | 0 | 0 | 0 |
| Not CLEAN MBA | 1230 | 163 | 104 | 35 |

The rate of contamination of NEO orbits by false positives is extremely low, despite the large numbers of false positives injected into the detection stream. As shown in Table 14, after filtering at $p_{val} > 25\%$, only 5 false detections appear in the NEO catalog. This can be compared to the total of 29k detections that form the NEO catalog and the 51M false detections present in the data stream. This result provides a clear indication that NEOs can be successfully linked with high efficiency and high accuracy when surveying with the baseline LSST cadence, even in the presence of significant numbers of false detections.

**Table 14.** Number of detections of various classifications from OC 28. The total number in the input detection list and the number that were linked into the derived NEO catalog are shown.

| | Total | —Derived NEO Catalog— | |
|---|---|---|---|
| | | All | $p_{val} < 25\%$ |
| Total | 65,900,928 | 39,188 | 29,288 |
| MBA | 14,899,279 | 20,680 | 11,868 |
| NEO | 48,628 | 18,446 | 18,060 |
| False | 50,953,021 | 62 | 5 |
| % False | 77.3% | 0.16% | 0.02% |

### 3.2.5. Confusion from MBAs

To better understand the issue of the large fraction of NEO orbits stemming from correctly linked non-NEO objects, we used systematic ranging to explore the full orbit determination problem for these cases. Systematic ranging is an orbit estimation technique designed to analyze poorly constrained orbits, typically with only one or a few nights of astrometry, for which the conventional least squares orbit determination can fail due to nonlinearity (Farnocchia et al. 2015). We tested hundreds of cases and found that nearly all showed a characteristic "V"-shaped orbital uncertainty pattern in $e$ vs. $q$ that allowed both NEO and MBA orbits (left panel, Figure 44). In some cases the "V" shape was broken at the vertex so that there were two distinct orbital solutions (center panel, Figure 44). The systematic ranging technique affords a statistically rigorous estimate of the probability that the track represents an NEO orbit, and for these correctly-linked MBAs that appear with NEO orbits, few have high NEO probabilities, reflective of the fact that the data are compatible with the non-NEO (truth) orbits (Figure 45). It is also important to note that most of these MBAs that appear as NEOs are detected far from opposition. Figure 46 shows that only ∼ 10% of these cases are found within 60° from opposition, and that about half are detected at 80° or farther from opposition. This result is merely reflecting the classical result that orbital ambiguities result from three-night orbits of objects far from opposition. It is an unavoidable feature of observing at low solar elongations, and is generally corrected after a fourth night of data is obtained. However, as described below, the current MOPS configuration does not efficiently attribute a fourth night of data to the already cataloged orbit, and so the ambiguity is often not resolved in our simulations. We note also that this confusion is an artifact of simulating only a single observing cycle. In actual operations, MBAs seen at low solar elongation would eventually move into the opposition region and appear even brighter there. These MBAs would be readily cataloged with their correct orbits because there is little ambiguity in the opposition region, at which point it becomes straightforward to link to the ambiguous orbits arising from near-sun detections. After concluding the analysis of the OC 28 linking products described here, we extended the high-density linking simulation to include OC 28-30 in order to understand the extent to which more dark runs alleviate the issue, but analysis of those products could not be completed within the schedule constraints of this study.

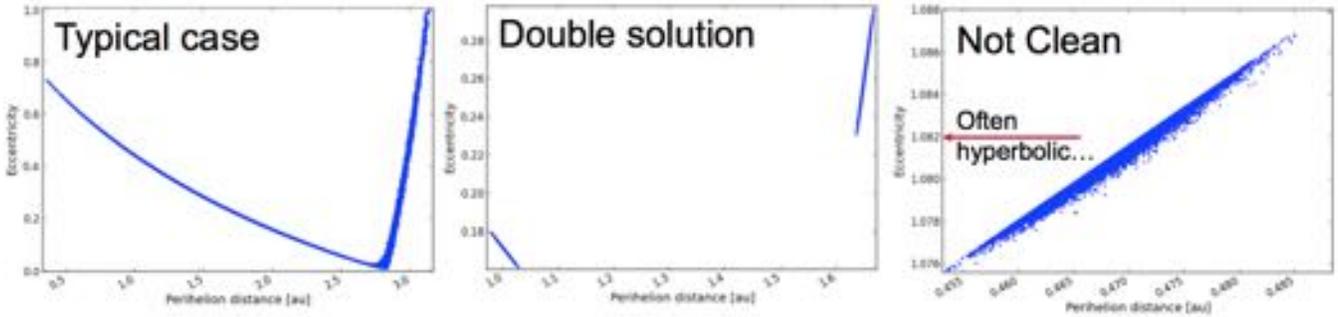

**Figure 44.** Examples of typical uncertainty regions for misclassified or erroneous linkages in the derived NEO orbit catalog. The plots depict Monte Carlo samples from systematic ranging that reflect the extent of possible solutions in perihelion distance $q$ and eccentricity $e$. The plots show the typical case of an MBA discovery (left) where the data are compatible with orbits spanning the NEO and MBA orbital regimes. In some such cases two disjoint solutions are present, one NEO and one MBA (center). Erroneous linkages of two different MBAs often lead to NEO orbits with a small uncertainty, though many such cases are hyperbolic.

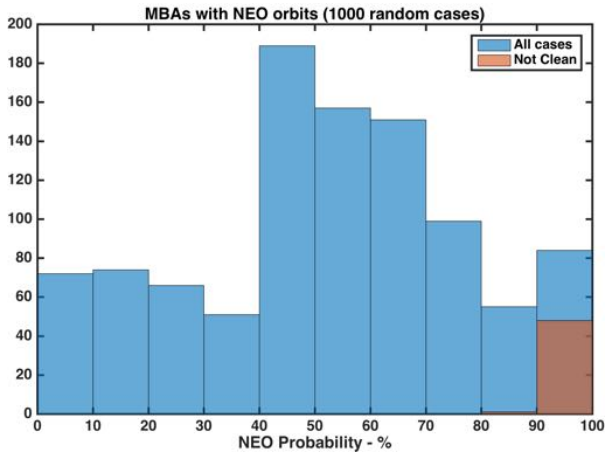

**Figure 45.** Histogram of computed probability that a track derived from MBA tracklets relates to an NEO orbit, as derived from systematic ranging analyses.

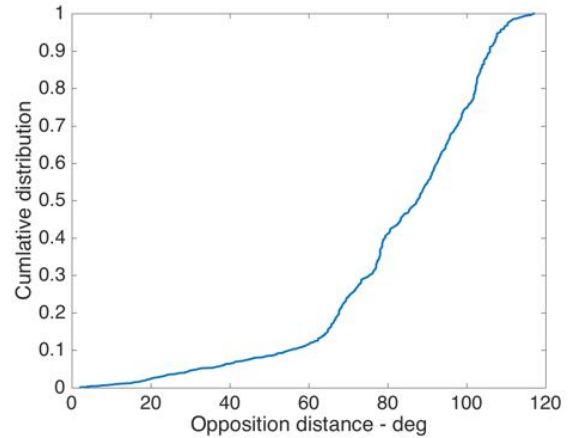

**Figure 46.** Cumulative distribution of opposition distance for MBAs that appear in the NEO orbit catalog with $p_{val} > 25\%$. The distribution shows that this main-belt confusion is largely limited to detections made far from opposition, i.e., with low solar elongation.

We also conducted systematic ranging analyses on some of the erroneous linkages leading to NEO orbits, almost all of which were erroneous MBA-MBA linkages, and these revealed a very different characteristic pattern in the $e$ vs. $q$ uncertainty space (right panel, Figure 44). The uncertainty region was typically very small, leading to a high computed probability that the orbit is of an NEO ("Not Clean" in Figure 45). In these cases, the uncertainty regions were also elongated and with one side having a sharp cutoff. In many such cases the heliocentric orbits were hyperbolic. This points to a likelihood that more effective screening tests can be developed to eliminate these false MBA-MBA linkages, despite the fact that some pass even strict orbit quality tests. For example, Table 13 shows that even for $p_{val} > 90\%$ a few dozen erroneous linkages remain in the NEO catalog. However, most of these erroneous MBA-MBA linkages are readily repaired when the individual MBAs are eventually re-observed at other epochs and correctly linked through other tracklets.

### 3.2.6. *Duplicate Orbits*

Table 12 indicates that there were 4118 clean linkages in the NEO catalog, but not all of these are unique. Table 15 shows that 8.7% are actually duplicate entries of the same object. In Figure 48 we see that the duplicate NEO entries are of almost identical orbits, with 95% of duplicates matching in both eccentricity and perihelion distance (in au) to within 0.02. The non-NEO catalog has an even greater rate of duplication (17.3%).

Virtually all of these duplicates are readily linked with standard orbit-to-orbit identifications techniques (which are already part of MOPS), and most duplicates can be avoided altogether with a more efficient application of the MOPS attribution algorithm. Figure 47 provides a schematic diagram of the linking process, where a tracklet is first checked to see if it is can be attributed to an object already in the catalog. If so then it is linked to that object and removed from the tracklet list so that it is not passed along to kd-tree linking. The fact that so many objects in our simulation are linked into multiple independent tracks in a single observing cycle implies, first, that there are at least six tracklets in the lunation, indicating a very solid discovery, and second, that the attribution algorithm can easily be tuned to catch these extra tracklets before they are even linked into tracks. Not only would such a re-tuning keep the orbit catalog cleaner, it would also cut down on the computational expense of kd-tree searches by removing tracklets from the search that are associated with already discovered objects. The problem of duplicate orbits is likely to be easily resolved through testing and tuning of existing MOPS functionality.

**Table 15.** Duplication among derived orbits.

| Class | Clean | Unique | Duplicates | Fraction |
|---|---|---|---|---|
| NEO | 4118 | 3758 | 360 | 8.7% |
| Non-NEO | 764,198 | 632,298 | 131,900 | 17.3% |

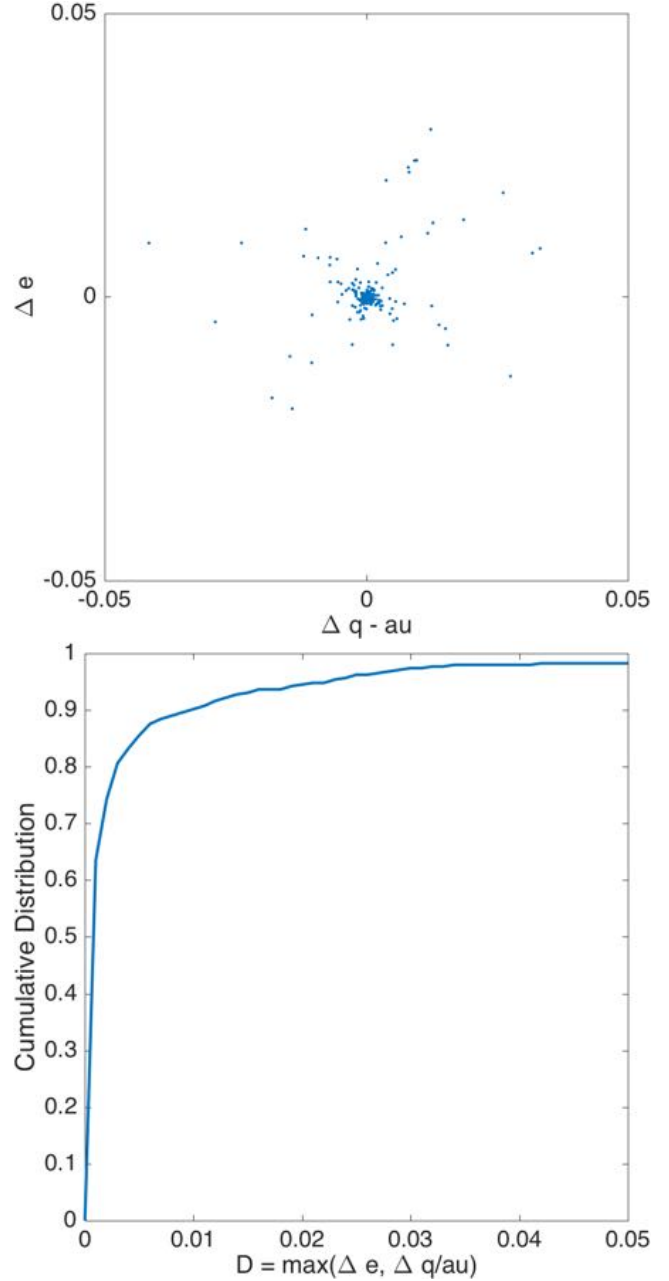

**Figure 48.** (top) Scatter plot of $\Delta q$ and $\Delta e$ between duplicate NEO orbits. (bottom) Cumulative distribution of duplicate separation in the $q$ and $e$ phase space.

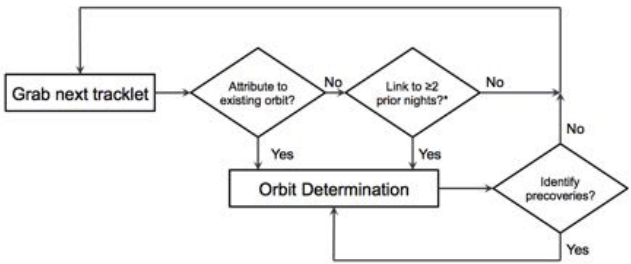

**Figure 47.** Flow chart of the high-density linkage algorithm, from tracklets to track generation and orbital determination.

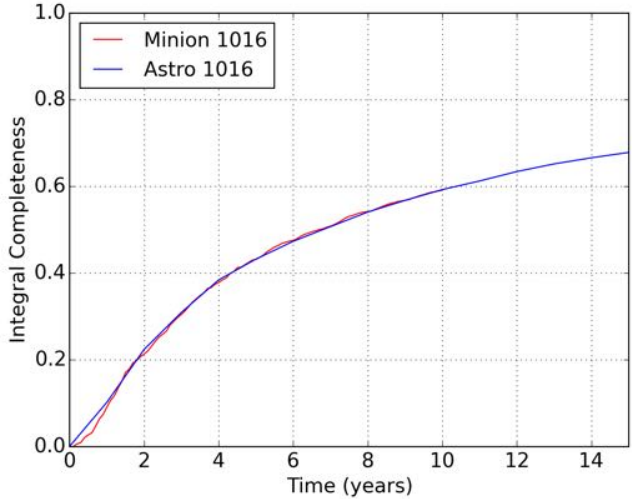

**Figure 49.** NEO completeness `minion_1016` and `astro_1016` surveys. Fill factor, fading and trailing loss were applied in low-density simulations with Bottke's NEO population.

### 3.3. *NEO-enhanced LSST Survey*

As described in Sec. 2.1, the `astro_1016` survey was designed to focus more heavily on the Northern Ecliptic Spur and to continue for 5 years beyond the baseline 10-year survey. Our simulation and analysis of the `astro_1016` survey reveals that at 10 years it provides a negligible improvement compared to the `minion_1016` survey, against which it should be compared (Figure 49). The only meaningful improvement from `astro_1016` is the extended survey, which leads to an additional 4.2% gain after 12 years and 8.7% after 15 years.

It is surprising that the `astro_1016` survey is not more productive. We speculate that this implies that extra survey time in the Northern Ecliptic Spur is an ineffective approach to increasing NEO productivity, and spending additional survey effort on different enhancements could prove more beneficial. However, to provide solid conclusions on this question would require a substantial effort that is beyond the intended scope of this study.

### 3.4. *Alternatives to Nightly Pairs*

Historically, the Minor Planet Center accepts only high-reliability tracklets from observers, and its internal linking processes assume that the false tracklet rate is low. This has been a reasonable assumption for past and current major surveys, which follow a cadence that naturally returns 3–5 detections per tracklet by repeatedly returning to the same field within a span of an hour or so. This survey approach is robust against false positives because the 3–5 detections in a tracklet must all be consistent with a common linear (or nearly linear) motion, substantially eliminating the possibility that one or more false detections could contaminate a tracklet.

LSST, on the other hand, is baselined to return only two detections per tracklet, which removes the possibility of checking for internal consistency among the elements of a tracklet. The result is a high rate of false tracklets (57% false in our full-density simulation) that is not suitable for submission to the MPC. LSST works past this by submitting only high-reliability three-night tracks to the MPC, which we have shown have a relatively high purity in our sims.

The LSST approach of obtaining nightly pairs is certainly more fragile for linking than a cadence that returns tracklets with three or more detections, but the fragility comes with the marked benefit of significantly increased sky coverage per night and hence a shorter return period, leading to more tracklets per observing cycle, which restores some measure of robustness and leads to increased discovery rates, so long as the linking problem can be managed. If, for whatever reason, and however unlikely, LSST cannot successfully link two-detection tracklets then it could conceivably be forced to alter its cadence to meet survey objectives.

Here we compare the performance of the pair-wise `enigma_1189` baseline survey with `enigma_1271` and `enigma_1266`, which are tuned to provide 3- and 4-visit cadences, respectively. For the 2-visit baseline survey, we required at least 2 detections for tracklet creation, in 3-visit baseline at least 3-detections per tracklet ("triples") and in 4-visit cadence at least 4 detections per tracklet ("quads").

We emphasize that the benefit of a cadence that produces triples or quads is that it eases the linkage challenge. It does not produce better orbits. Tracklets formed from three or more detections have far higher confidence than those obtained from pairs because with only two positions there is no independent corroboration of linear motion. Thus for pairs, the idea that a tracklet's detections are associated with a single moving object is a hypothesis to be tested by the linking engine, whereas with three or more detections the tracklet has a high likelihood of being real. Linking is challenging for pairs because of the high false tracklet rate and easy for triples or quads because there is no hypothesis testing involved.

In terms of orbit quality there is no appreciable difference between orbits derived from pairs or triples or quads. Each tracklet provides a position and rate of motion on the plane of sky, with no information on plane of sky acceleration (except for very close objects, which are rare enough to be ignored in this context). The orbit quality depends primarily on how many distinct nights

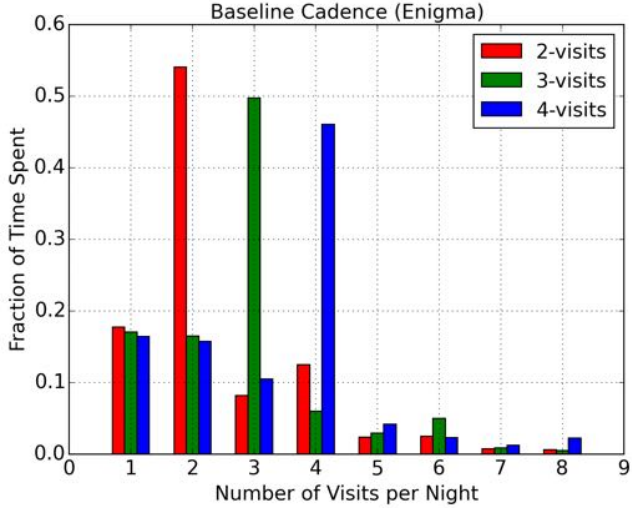

**Figure 50.** Baseline surveys optimized for 2, 3 and 4 visits and the histogram of time spent on fields visited N-times per night.

the object has been observed and the time interval between the first and last night.

Table 16 shows that visiting the same field more often per night predictably decreases the effective areal search coverage significantly, even though the alternate surveys have a similar number of fields observed in 10 years, similar limiting magnitudes, and similar inter-night survey patterns. Figure 50 shows that all three surveys also contain some fields that are visited fewer or more than the target number of visits per night. As mentioned earlier, singleton and Deep Drilling fields are not used in our simulations.

Figure 51 and Table 17 show the direct impact of the 3 and 4-visits cadence approach on NEO completeness. The completeness penalty is not as dramatic on the single tracklet level; however, due to reduced sky coverage the tracks are dramatically affected. Three and four-visit cadences could decrease the number of false tracklets significantly, but at the cost of a steep reduction in $C_{H<22}$. The figure shows that 3- and 4-visit cadences have a severe impact on NEO completeness in all stages, from detections to tracklets to tracks. At $H < 22$, the completeness for 12-day tracks falls from 58.6% for nightly pairs to 36.9% for 3-visit cadence to 19.0% for 4-visit cadence. The 4-visit cadence could be improved by accepting tracklets with detections on three out of the four visits, which would provide a performance somewhere in between than tabulated (and plotted) for the 3- and 4-visit cadences.

We note that triples or quads should dramatically reduce the false tracklet rate so that reliable tracks could likely be assembled from tracklets on only two nights.

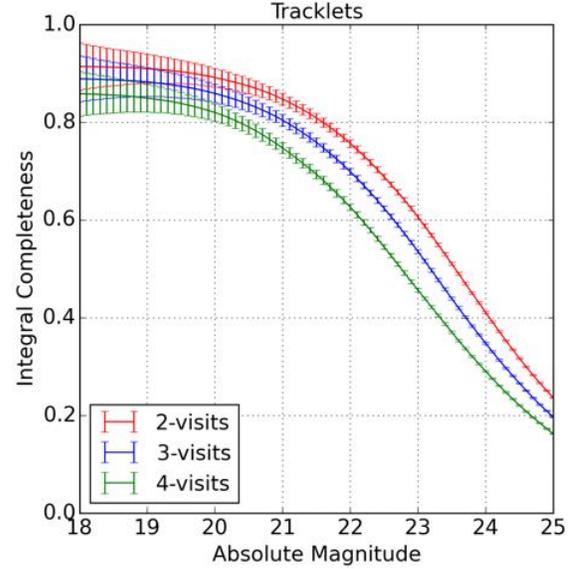

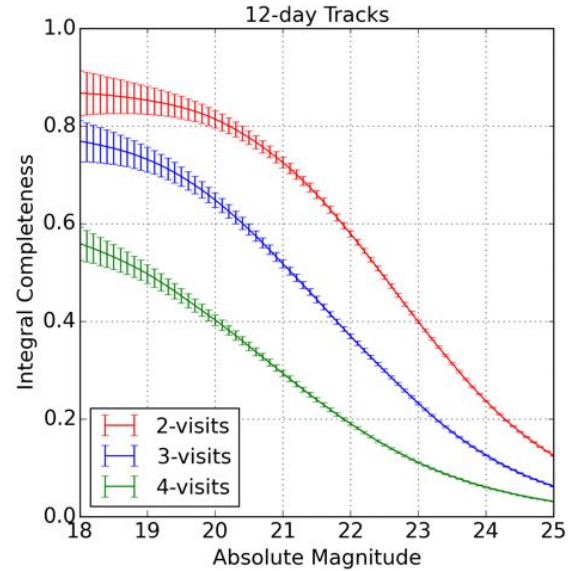

**Figure 51.** Integral completeness for 2, 3 and 4-visit optimized baseline surveys as a function of absolute magnitude. The plots show the completeness over 10 years for NEOs appearing in at least one tracklet (top) and in at least one 12-day, three-night track (bottom). Parameters applied: fading, fill factor=0.89, trailing loss, vignetting, colors, light curves.

This approach would lead to increased completeness; however, such two night tracks would inevitably lead to weak orbits that are not likely to meet survey objectives. In particular, two-night orbits are generally uncertain enough that MBAs and NEOs cannot be distinguished in general. As discussed in Sec. 3.2.5, this is already an issue for some three-night orbits, but for two-night orbits it is the norm. Table 18 indicates the level of NEO completeness that would be obtained for the four-visit enigma_1266 survey, assuming that a minimum of either three or four detections are available for each tracklet. We did not analyze the orbit quality of two night tracks, but the challenge here is that the shortest arcs (e.g., 3 days) may be readily linked but the orbits are likely to be highly uncertain, while for the longer arcs (e.g., 12 days) the orbit may still be unacceptably weak but the linkage problem may be difficult. The numbers in Table 18 should be compared with the 12-day, 3-night tracks completeness for enigma_1266, which is 19.1%. Even the most optimistic scenario in Table 18 significantly underperforms compared to the baseline enigma_1189.

**Table 16.** Comparison of 2, 3 and 4-visit optimized survey cadences.

| OpSim Desig. | No. visits | No. fields | Used fields | Mean $m_5$ in V-band | Effective sky coverage ($deg^2$) |
|---|---|---|---|---|---|
| enigma_1189 | 2 | 2,469,307 | 1,943,901 | 23.63 | 805,128 |
| enigma_1271 | 3 | 2,438,990 | 1,932,391 | 23.65 | 463,636 |
| enigma_1266 | 4 | 2,417,999 | 1,928,921 | 23.65 | 309,154 |

**Table 17.** NEO completeness of 2, 3 and 4-visit optimized surveys based on the enigma_1189 baseline, with fading, fill factor equal to 0.90, trailing losses, vignetting, S+C NEO colors, light curves and bright source removal represented by 1% penalty in fill factor.

| | | $C_{H<22}$ | | | | |
|---|---|---|---|---|---|---|
| OpSim Desig. | No. visits | Detection | Tracklet | 3 Tracklets | 12-day track | 20-day track |
| enigma_1189 | 2 | 81.6 | 75.7 | 67.9 | 58.0 | 61.0 |
| enigma_1271 | 3 | 81.3 | 69.9 | 57.1 | 37.0 | 43.1 |
| enigma_1266 | 4 | 80.6 | 62.6 | 45.2 | 19.1 | 25.7 |

**Table 18.** NEO completeness $C_{H<22}$ for two-night tracks based on the four-visit enigma_1266 survey, for a range of arc length requirements and for a requirement of either $\geq 3$ and $\geq 4$ detections per tracklet.

| Detections per | Arc Length | | | |
|---|---|---|---|---|
| tracklet | 3 days | 6 days | 9 days | 12 days |
| $\geq 3$ | 23.3 | 38.3 | 46.1 | 49.9 |
| $\geq 4$ | 11.2 | 22.3 | 29.9 | 35.5 |

### 3.5. Comparison with other results

#### 3.5.1. Grav et al. 2016

Grav et al. (2016) simulated the NEO and PHA performance of LSST for the baseline `enigma_1189` and 4-visit `enigma_1266` surveys. Additionally, they discussed the discovery performance of LSST combined with existing surveys and the proposed NEOCAM space-based survey. They used the same fields as this study (except that they did not remove singletons and Deep Drilling fields), the same limiting magnitudes and a 95% fill factor, which is higher than the currently anticipated 90.8%. The synthetic population was different, consisting of 20,000 NEOs down to a size threshold of 100 m. Also, Grav et al. (2016) allowed creation of tracklets in a velocity range between 0.011–48°/day, which is significantly broader than our range (0.05–2.0°/day). The spectral distribution of synthetic objects had an equal balance between C and S types, however, the color transformation to V-band was derived by slightly outdated specifications from SDSS (Ivezić et al. 2001). Their simulation process was similar to our low-density simulations where detections and tracklets were assembled into lists and built into assumed tracks through postprocessing. A track was *created* when 3 tracklets were detected within 12 days, with a maximum separation of 6 days from two of them, but for the 4-visit case, `enigma_1266`, only two 4-detection tracklets were required to build a track within 12 days. This is a significant difference from our three tracklets on three distinct nights over at most 12 or 20-days tracks. Even though Grav et al. (2016) were skeptical of the 2-visit cadence, given that it has never been tested in a high-density, complete pipeline, their 4-visit alternative only required two tracklets for a track. They also did not include trailing losses, vignetting, fading, light curve variation, and they assumed the linking efficiency to be 100%. Grav et al. (2016) derived that for the 2-visit cadence LSST will have a 63% completion of NEOs larger than 140 m and for PHAs they find $C_{D>140\,m} = 62\%$. In the 4-visit cadence, they reported $C_{D>140\,m} = 59\%$ for NEOs and 58% for PHAs. The completeness was presented with a ±1% uncertainty.

To make a robust comparison with this work, we ran a simulation with a Grav et al. (2016) population of NEOs (19,597) and PHAs (2,346). Our model included fading, trailing losses, vignetting and 90% fill factor. Table 19 and Figure 52 present the comparison. Our results are substantially consistent with Grav et al. (2016); however, we find that PHA completeness is higher than that of NEOs, while the converse is true for Grav et al. (2016).

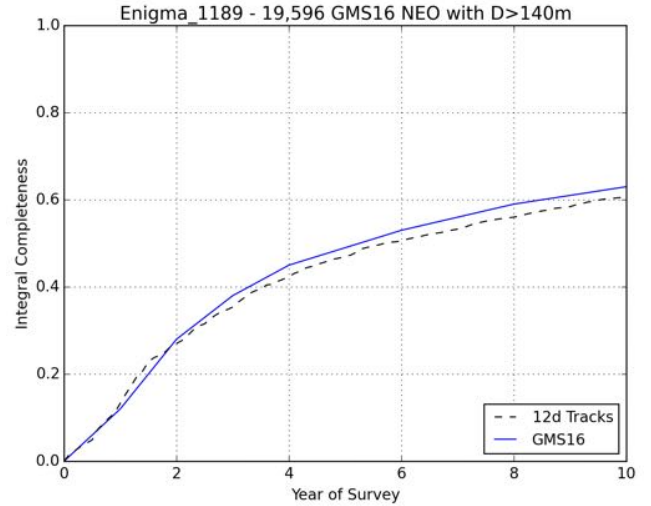

**Figure 52.** Comparison of 12-day tracks from the present study with the published Grav et al. (2016) results. Both simulations use `enigma_1189` and a compatible NEO population. The integral completeness is computed for $D > 140\,m$, rather than $H < 22$, which is used everywhere else in this report.

**Table 19.** 140 m integral completeness of NEAs and PHAs according to Grav et al. (2016) and the present study. The columns marked "w/others" incorporated discoveries from other (past and future) survey activity in the completeness estimate.

|       | This work | Grav et al. (2016) | | | |
|       | 2-det. | 2-det. | 2-det. w/others | 4-det. | 4-det. w/others |
|-------|--------|--------|-----------------|--------|-----------------|
| NEO   | 62.6%  | 63%    | 67%             | 59%    | 67%             |
| PHA   | 63.8%  | 62%    | 73%             | 58%    | 71%             |

### 3.5.2. *University of Washington/LSST*

Predictions of NEO and PHA discovery and detection rates have been done for LSST at the University of Washington as a part of the development and testing of the OpSim and the LSST MOPS. Ivezić et al. (2014) used a list of LSST pointings and a higher density model for the survey, using a diameter-limited PHA population of 800 objects. This analysis found 82% PHA completeness ($D > 140m$) that could be improved to 90% with the same cadence if stretched to 12 years.

Jones et al. (2017) performed a detailed computation of LSST performance using multiple parameters mentioned in this work as well as identical survey fields and patterns (`enigma_1189`, `minion_1016` and `astro_1016`). Their pipeline based on MOPS was working on the detection and tracklet level, predicting tracks in post-processing of the tracklet list, similar to our low-density simulations and providing metrics within their Metrics Analysis Framework (MAF). They used an H-delimited population based on Bottke et al. (2002). Their results are compared to ours in Table 20, where we see an excellent match, within a percentage point, for `minion_1016`. For `astro_1016`, the UW results show a $C_{H<22}$ slightly increased ($\sim 3\%$) completeness compared to the present study.

There were several modeling differences between the present report and Jones et al. (2017). For example, Jones et al. (2017) did not apply a rate-of-motion cutoff, which should lead to an increase in computed completeness. Also, they used all OpSim fields, without rejecting Deep Drilling and singletons, which should again lead to a slight increase in their modeled completeness. However, they considered all NEOs to be C type, and had a 1-hour restriction for a maximum tracklet duration which is half of our maximum duration. These two modeling effects could drive their estimated completeness down. Also, Jones et al. (2017) used a more sophisticated model for chip and raft gaps, masking the exact locations in the focal plane, whereas our work used only a random number as a fill factor to assign a detection probability to every detection. Collectively, these model variations readily explain the slightly different NEO completeness estimates reported by Jones et al. (2017).

### 3.6. *Overall LSST Performance and Uncertainties*

Throughout this study we have tested and analyzed an array of different survey models, including various OpSim runs, NEO population models and detection models. For what we consider our final result we select the most current LSST baseline survey, `minion_1016`, and the latest debiased NEO population estimate, namely

**Table 20.** Comparison of LSST NEO and PHA completeness $C_{H<22}$ in percentage for the 10-year `minion_1016` and the 15-year `astro_1016` surveys for Jones et al. (2017) and this work. For consistency, tabulated results are for 15-day tracks, and so the results differ from those in Table 21. Other aspects of the results from this study are 90.8% fill factor, with fading, trailing and the Bottke NEO population.

| Pop. | Survey | Jones et al. (2017) | This work |
|------|--------|---------------------|-----------|
| NEO | `minion_1016` | 60.7% | 61.6% |
| PHA | `minion_1016` | 65.6% | 64.9% |
| NEO | `astro_1016` | 69.1% | 65.6% |
| PHA | `astro_1016` | 74.3% | 71.5% |

Granvik's NEO model. Neither of these elements were available until our study was well underway. This "new" approach can be compared with the earlier alternative of `enigma_1189` with Bottke's NEO model, where we find that there is very little difference for $C_{H<22}$ performance. This is because of the compensating factors of Granvik's steeper size distribution, which provide a 2% increase in $C_{H<22}$, and the $\sim 0.25$ mag reduction in faint limit seen for `minion_1016`, which leads to a somewhat smaller drop in performance.

The comparison shows that `minion_1016` with Granvik's NEOS is slightly better than `enigma_1189` with Bottke's NEOS, by about 1.5% in $C_{H<22}$ after 10-years (Figure 53). The effect of using two different populations is more clear at different $H$ limits. Specifically, Granvik's population is more numerous at the small end, therefore, the `minion_1016` completeness is lower for $H < 25$ when compared to `enigma_1189`. Also, `enigma_1189` seems to be slightly more productive in the early years, while `minion_1016` catches up and passes `enigma_1189` later in the survey.

The completeness derived from our simulations has numerous error sources. The easiest to define are those associated with sampling, as shown by the error bars in in Figure 53. Other contributions from the various individual modeling details were derived step-by-step, and we discuss them below and summarize them in Table 21. The modeling details and our ad hoc approach for understanding their potential effect on $C_{H<22}$ are as follows:

- By using low-density simulations, the statistical sampling error is about 0.6%.

- Changing the fill factor by $+0/-2\%$ leads to a shift of $+0/-0.8\%$ in $C_{H<22}$.

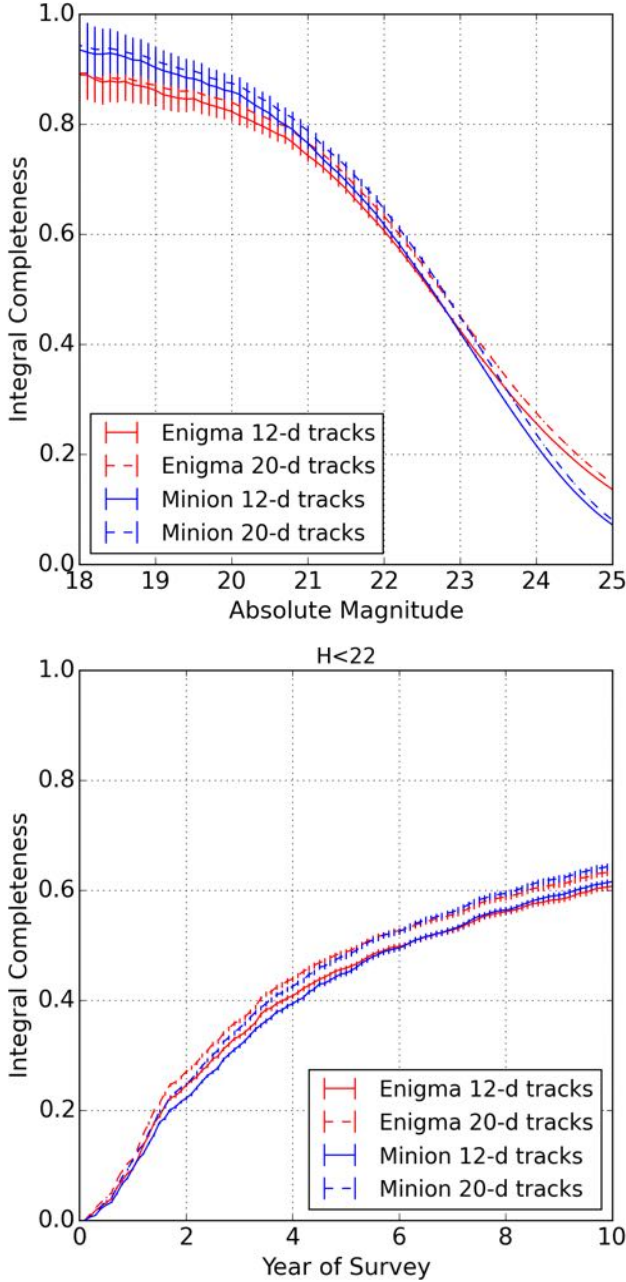

**Figure 53.** Integral NEO completeness for the old baseline (`enigma_1189`) with Bottke's NEO model versus the new baseline (`minion_1016`) with Granvik's NEO model as a function of H for 10 years (top) and time for $H < 22$ (bottom). These results reflect full-density, NEO-only simulations with 90.8% fill factor, fading included and detection trailing losses.

- The assumed width for detection fading was a $w = 0.1$. If this value is altered by 32%, then the completeness error from this source is 0.2%.

- Trailing losses are described by a detection loss function that is theoretical and not operationally

validated. If the detection loss penalties varies by 32% it leads to a 0.7% error in completeness.

- For vignetting, 32% variation in the magnitude loss leads to a 0.1% variation in completeness.

- We allow the color indices taken from SDSS and assumed in the study to vary by $1\sigma$ of their values, which affects completeness by 0.1%.

- We allow a 32% variation in light curve amplitude, which leads to a 0.2% effect on completeness.

- Bright source masking to reduce false positives from image differencing is expected to reduce the fill factor by 1%. If this value varies by $-0.5/+$1.0%, then the completeness error from this source is +0.2/−0.4%.

- The demonstrated linking efficiency for $H < 22$ NEOs was 93.6%, but we believe that much higher efficiencies are possible with the application of more computational resources and careful MOPS optimization. We suppose that the linking efficiency can vary by +6/−1% leading to +3.3/−0.6% variation in completeness.

- Using Granvik's NEO population instead of Bottke's leads to a 2% increase in $C_{H<22}$, and we take this as proxy for the uncertainty due to uncertainties in the NEO size distribution.

- The risk that the limiting magnitude will not be as good as the LSST models predict largely compensates the significant positive uncertainty associated with linking efficiency. We take the somewhat pessimistic view that the actual operational LSST $m_5$ can vary by +0.1/−0.25 leading the completeness to vary by +1.8/−4.4%.

We note that many of the foregoing stated uncertainties are more akin to sensitivity exercises than uncertainty estimates. In many cases we have no good statistical footing from which to infer the uncertainty in the inputs and so the corresponding uncertainty estimate relies heavily on judgement. Nonetheless, taken all together, these modeling effects lead to a $\pm 5\%$ uncertainty in the predicted value of $C_{H<22}$, and so, in light of this uncertainty, the difference between the old (`enigma_1189`/Bottke) and new (`minion_1016`/Granvik) simulations is negligible. Therefore we report our final LSST performance result as $C_{H<22} = 55 \pm 5\%$. Here we emphasize that the stated uncertainty is not a Gaussian 1-sigma error bar, but rather reflects the possibility of modeling systematics that could compromise the result by *up to* 5%.

**Table 21.** NEO completeness in percentage points as multiple parameters are being applied one-by-one (right column). $\Delta C$ denotes the difference between consecutive steps. The associated error in percentage points coming from uncertainty of the model is listed under the heading "Uncert." The overall uncertainty on completeness is about 5%. The `enigma_1189` results used the Bottke NEO model, and the `minion_1016` results are for Granvik's NEO model.

| Model Variation | enigma_1189 | | minion_1016 | | Uncert. | Remarks |
|---|---|---|---|---|---|---|
| | $C_{H<22}$ | $\Delta C$ | $C_{H<22}$ | $\Delta C$ | | |
| None | 65.87 | | 67.66 | | $\pm0.6$ | Assumes 100% fill factor, hard $m_5$ cutoff, no trailing losses, etc. |
| + fill factor (90.8%) | 63.18 | $-2.70$ | 65.07 | $-2.60$ | $^{+0.0}_{-0.8}$ | $90.8^{+0}_{-2}$% fill factor |
| + fading | 62.41 | $-0.76$ | 64.29 | $-0.78$ | $\pm0.2$ | fading width $0.1 \pm 0.032$ mag |
| + trailing loss | 60.30 | $-2.12$ | 61.75 | $-2.54$ | $\pm0.7$ | $\pm32$% of $\Delta m_{\text{trail}}$ |
| + vignetting | 59.91 | $-0.38$ | 61.29 | $-0.46$ | $\pm0.1$ | $\pm32$% of $\Delta m_{\text{vignette}}$ |
| + colors | 58.65 | $-1.26$ | 59.92 | $-1.37$ | $\pm0.1$ | 50-50 S vs. C, $\pm1\sigma$ in SDSS color indices |
| + light curves | 58.31 | $-0.35$ | 59.58 | $-0.35$ | $\pm0.2$ | $\pm32$% of light curve amplitude |
| + bright source removal | 57.96 | $-0.69$ | 59.24 | $-0.68$ | $^{+0.2}_{-0.4}$ | $1.0^{+1.0}_{-0.5}$% masked |
| + Linking efficiency | 54.25 | $-3.71$ | 55.45 | $-3.79$ | $^{+3.3}_{-0.6}$ | $93.6^{+6}_{-1}$% linking eff. |
| Population Model | | | | | $\pm2.0$ | Granvik vs. Bottke NEO models |
| Variation in faint limit | | | | | $^{+1.8}_{-4.4}$ | $m_5$ variation of $^{+0.10}_{-0.25}$ mag |
| OVERALL | 54.25 | $-11.62$ | 55.45 | $-12.21$ | $^{+4.3}_{-5.0}$ | $55 \pm 5$% adequately describes final result |

### 3.7. Including Prior and Ongoing Surveys

In all of the analyses presented so far, we have implicitly assumed that no NEOs have been discovered prior to the LSST survey. However, based on current NEO discovery statistics and published population models (e.g., Bottke et al. (2002); Harris & D'Abramo (2015); Granvik et al. (2016)), the population of NEOs with $H < 22$ is currently already complete to a level of approximately 30%. It is expected that this number will continue to increase until LSST becomes operational in 2022 and that at least some current or future NEO survey assets will continue to operate during the LSST mission. Therefore, some fraction of potential LSST NEO discoveries will have already been discovered by other surveys. Similarly, some fraction of objects missed by LSST will also have already been discovered by LSST. To make some prediction of what the completeness will be after the ten-year LSST survey we must make an accounting for the contributions of other surveys.

Spacewatch (McMillan & Spacewatch Team 2006) was the first CCD-based NEO search program, but the era of dedicated wide-field NEA surveys began approximately 18 years ago when LINEAR (Stokes et al. 2000) became operational. Since then improvements in instrumentation and techniques allowed fielding of other advanced ground based surveys like NEAT (Pravdo et al. 1999), LONEOS (Koehn & Bowell 1999), the Catalina Sky Survey and the Mt. Lemmon Survey (Larson et al. 2003; Christensen et al. 2012) and Pan-STARRS (Hodapp et al. 2004; Kaiser et al. 2010). The space-based NEO-WISE program (Wright et al. 2010) has also made significant contributions to cataloging NEOs.

We simulated past, current and presumed future NEO surveys starting 15 years in the past to 15 years in the future (2002–2032). In this simulation LSST starts operation 5 years from now, in 2022. Ephemerides of all objects where calculated once per day and an object was considered discoverable when all of the following criteria were met:

- Ecliptic latitude was between $\pm60°$
- Geocentric opposition-centered ecliptic longitude was between $\pm90°$
- Declination from $-30°$ to $+75°$
- Lunar elongation $> 90°$

Ground-based surveys are limited by weather and cannot cover the entire sky per night, therefore only a fraction $F_{\text{disc}}$ of discoverable objects were added to the catalog. Surveys have improved in time and so we slowly improve the detection model in 5-years steps:

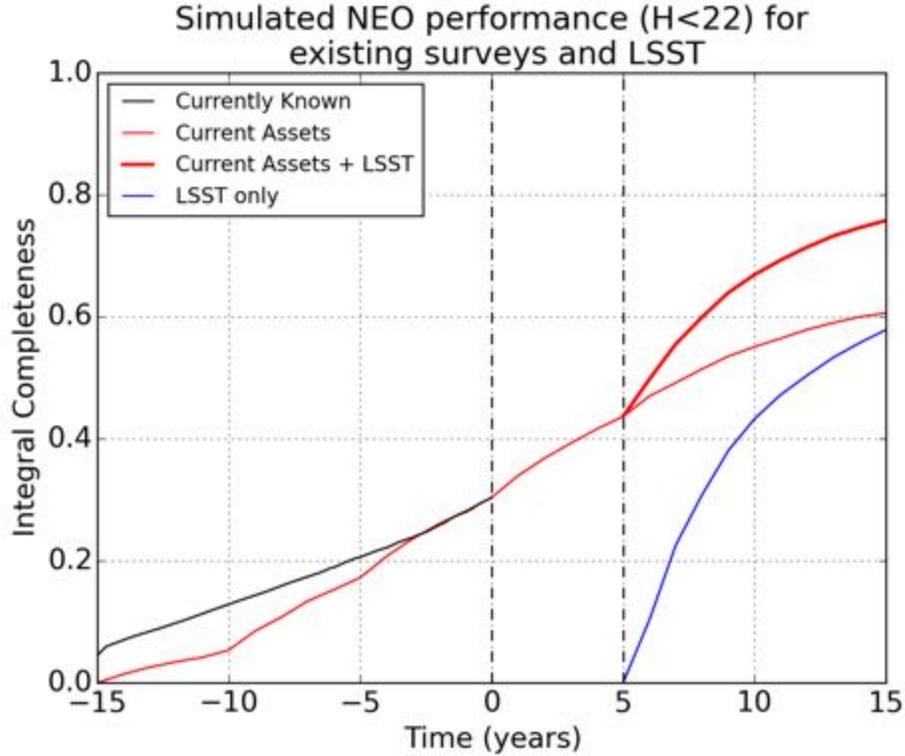

**Figure 54.** Estimate of integral NEO completeness ($H < 22$) for past, existing and future NEO survey assets combined. The black curve represents the real-world NEO discovery rate, the red curve the modeled assets, the blue curve the LSST performance as if no NEO are known at the beginning of LSST survey and the thick red curve represents the combined performance with LSST. The present date is at zero on the time axis, and LSST starts to operate in year 5. The LSST simulations are for `enigma_1189` with the Bottke NEO model, 90.8% fill factor, fading, trailing, 50-50 C & S type asteroids and vignetting. Linking losses (bright source masking and linking efficiency) are neglected.

I. A LINEAR-like era from 15–10 years ago with limiting magnitude $V = 19.5$ and $F_{disc} = 0.5$

II. A Catalina era with limiting magnitude of $V = 21.0$ the next 5 years and $F_{disc} = 0.6$

III. For the past 5 years Pan-STARRS1 and Mt. Lemmon Survey have operated at the limiting magnitude of $V = 21.5$ and $F_{disc} = 0.7$

IV. For the next 5 years from present, we expect the limiting magnitude to be $V = 22.0$ and we increase $F_{disc} = 0.8$ to account for improvements in the combination of Pan-STARRS1, Pan-STARRS2 and both Catalina surveys. Also, the southern declination limit is extended to $-45°$.

V. Starting 5 years from present we augment the previous search interval with the LSST survey and increase $F_{disc} = 0.9$ to account for continuing improvements for the other surveys.

Figure 54 shows the outcome of the rudimentary simulation, which is deliberately tuned to match the current estimated completeness at the current time. Our simple model predicts that 42% of NEOs with $H < 22$ will be discovered before LSST becomes operational, and that without LSST the current NEO surveys alone could achieve $C_{H<22} = 61\%$ by 2032, when the LSST survey is planned to conclude. We have shown above that LSST acting alone will achieve completeness of about 58% by itself (neglecting linking efficiency here), but when combined with past and other expected NEO search efforts, $C_{H<22}$ rises to 77%. This is not a high-fidelity analysis, but it shows that the combination of LSST with other ground-based search activity will increase $C_{H<22}$ by about 20% compared with the naive assumption of LSST starting with an empty catalog. Put another way, we project that LSST will provide a 16% increase in $C_{H<22}$ compared with the anticipated efforts of the existing NEO search programs.

## 4. CONCLUSIONS

This study report represents a major undertaking, collectively nearly two years of effort between the authors, as well as the support of colleagues at LSST and IPAC. And yet there remain open questions and potentially useful avenues for further investigation. We conclude the report by noting the study's the key findings and listing some areas worthy of additional investigation.

### 4.1. *Findings*

- Among the numerous modeling details that we investigated, few result in a significant effect on survey performance if reasonable care is take in developing the nominal survey model. A 2% reduction in fill factor or a major increase in trailing losses could lead to nearly a 1% reduction in $C_{H<22}$ for NEOs.

- We found a 2% difference in $C_{H<22}$ between the Bottke and Granvik NEO models, suggesting some dependence on the population size distribution. However, the Granvik model is based on vastly superior input dataset compared to the 15-year old Bottke model. One should not expect future population models will lead to such large swings in the size distribution, and thus $C_{H<22}$ should be more stable in this regard.

- The effective, operational limiting magnitude of the LSST survey is a crucial parameter. We find that $C_{H<22}$ degrades by $\sim 1.8\%$ for every 0.1 mag loss in sensitivity, making this the single largest source of model uncertainty in our completeness estimates.

- On a single lunation, full-density simulation, with NEOs, MBAs and false detections, we obtained a linking efficiency of 93.6% for $H < 22$ NEOs. Linking efficiency on the full population down to $H < 25$ was lower. If realized, this linking efficiency would represent a $\sim 4\%$ cut in $C_{H<22}$. However, we believe that, with modest revision and tuning of the MOPS linking algorithms and an appropriate allocation of computational resources, the linking efficiency can be significantly increased, probably to 99% or more.

- On the same simulation, after rudimentary orbit quality filtering, the derived NEO catalog was comprised of 96% correct linkages. The remaining 4% of linkages were almost exclusively incorrect MBA-MBA links, most of which should be eliminated over a longer duration simulation. Only 0.1% of orbits in the derived NEO catalog included false detections.

- We find that $C_{H<22} \simeq 55 \pm 5.0\%$ for the baseline `minion_1016` survey operating alone. This result assumes 12-day, 3-night tracks, and accounts for all of the considered modeling features and sources of error, including linking efficiency. 12-day track linking is very conservative; $C_{H<22}$ increases by 2–3% if the linking uses 20-day tracks in the baseline cadence.

- In agreement with Jones et al. (2017), we find that $C_{H<22}$ for PHAs is generally 3–4% greater than for NEOs. However, for a diameter-limited population provided by Grav et al. (2016) the difference between PHAs and NEOs appears minor, about +1% for this study and −1% for Grav et al. (2016).

- The `astro_1016` survey was developed by LSST as a special-purpose survey designed to enhance the NEO discovery rate, but our results show little improvement over the baseline for a ten-year survey. The old and new LSST baselines (`enigma_1189` and `minion_1016`, respectively) and the NEO-enhanced scenario (`astro_1016`) all provide similar NEO detection efficiencies, to within 1%, for a 10-year survey. Surveying for longer than 10 years increases $C_{H<22}$ by about 2% per year (1% per year if other surveys are taken into consideration).

- The three- and four-visit optimized LSST cadences that we tested had a dramatically reduced NEO completeness relative to that obtained for the two-visit baseline cadences. This result relies on the modeling hypotheses assumed throughout this study and assumes that tracklets in three distinct nights within 12 days are required for cataloging. The performance loss associated with these alternate cadences could be significantly eased if the cataloging requirement was for tracklets on only two nights, rather than three, but it is doubtful that such two-nighters would have high enough orbit quality to meet cataloging objectives.

- When LSST becomes operational in 2022, about 42% of NEOs with $H < 22$ will have been discovered with the current assets. Without LSST, current assets could discover 61% of the catalog during the LSST era. With LSST and other surveys combined, $C_{H<22}$ should reach 77% by the end of 10-year mission in 2032, assuming 12-day tracks with `enigma_1189` and neglecting linking losses.

- Assembling the foregoing completeness results, including the contribution of other surveys, the post-LSST $C_{H<22}$ should reach 80% for PHAs in ten years, and slightly more if 20-day tracks are linked. In our judgement, the $H < 22$ PHA catalog is likely to approach $\sim 85\%$ completeness, but probably not 90%, after 12-15 years of LSST operation. The NEO-enhanced cadence that we tested (`astro_1016`) did not provide a meaningful performance improvement relative to the baseline.

- Some enhancements to MOPS are needed in the linking stage to eliminate duplicate and false orbits. This includes improving the orbit quality filter and tuning of the attribution, precovery and orbit-orbit identification modules. Of particular importance is a thoughtful optimization of kd-tree track building. This would increase the linking efficiency and thus increase the number of cataloged NEOs.

## 4.2. *Future Work*

The following items could benefit from further investigation and analysis.

- It may be worthwhile to test higher rates of false positives. The upper bound from the Slater et al. (2016) analysis is only $2\times$ greater than that used here. While we do not believe that factors of a few in the false detection rate would alter the fundamental conclusions of our study, it could be interesting—though computationally expensive—to quantify the false detection rate that would actually compromise linking in the baseline survey.

- To meet the goal of demonstrating successful linking this study deliberately set a low bar, 12 days, for track generation. It would be worthwhile to measure the linking efficiency for 15- and 20-day tracks, in particular after a careful tuning of MOPS.

- We have assumed realistic levels of astrometric error in the linking, but these are somewhat lower than has been delivered so far by other NEO surveys. Larger errors increase the linkage challenge and may reduce the accuracy of the NEO catalog. This issue could be explored by testing more conservative astrometric error models, but again we do not believe our findings are sensitive to modest changes in the assumptions.

- We assumed a moderate velocity cut off of $2°$/day in order to avoid linking difficulties. However, by making use of the detection trailing information in the tracklet generation process there is, in principle, no reason to limit the rate at all. If demonstrated, this enhancement should increase completeness by $\sim 1\%$.

- On a related topic, our key linking simulation used the exact velocity of a trailed detection when the velocity was $1.2$–$2°$/day. Velocity measurement errors were not included, which is not wholly realistic. At those velocities, which are the worst case, the trail length is $1.6$–$2.7$ arcsec, or $2$–$3\times$ the typical PSF, and so we believe that velocity errors should not compromise the tracklet generation process. Even so, future simulations could verify that this issue does not impact linking efficiency.

- A large and important task is the overall tuning of MOPS for linking, including tracklet generation, attribution, precovery, kd-tree linking, initial orbit determination, and orbit quality filtering. While this is well beyond the scope of this study, the experienced gained as a result should prove beneficial in completing this task.

- This study linked a full-density detection list over only a single observing cycle. One of the results was that main belt confusion led to a minor degradation in linking accuracy and also some short-arc MBA interlopers appearing in the NEO catalog. Longer linking runs should reduce the confusion, first by eliminating erroneous MBA-MBA linkages and second by extending the MBA arcs, which will improve the correctly-linked but weak orbits. Longer linking simulations, covering a few months to up to 15 months (roughly the MBA synodic period with Earth) would reveal the extent that main-belt confusion would automatically resolve itself, and may prove informative for MOPS tuning. Late in this study we completed a run of three observing cycles but have so far not analyzed the results.

*Acknowledgments*

The Moving Object Processing System was crucial to the successful completion of this study. This tool is the product of a massive development effort involving many contributors, whom we do not list here but may be found in the author list of Denneau et al. (2013). This report identifies a few deficiencies in MOPS, but our remarks should not be viewed in a pejorative sense. The software has so far never been fielded for its designed purpose, and we expect that minor improvements and tuning can resolve the issues that we have mentioned. We thank Larry Denneau (IfA, Univ. Hawaii) for his tremendous support in installing and running the MOPS software.

This study benefited from extensive interactions with Zeljko Ivezic, Lynne Jones and Mario Juric, all from the University of Washington. As members of the LSST project, they provided vital guidance in understanding the performance and operation of LSST. They also provided important insight into the expected interpretation and reliability of LSST data. They ensured that some of the OpSim runs required to fulfill study objectives were generated by LSST. And they reviewed with us their early results on DECam image processing, which allowed us to include credible image differencing artifacts in the simulated LSST detection stream.

Tom Statler (NASA HQ) provided valuable guidance in the development and documentation of this study.

Davide Farnocchia (JPL) supported the systematic ranging analyses of linking products described in this report.

George Helou and Frank Masci (IPAC, Caltech) collaborated in the study design and independently tested some of the MOPS output data products.

Mikael Granvik (Univ. Helsinki) kindly provided an early version of the Granvik et al. (2016) NEO population model, which was used extensively in this study.

Tommy Grav (PSI) provided an instantiation of the Grav et al. (2016) NEO population, which allowed us to conduct a direct comparison with their results.

## APPENDIX

### A. LIST OF SIMULATIONS

### B. PARAMETER SENSITIVITY IN $KD$-TREE LINKING

Figures 55–62 depict the sensitivity of several key linking performance parameters to the selection of three key kd-tree linking parameters, namely `vtree_thresh`, `pred_thresh`, and `plate_width`. The various plots show the results for `linkTracklets` runs on a single dense field in OC28 from the `enigma_1189` survey. The following list itemizes the various plots.

- Total number of candidate tracks (Figure 55).

- Total number of clean NEO tracks (Figure 56).

- Ratio of all candidate tracks to clean (MBA and NEO) tracks (Figure 57).

- Ratio of Bad candidate tracks (those with at least one false detection) to Clean tracks (Figure 58).

- Ratio of Non-synthetic candidate tracks (those with only false detections) to Clean tracks (Figure 59).

- Ratio of Mixed candidate tracks (those with more than one distinct object) to Clean tracks (Figure 60).

- CPU load for the `linkTracklets` run (Figure 61).

- Memory load for the `linkTracklets` run (Figure 62).

**Table 22.** List of performed simulations. NEO and PHA populations are based on the Bottke model unless specified otherwise.

| name | density | orbit type | count | survey type | visits | length | MOPS stage |
|---|---|---|---|---|---|---|---|
| astro_pha | low | PHA | 3,000 | astro_1016 | 2 | 15 years | TRACKLET |
| minion_pha | low | PHA | 3,000 | minion_1016 | 2 | 10 years | TRACKLET |
| 3oc | high | NEO | 250,000 | enigma_1189 | 2 | 3 months | LINKOD |
| astro | low | NEO | 3,000 | astro_1016 | 2 | 15 years | TRACKLET |
| comp | high | NEO (Granvik) | 800,000 | enigma_1189 | 2 | 1 month | LINKOD |
| granvik | low | NEO (Granvik) | 800,000 | minion_1016 | 2 | 10 years | TRACKLET |
| grav | low | NEO (Grav) | 20,000 | enigma_1189 | 2 | 10 years | TRACKLET |
| L | high | MB+NEO (Granvik)+noise | 14 million | enigma_1189 | 2 | 3 months | LINKOD |
| large | high | MB+NEO | 14 million | enigma_1189 | 2 | 1 month | TRACKLET |
| NEO_1 | low | NEO | 3,000 | enigma_1189 | 2 | 10 years | TRACKLET |
| NEO_10y | high | NEO | 250,000 | enigma_1189 | 2 | 10 years | LINKOD |
| NEO_3f | high | NEO | 250,000 | enigma_1266 | 3 | 10 years | TRACKLET |
| NEO_4 | low | NEO | 3,000 | enigma_1271 | 4 | 10 years | TRACKLET |
| NEO_H | low | NEO | 3,000 | enigma_1189 | 2 | 10 years | TRACKLET |
| NEO_H3 | low | NEO | 3,000 | enigma_1271 | 3 | 10 years | TRACKLET |
| newminion | low | NEO | 3,000 | minion_1016 | 2 | 10 years | TRACKLET |
| nminion | low | NEO | 3,000 | minion_1016 | 2 | 10 years | TRACKLET |
| oc28_10 | high | NEO+0.1 MB | 1.6 million | enigma_1189 | 2 | 3 months | LINKOD |
| oc28_100 | high | NEO+MB | 14 million | enigma_1189 | 2 | 3 months | LINKOD |
| oc60 | high | NEO+0.1 MB | 1.6 million | enigma_1189 | 2 | 1 month | TRACKLET |
| pha | low | PHA | 3,000 | enigma_1189 | 2 | 10 years | TRACKLET |

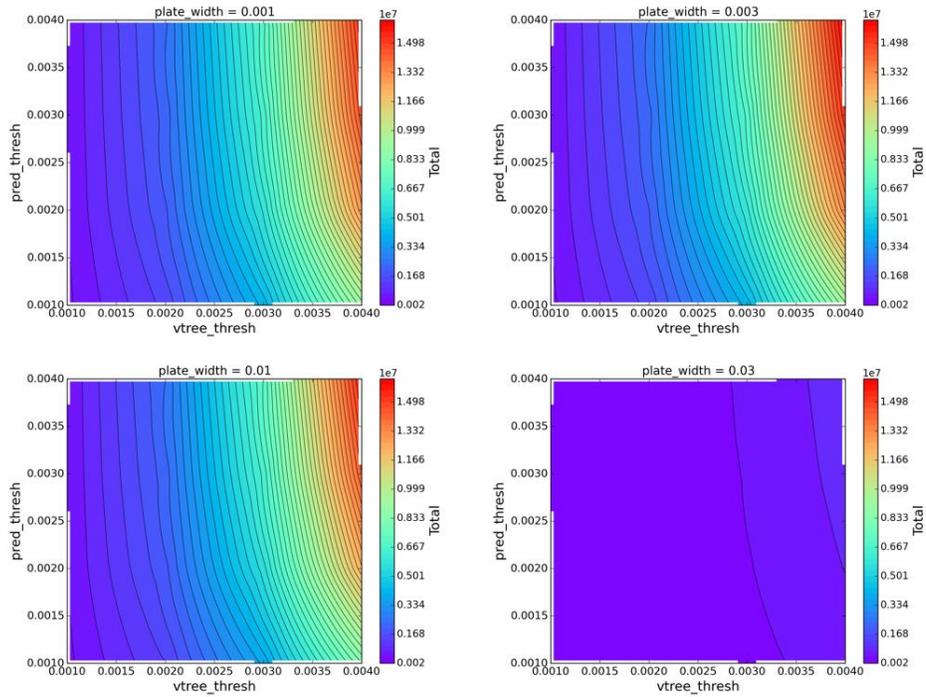

**Figure 55.** Total number of candidate tracks derived for a single, dense field as a function of the `vtree_thresh`, `pred_thresh` and `plate_width` kd-tree linking parameters.

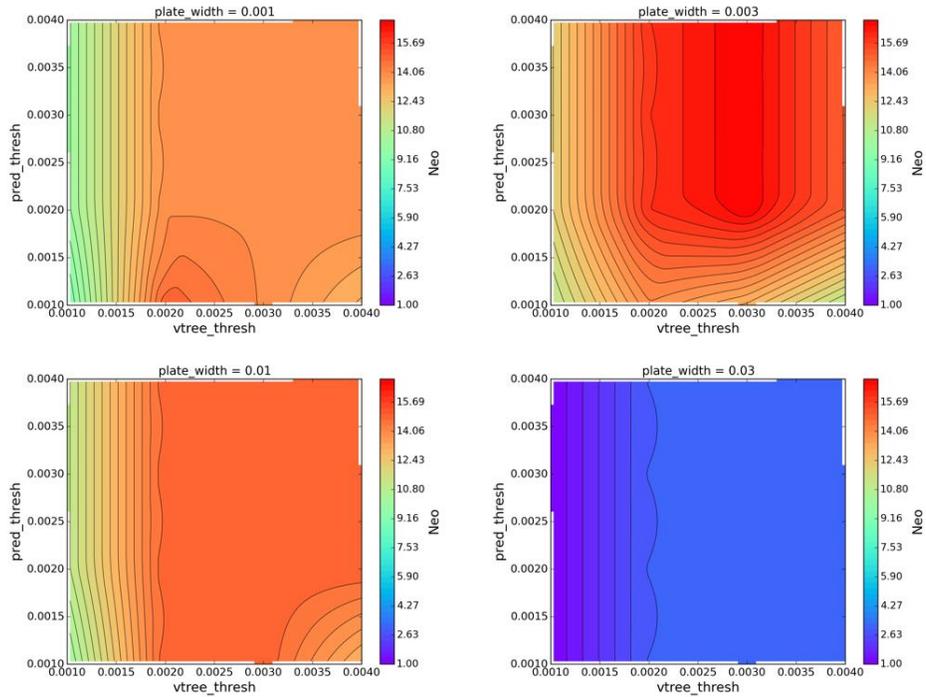

**Figure 56.** Total number of clean NEO tracks derived for a single, dense field as a function of the `vtree_thresh`, `pred_thresh` and `plate_width` kd-tree linking parameters.

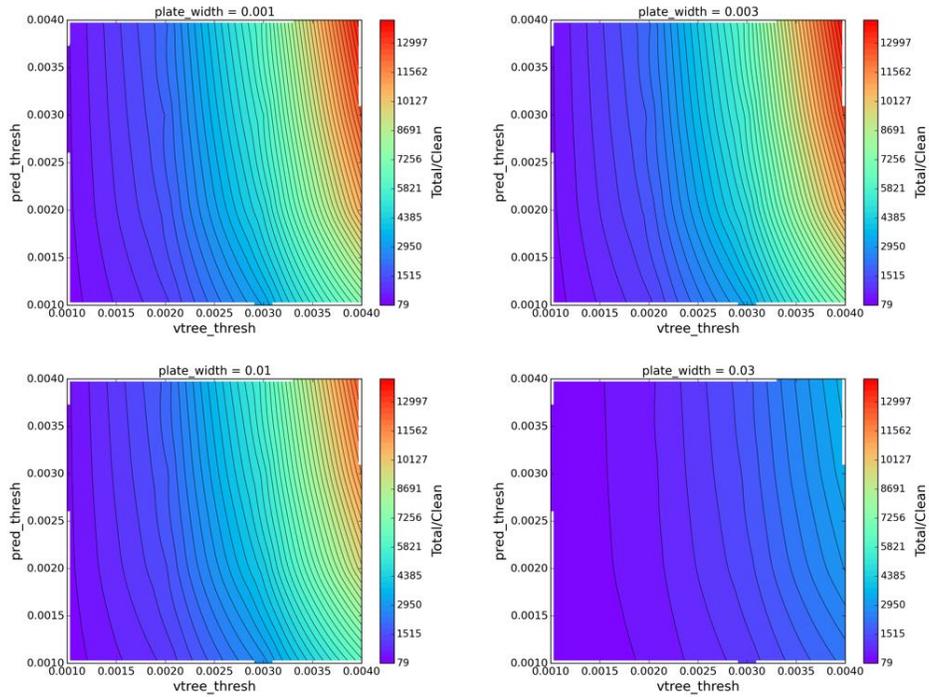

**Figure 57.** Ratio total number of tracks to number of CLEAN tracks derived for a single, dense field as a function of the `vtree_thresh`, `pred_thresh` and `plate_width` kd-tree linking parameters.

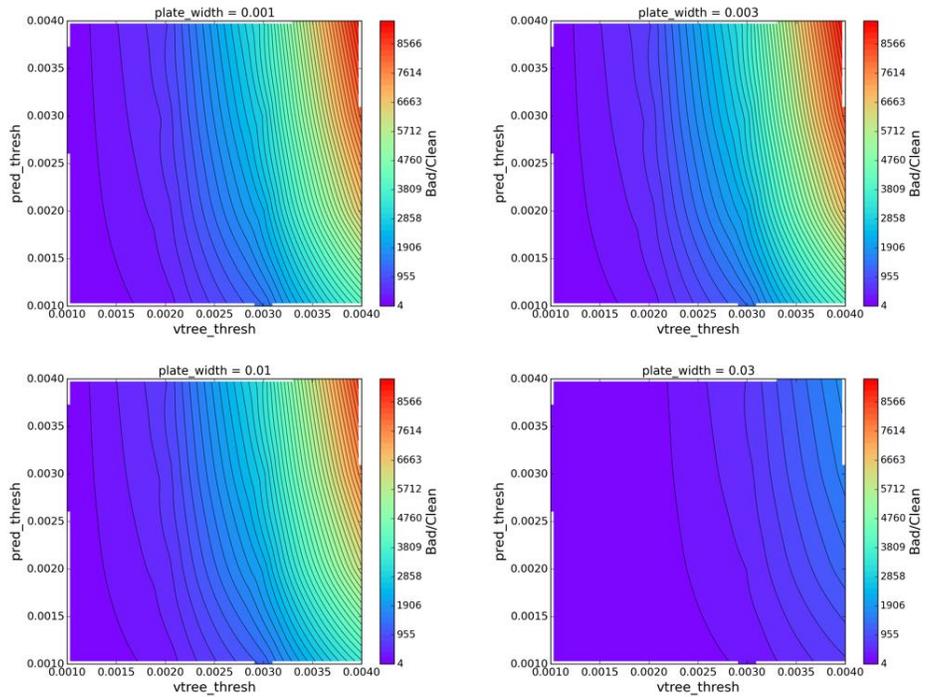

**Figure 58.** Ratio number of BAD tracks to number of CLEAN tracks derived for a single, dense field as a function of the `vtree_thresh`, `pred_thresh` and `plate_width` kd-tree linking parameters.

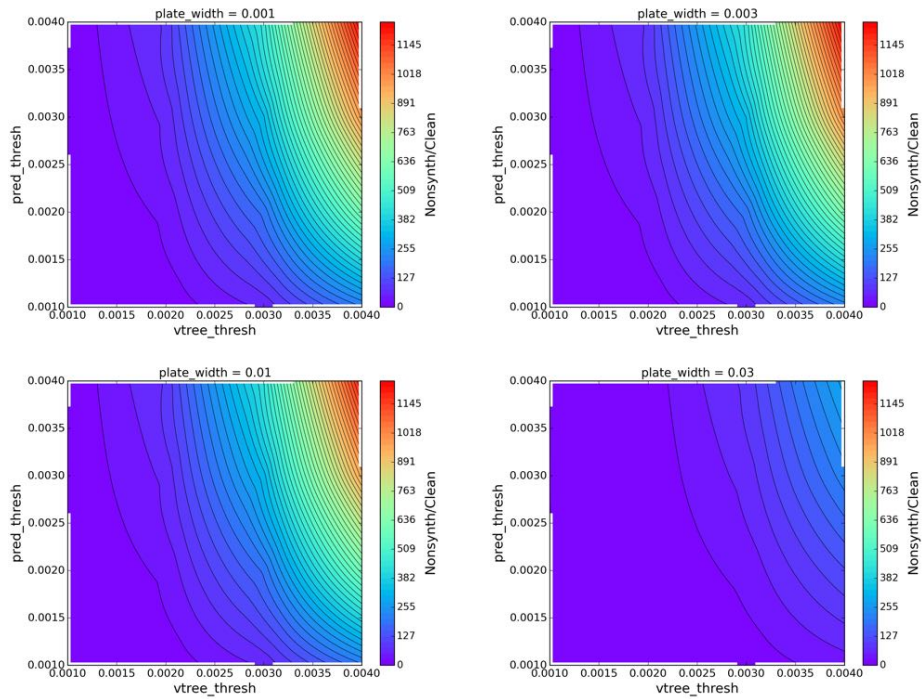

**Figure 59.** Ratio number of NONSYNTH tracks to number of CLEAN tracks derived for a single, dense field as a function of the `vtree_thresh`, `pred_thresh` and `plate_width` kd-tree linking parameters.

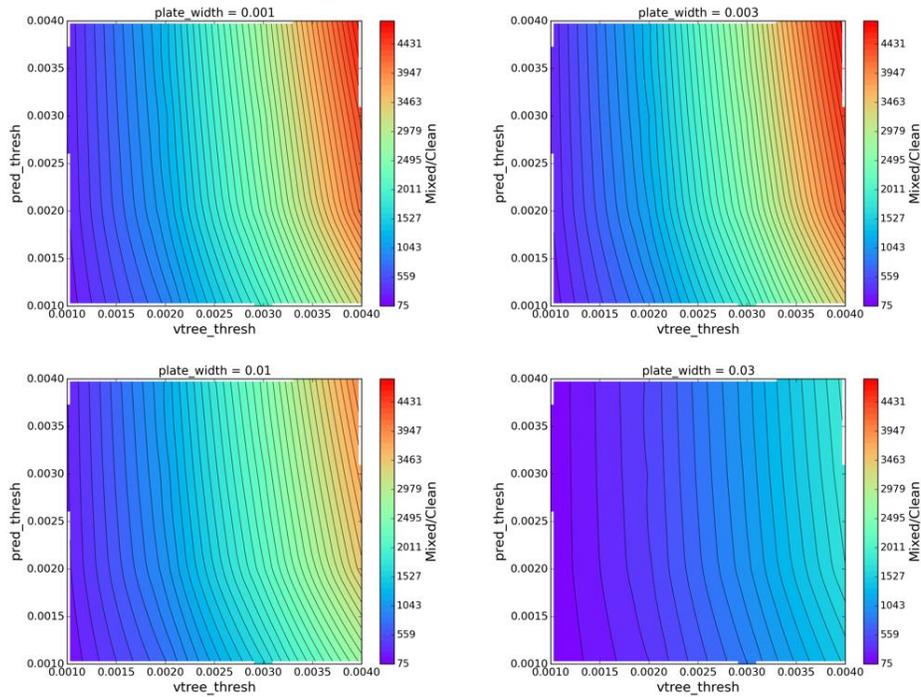

**Figure 60.** Ratio number of MIXED tracks to number of CLEAN tracks derived for a single, dense field as a function of the `vtree_thresh`, `pred_thresh` and `plate_width` kd-tree linking parameters.

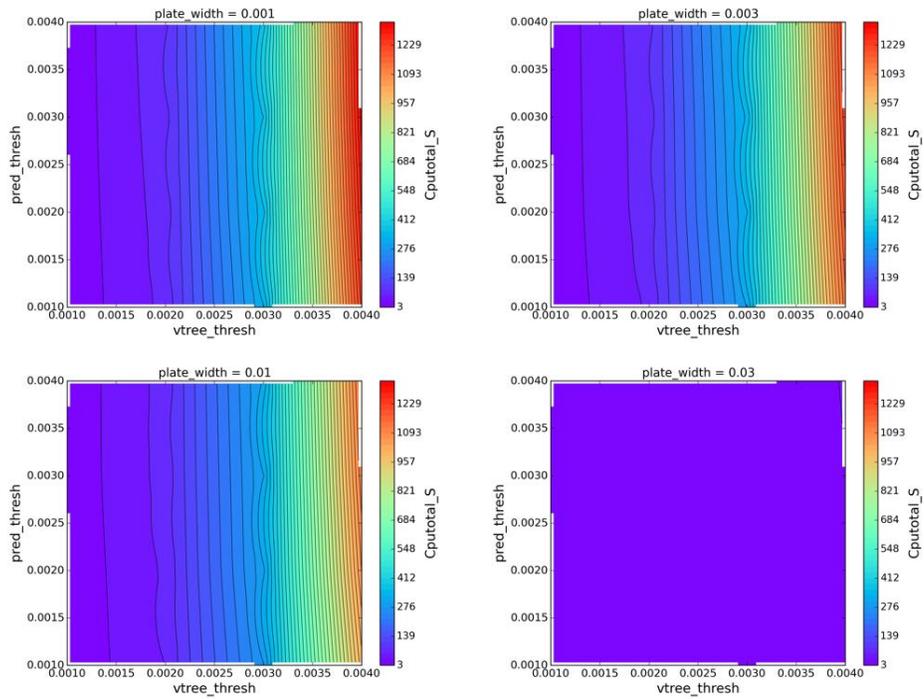

**Figure 61.** CPU time for running `linkTracklets` on a single, dense field as a function of the `vtree_thresh`, `pred_thresh` and `plate_width` kd-tree linking parameters.

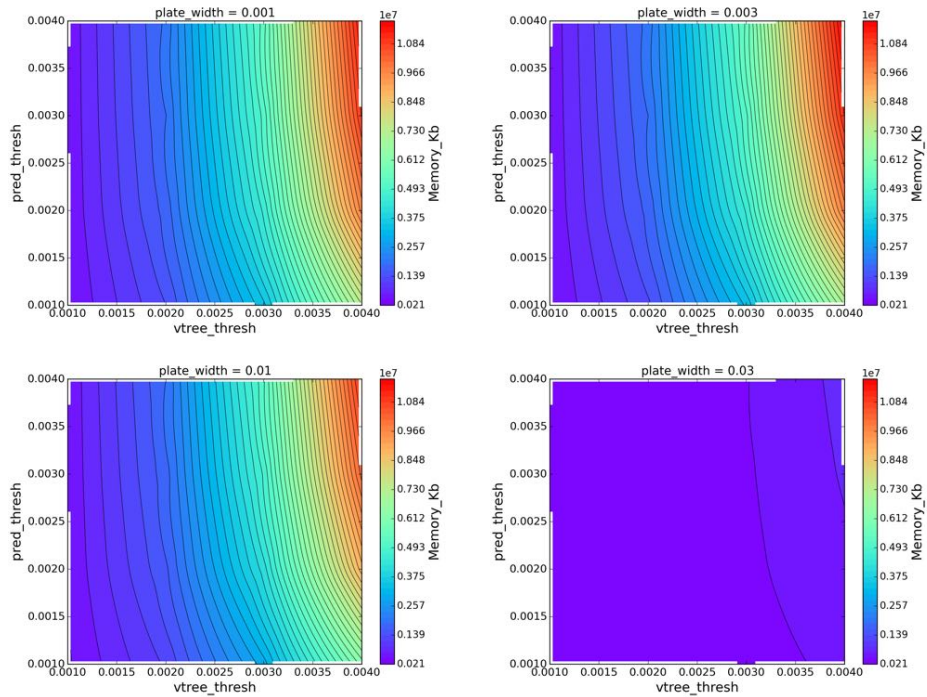

**Figure 62.** Memory usage of `linkTracklets` for a single, dense field as a function of the `vtree_thresh`, `pred_thresh` and `plate_width` kd-tree linking parameters.